\documentclass[showpacs,prc,nofootinbib,floatfix]{revtex4}

\usepackage{graphicx}
\usepackage{latexsym}
\usepackage{hhline}
\usepackage{amssymb}
\usepackage{wasysym}

\sloppypar

\newcommand{\be}{\begin{equation}}
\newcommand{\ee}{\end{equation}}
\newcommand{\bea}{\begin{eqnarray}}
\newcommand{\eea}{\end{eqnarray}}
\newcommand{\ba}{\begin{array}}
\newcommand{\ea}{\end{array}}
\newcommand{\bi}{\begin{itemize}}
\newcommand{\ei}{\end{itemize}}
\newcommand{\mi}{\mbox i}

\newcommand{\refe}[1]{(\ref{#1})}

\newcommand{\mcf}{{\mathcal F}}
\newcommand{\mci}{{\mathcal I}}
\newcommand{\mck}{{\mathcal K}}
\newcommand{\mcl}{{\mathcal L}}

\newcommand{\mcp}{{\mathcal P}}

\newcommand{\mct}{{\mathcal T}}
\newcommand{\mcv}{{\mathcal V}}
\newcommand{\difd}{\mathrm d}
\newcommand{\vt}{\vartheta}
\newcommand{\vp}{\varphi}
\newcommand{\ve}{\varepsilon}
\newcommand{\ra}{\rightarrow}

\renewcommand{\vec}[1]{\mbox{\boldmath $#1 \!\!$ \unboldmath}}
\renewcommand{\slash}{/ \!\!\!\!\,}
\newcommand{\smallfrac}[2]{\mbox {$\frac{#1}{#2}$}}
\newcommand{\avec}[1]{\: \: \! \! {\mathrm #1}}
\newcommand{\foh}{\frac{1}{2}}
\newcommand{\fot}{\frac{1}{3}}
\newcommand{\fof}{\frac{1}{4}}

\newcommand{\fth}{\frac{3}{2}}
\newcommand{\ffh}{\frac{5}{2}}
\newcommand{\fvh}{\frac{\vartheta}{2}}
\newcommand{\sqt}{\sqrt 2}
\newcommand{\sqth}{\sqrt 3}
\newcommand{\sfoh}{\smallfrac{1}{2}}
\newcommand{\sfot}{\smallfrac{1}{3}}

\newcommand{\sfth}{\smallfrac{3}{2}}
\newcommand{\sftt}{\smallfrac{2}{3}}
\newcommand{\umat}{1 \! \! 1}

\newcommand{\etp}{\, \varepsilon \! \cdot \! p \,}
\newcommand{\etpp}{\, \varepsilon \! \cdot \! p' \,}

\newcommand{\JPR}[3]{Phys. Rev. {\bf #1}, #2 (#3)}
\newcommand{\JPS}[3]{Phys. Scr. {\bf #1}, #2 (#3)}
\newcommand{\JPRL}[3]{Phys. Rev. Lett. {\bf #1}, #2 (#3)}
\newcommand{\JPL}[3]{Phys. Lett. {\bf #1}, #2 (#3)}
\newcommand{\JPRC}[3]{Phys. Rev. C {\bf #1}, #2 (#3)}
\newcommand{\JPRD}[3]{Phys. Rev. D {\bf #1}, #2 (#3)}
\newcommand{\JNP}[3]{Nucl. Phys. {\bf #1}, #2 (#3)}
\newcommand{\JNPP}[3]{Nucl. Phys. B, Proc. Suppl. {\bf #1}, #2 (#3)}
\newcommand{\JSP}[3]{Sov. J. Nucl. Phys. {\bf #1}, #2 (#3)}

\newcommand{\JAPP}[3]{Acta Phys. Pol. {\bf #1}, #2 (#3)}
\newcommand{\JEPJA}[3]{Eur. Phys. J. A {\bf #1}, #2 (#3)}
\newcommand{\JEPJC}[3]{Eur. Phys. J. C {\bf #1}, #2 (#3)}

\newcommand{\JZP}[3]{Z. Phys. {\bf #1}, #2 (#3)}
\newcommand{\JJP}[3]{J. Phys. {\bf #1}, #2 (#3)}
\newcommand{\Jpin}[3]{$\pi N$-Newsletter {\bf #1}, #2 (#3)}
\newcommand{\JPPNP}[3]{Prog. Part. Nucl. Phys. {\bf #1}, #2 (#3)}
\newcommand{\ibid}[3]{{\it ibid.} {\bf #1}, #2 (#3)}

\newcommand{\JYF}[3]{Yad. Fiz. {\bf #1}, #2 (#3)}
\newcommand{\JPTP}[3]{Prog. Theor. Phys. {\bf #1}, #2 (#3)}

\begin{document}

\title{Vector meson production and nucleon resonance analysis in a
  coupled-channel approach for energies $m_N < \sqrt s < 2$ GeV \\
  II: photon-induced results}

\author{G. Penner}
\email{gregor.penner@theo.physik.uni-giessen.de}
\author{U. Mosel}
\affiliation{Institut f\"ur Theoretische Physik, Universit\"at Giessen, D-35392
Giessen, Germany}


\begin{abstract}
We present a nucleon resonance analysis by simultaneously considering
all pion- and photon-induced experimental data on the final states
$\gamma N$, $\pi N$, $2\pi N$, $\eta N$, $K\Lambda$, $K\Sigma$, and
$\omega N$ for energies from the nucleon mass up to $\sqrt s = 2$
GeV. In this analysis we find strong evidence for the resonances
$P_{31}(1750)$, $P_{13}(1900)$, $P_{33}(1920)$, and
$D_{13}(1950)$. The $\omega N$ production mechanism is dominated by
large $P_{11}(1710)$ and $P_{13}(1900)$ contributions. In this second
part we present the results on the photoproduction reactions and the
electromagnetic properties of the resonances. The inclusion
of all important final states up to $\sqrt s = 2$ GeV allows for
estimates on the importance of the individual states for the GDH sum
rule.
\end{abstract}


\maketitle

\section{Introduction}

The reliable extraction of nucleon resonance properties from experiments 
where the nucleon is excited via either hadronic or electromagnetic 
probes is one of the major issues of hadron physics. The goal is 
to be finally able to compare the extracted masses and partial-decay widths 
to predictions from lattice QCD (e.g., Ref. \cite{flee}) and/or quark models 
(e.g., Refs. \cite{capstick,riska}).

With this aim in mind, in Refs. \cite{feusti98,feusti99} we developed a
unitary coupled-channel effective Lagrangian model that 
incorporated the final states $\gamma N$, $\pi N$, $2\pi N$, $\eta N$,
and $K \Lambda$, and was used for a simultaneous analysis of all
avaible experimental data on photon- and pion-induced reactions on the 
nucleon. The premise is to use the \textit{same Lagrangians} for the
pion- and photon-induced reactions, allowing for a consistent analysis,
thereby generating the background dynamically from $u$ and
$t$ channel contributions without new parameters. In the preceding
paper \cite{pm1}, called PMI in the following, we presented 
the results of the extension of the model space to center-of-mass
energies of $\sqrt s = 2$ GeV, which requires the additional
incorporation of the final states $\omega N$ and $K\Sigma$. 
The ingredients mandatory for a unitary description of all the above 
final states and the results on the pion-induced reactions have been
discussed both for calculations where only the pion-induced reactions
were considered and calculations where pion- and photon-induced
reactions were considered. In this paper, we concentrate on the
photoproduction reactions. 

For the photoproduction of the newly incorporated channels $\omega N$
and $K\Sigma$, almost all models in the literature are based on
single-channel effective
Lagrangian calculations ignoring rescattering effects (often called
``$T$-matrix models''). Especially the inclusion of nucleon Born
contributions for the $\omega$ production mechanism in these models
has led to an overestimation of the data for energies above $\sim
1.77$ GeV, and only either the neglect of these diagrams or very soft
form factors has resulted in a rough description of the
experimental data. In the first calculation on $\omega$
photoproduction, performed by Friman and Soyeur
\cite{frimansoyeur}, a rough description of the experimental data was
achieved by only including $\pi$ and $\sigma$ $t$-channel 
exchange. In the model of Oh and co-workers \cite{oh,oh01} the nucleon
contributions are damped by rather soft form factors [$\Lambda_N =
0.5-0.7$ GeV using $F_p$, Eq. \refe{formfacp}]. A similar
observation was made in the model of Babacan {\it et al.}
\cite{babacan}, where the Born contributions were not damped 
by soft form factors, but a very small $\omega NN$ coupling constant
was extracted ($g_{\omega NN} \leq 1$). Hence in both models,
the Born contributions are effectively neglected. Since Babacan {\it et
al.} did not include any baryon resonances, the effective reaction
process is almost purely given by $t$-channel exchanges, and is thus close
to the model of Friman and Soyeur. Oh {\it et al.}, however, included
baryon resonances by using nonrelativistic Breit-Wigner descriptions with vertex
functions taken from the quark model of Capstick \cite{capstick} and
thus did not consistently generate a $u$-channel background. An
imaginary part of the amplitude was only taken into account via
total widths in the denominator of the implemented Breit-Wigner
resonance description. In a similar way resonances were also 
included in the effective Lagrangian quark model of Zhao and co-workers
\cite{zhao,zhao01,zhao02} on $\omega$ photoproduction. However, none
of these models 
on $\omega$ photoproduction included rescattering effects. Only in
the most recent two works of Oh and co-workers \cite{oh} did the authors
start to consider the coupled-channel effects of intermediate $\pi N$
and $\rho N$ states. 

This restriction to a single-channel analysis is a fundamental 
weakness of all the $T$-matrix models. Although the above models
on $\omega N$ and also single-channel analyses on $K\Lambda$ or
$K\Sigma$ \cite{adelseck,martbenn,mart,lee01,cheoun} photoproduction
aim to provide a tool for the search and identification of missing
resonances, an inherent problem of such an extraction is 
ignored: Due to the restriction on one single reaction channel,
rescattering effects can only be incorporated in those models by
putting in by hand a total width in the denominator of the included
resonances. It often cannot be examined whether the applied resonance 
parameters are compatible with other reaction channels. Thus the
``hunt for hidden resonances'' by single channel analyses becomes
questionable.

This problem can only be circumvented if all channels are compared
simultaneously to experimental data, thereby restricting the freedom
severely; this is done in the model underlying the present
calculation. The aim of this paper is to discuss the results of the
photoproduction reactions. We start in Sec. \ref{model} with a
review of the necessary extension of the model for the inclusion of
photoproduction reactions. In Sec. \ref{expdata} the implemented
data base is discussed and the changes with respect to Ref. \cite{feusti99}
are pointed out. In Sec. \ref{secphotoresults} our calculations are
compared to the available experimental data, and we conclude with a
summary. In the Appendices, we give a summary of the extensions of
the formalism underlying the present calculations necessary for the
inclusion of photoproduction reactions; more details can be found in
PMI \cite{pm1} and Ref. \cite{gregiphd}.

\section{\label{model}Inclusion of Photoproduction in the Giessen
  Model}

For the inclusion of photon-induced reactions in the Bethe-Salpeter
equation (see Appendix \ref{appampli} and PMI \cite{pm1}),
\bea
\mct^{fi}_{\lambda' \lambda} =
\mck^{fi}_{\lambda' \lambda} + \mi 
\int \difd \Omega_a \sum_a
\sum_{\lambda_a} \mct^{fa}_{\lambda' \lambda_a} 
\mck^{ai}_{\lambda_a \lambda} \; ,
\label{tkrelwint}
\eea
a full isospin decomposition of the photon-induced reactions including 
Compton scattering has to be performed. In Eq. \refe{tkrelwint}, $a$
represents the intermediate two-particle state. Although this
decomposition can in principle be easily 
achieved (see Ref. \cite{gregiphd}), one runs into problems concerning
gauge invariance of Compton scattering. This is due to the fact that
the rescattering takes place via the $I=\foh$ and $\fth$ amplitudes,
thus weighing the Compton isospin amplitudes $T^{11,\foh}_{\gamma
  \gamma}$ with $I=\foh$ and $T^{11,\fth}_{\gamma \gamma}$ with
$I=\fth$ of Eq. \refe{isocompton} 
differently, while gauge invariance for the nucleon contributions is
only fulfilled for the proton and neutron amplitude (more precisely,
for the combination $T^{11,\foh}_{\gamma \gamma} + 2
T^{11,\fth}_{\gamma \gamma}$). This is related to the fact that only
two physical amplitudes for Compton scattering exist ($\gamma p \ra
\gamma p$, $\gamma n \ra \gamma n$) and rescattering effects are
usually calculated in a basis using physical ($\pi^0 p$, $\pi^+ n$,
$\pi^- p$, $\pi^0 n$), not isospin states \cite{ben92}. Consequently,
the electromagnetic interaction is included only perturbatively in the 
present calculation model\footnote{The difference between the full
  rescattering calculation and the perturbative calculation have been
  checked and found to be less than 1 per mill.}. The perturbative
inclusion is equivalent to neglecting all 
intermediate electromagnetic states $a$ in the rescattering part of 
Eq. \refe{tkrelwint}. Due to the smallness of the fine structure
constant $\alpha$, this approximation is reasonable. The consequence
is that the calculation of the hadronic reactions decouples from the
electromagnetic ones and can be extracted independently. Hence the
full partial-wave decomposed $K$-matrix equation
\bea
\mct^{IJ\pm}_{fi} =
\left[ \frac{\mck^{IJ\pm}}{1 - \mi \mck^{IJ\pm}} \right]_{fi} \; ,
\label{bsematinv}
\eea
where $\mck \propto \langle f|K|i \rangle = \langle f|V|i \rangle$
[see Eq. \refe{kandv} in Appendix \ref{appampli}], is only solved for
the hadronic states. In the second step, the meson photoproduction
amplitudes can be extracted via
\bea
\mct^{IJ\pm}_{f\gamma} =
\mck^{IJ\pm}_{f\gamma} + \mi 
\sum\limits_a
\mct^{IJ\pm}_{fa}
\mck^{IJ\pm}_{a\gamma} \; ,
\eea
where the helicity indices are omitted. The sum runs only over 
hadronic states. Finally, the Compton amplitudes result from 
\bea
\mct^{IJ\pm}_{\gamma \gamma} =
\mck^{IJ\pm}_{\gamma \gamma} + \mi 
\sum\limits_a
\mct^{IJ\pm}_{\gamma a}
\mck^{IJ\pm}_{a \gamma}
\eea
with $a$ running again only over hadronic states. Since the Compton
isospin amplitudes of the potential only enter in the direct
contribution $\mck^{IJ\pm}_{\gamma \gamma}$ and only the proton and
neutron Compton amplitudes of Eq. \refe{isocomptonpn} are of interest,
gauge invariance is fulfilled.

\subsection{\label{secelmgpot}Electromagnetic part of the potential}

The contributions to the potential $V$ in the case of photon-induced
reactions come from bremsstrahlung of asymptotic particles
($N$, $\Sigma$, $\pi$, $K$), electromagnetic decays of nucleon
resonances, and intermediate (vector) mesons. Since the corresponding
Lagrangians have already been given in Ref.
\cite{feusti99}, we only present a short summary of the
electromagnetic part of the interaction, which is added to the
hadronic part specified in PMI \cite{pm1}. The $s$-, $u$-,
and $t$-channel Born contributions and the Kroll-Rudermann
term are generated by
\bea
\mcl &=& - e \bar u_{B'} (p') \left[ 
\left( \hat e \gamma_\mu A^\mu + 
  \frac{\kappa}{2m_N} \sigma_{\mu \nu} F^{\mu \nu} \right) +
\frac{g_\varphi}{m_B+m_{B'}} \gamma_5 \gamma_\mu A^\mu
\right] u_B (p) 
\nonumber \\
&& -\mi \hat e e \varphi^* \left( 
\partial_\mu^\varphi - \partial_\mu^{(\varphi^*)} 
\right) \varphi A^\mu
-e \frac{g_\varphi}{m_B+m_{B'}}
\bar u_{B'} (p') \gamma_5 \gamma_\mu u_{B} (p) A^\mu
\label{elmgborn}
\eea
with the asymptotic baryons $B,B'=(N,\Lambda,\Sigma)$, the
pseudoscalar mesons $(\varphi,\varphi')=(\pi,K)$ and 
$F^{\mu \nu} = \partial^\mu A^\nu - \partial^\nu A^\mu$.

For the intermediate ($t$-channel) mesons, which are summarized in
Table \ref{tabparticleprops},
\begin{table}
  \begin{center}
    \begin{tabular}
      {l|c|r|r|r|c}
      \hhline{======}
      & mass [GeV] & $S$ & $P$ & $I$ & $t$-channel contributions \\ 
      \hhline{======}
      $\pi$ & 0.138 & $0$ & $-$ & $1$ & $(\gamma,\gamma),(\gamma,\pi),(\gamma,\omega)$ \\
      $K$ & 0.496 & $0$ & $-$ & $\foh$ & $(\gamma,\Lambda),(\gamma,\Sigma)$ \\
      $\eta$ & 0.547 & $0$ & $-$ & $0$ & $(\gamma,\gamma),(\gamma,\omega)$ \\
      $\omega$ & 0.783 & $1$ & $-$ & $0$ & $(\gamma,\pi),(\gamma,\eta)$ \\
      \hline
      $\rho$ & 0.769 & $1$ & $-$ & $1$ & $(\pi,\pi),(\pi,\omega),(\gamma,\pi),(\gamma,\eta)$ \\
      $K^*$ & 0.894 & $1$ & $-$ & $\foh$ & $(\pi,\Lambda),(\pi,\Sigma),(\gamma,\Lambda),(\gamma,\Sigma)$ \\
      $K_1$ & 1.273 & $1$ & $+$ & $\foh$ & $(\gamma,\Lambda),(\gamma,\Sigma)$ \\
      \hhline{======}
    \end{tabular}
  \end{center}
  \caption{Properties of asymptotic and intermediate $t$-channel
    mesons entering the potential for the photon-induced
    reactions. For those particles, 
    that appear in several charge states, averaged masses are
    used. In the last column, all reaction channels (including pion-
    induced), to which the mesons contribute, are given.
    \label{tabparticleprops}} 
\end{table}
the additional Lagrangian
\bea
\mcl = 
- \mi g_{K_1} \left( \gamma_\mu K_1^\mu + 
  \frac{\kappa_{K_1}}{2m_N} \sigma_{\mu \nu} K_1^{\mu \nu} \right) \gamma_5
u_B (p)
- \frac{g}{4 m_\varphi} \ve_{\mu \nu \rho \sigma} V^{\mu \nu}
{V'}^{\rho \sigma} \varphi
+ e \frac{g_{K_1 K \gamma}}{2 m_K} K
F_{\mu \nu} K_1^{\mu \nu}
\label{elmgintermed}
\eea
is taken. $V^{\mu \nu}$ and $K_1^{\mu \nu}$ are defined in analogy to
$F^{\mu \nu}$. Note that the second term in Eq. \refe{elmgintermed}
summarizes all processes as, e.g., $\omega \ra \gamma \pi$ and $\pi /
\eta \ra \gamma \gamma$. The meson and baryon coupling constants
entering Eqs. \refe{elmgborn} and \refe{elmgintermed} are summarized
in Appendix \ref{applagr}.

The radiative decay of the spin-$\foh$ resonances is described by
\bea
\mcl_{\foh N\gamma} 
= - e \frac{g_1}{4 m_N} \bar u_R 
\left( \begin{array}{c} 1 \\ - \mi \gamma_5 \end{array} \right)
\sigma_{\mu \nu} u_N F^{\mu \nu}
\label{reselmg12}
\eea
and for the spin-$\fth$ resonances by
\bea
\mcl_{\fth N\gamma} = 
\bar u_R^\mu e
\left( \begin{array}{c} \mi \gamma_5 \\ 1 \end{array} \right)
\left( \frac{g_1}{2m_N} \gamma^\nu + \mi \frac{g_2}{4 m_N^2} 
  \partial^\nu_N \right) u_N F_{\mu \nu} \; .
\label{reselmg32}
\eea
In Eqs. \refe{reselmg12} and \refe{reselmg32}, the upper (lower)
factor corresponds to positive- (negative-) parity resonances. Note 
that in the spin-$\fth$ case, 
both couplings are in addition contracted by an off-shell projector
$\Theta_{\mu \nu} (a) = g_{\mu \nu} - a \gamma_\mu \gamma_\nu$, 
where $a$ is related to the commonly used off-shell parameter $z$
by $a = (z + \sfoh )$ (see PMI \cite{pm1} for more details).

The calculation of the amplitudes, the extraction of electromagnetic
multipoles from partial waves, the isospin decomposition, and the
calculation of observables are given in Appendices \ref{applagr},
\ref{appampli}, \ref{apppartials}, \ref{appiso}, and \ref{appobs},
respectively.

\subsection{Form factors and gauge invariance}
\label{secformgauge}

To account for the internal structure of the mesons and baryons, as in
Refs. \cite{feusti98,feusti99}, the following form factors are
introduced at the vertices:
\bea
F_p (q^2,m^2) &=& \frac{\Lambda^4}{\Lambda^4 +(q^2-m^2)^2} \; ,
\label{formfacp} \\
F_t (q^2,m^2) &=& 
\frac{\Lambda^4 +\fof (q_t^2-m^2)^2}
{\Lambda^4 + \left(q^2-\foh (q_t^2+m^2) \right)^2} \; .
\label{formfact}
\eea
Here $q_t^2$ denotes the value of $q^2$ at the kinematical threshold
of the corresponding $s$, $u$, or $t$ channel. As in Ref.
\cite{feusti98}, the form factor $F_p$ is applied to all $s$- and
$u$-channel baryon resonance vertices and to all hadronic $s$- and
$u$-channel Born vertices. Only in the $t$-channel diagrams 
have calculations been performed using either $F_p$ or $F_t$ at the
meson-baryon-baryon vertex, see Sec. \ref{secphotoresults}.

In an effective Lagrangian model, the question of gauge invariance can 
be addressed on a fundamental level. Since the above resonance and
intermediate meson electromagnetic decay vertices fulfill gauge
invariance by construction ($\Gamma_\mu k^\mu = 0$, where $k_\mu$ is
the photon momentum), these vertices and the 
corresponding hadronic vertices can be independently multiplied by
form factors. However, it is well known that the inclusion of hadronic
form factors in the Born diagrams of photoproduction reactions leads
to problems, because only the sum of all charge contributions of the
Born diagrams contributing to one specific reaction is gauge
invariant. Form factors at the hadronic vertices of these diagrams
lead to putting $q^2$-dependent weights on the different diagrams;
thus the sum becomes misbalanced and gauge invariance is violated. In
order to restore gauge invariance, one needs to construct additional
current contributions beyond the usual Feynman diagrams (contact
diagrams) to cancel the gauge violating terms. As pointed out by
Haberzettl \cite{haberzettl} (see also the detailed discussion in Ref.
\cite{feusti99}) the effect of the additional current contributions is 
to replace the hadronic form factors multiplying the charge
contributions of the Born diagrams in pion photoproduction by a common 
form factor $\hat F(s,u,t)$. In the present model we follow Davidson
and Workman \cite{davidson01}, who proposed a crossing symmetric
shape for this form factor $\hat F(s,u,t)$, which ensures that the
additional current contributions are pole-free:
\bea
\hat F(s,u,t) = F_1 (s) + F_2 (u) + F_3 (t) - 
F_1 (s) F_2 (u) - F_1 (s) F_3 (t) - F_2 (u) F_3 (t) + 
F_1 (s) F_2 (u) F_3 (t)
\; .
\label{davidsongauge}
\eea
This form can also be applied easily to $\eta$ and $\omega$
photoproduction by setting $F_3(t) = 0$ and to $K \Lambda$
photoproduction by setting $F_2(u) = 0$, since the corresponding Born
diagrams are absent. Furthermore, no form factors are used at the 
electromagnetic vertices of the Born diagrams, see
Refs. \cite{feusti99} and \cite{gregiphd}. Note that Feuster and Mosel
\cite{feusti99} used the Haberzettl suggestion for the common form
factor multiplying the charge contributions of the Born diagrams,
\bea
\hat F(s,u,t) = a_1 F_1 (s) + a_2 F_2 (u) + a_3 F_3 (t) \; .
\label{habergauge}
\eea
with $a_1 = a_2 = a_3 = \fot$.

\section{\label{expdata}Experimental Data Base}

In this section, the implemented experimental photoproduction data
base is presented, especially in view of changes and extensions as
compared to Ref. \cite{feusti99}. A summary of all references and more 
details on data base weighing and error treatment are given in Ref.
\cite{gregiphd}. \par 

\noindent $\mathbf{\gamma N \ra \pi N}$: \\
For pion photoproduction we have implemented the 
continuously updated single-energy multipole analysis of the Virginia
Polytechnic Institute (VPI) group \cite{SP01}, which greatly
simplifies the analysis of experimental 
data within the coupled-channel formalism. For those energies, where
the single-energy solutions have not been available, the gaps have
been filled with the energy-dependent solution of the VPI group. Since 
the latter data are model dependent, they enter the fitting procedure
only with enlarged error bars.
\par

\noindent $\mathbf{\gamma N \ra 2 \pi N}$: \\
As discussed in PMI \cite{pm1}, for simplicity we continue to
parametrize the $2\pi N$ final state by an effective $\zeta N$ state,
where the $\zeta$ is an isovector scalar meson with mass $m_\zeta =
2m_\pi$. A consequence is that the $\zeta N$ state is only allowed to 
couple to baryon resonances, since only in this case the decay of the
resonance into $\zeta N$ can be interpreted as the total ($\sigma N +
\pi \Delta + \rho N + \dots$) $2\pi N$ width. As it turns out in the
pion-induced calculations, a qualitative description of the $\pi N \ra
2 \pi N$ partial waves extracted by Manley {\it et al.} \cite{manley84} up
to $J=\fth$ is possible. The same agreement as in the pion-induced
$2\pi N$ production, however, cannot be expected in the $2\pi N$
photoproduction reaction. It has been shown
(e.g., Refs. \cite{hirata,murphy,oset}) that 
the $\gamma N \ra 2 \pi N$ reactions require strong background
contributions from, e.g, $\rho$ contact interactions, which
can only be included in the present model by the introduction of
separate $2\pi N$ final states. Furthermore, there is no partial-wave
decomposition of this reaction as the one by Manley {\it et al.} for $\pi N
\ra 2\pi N$ \cite{manley84}, which is the only way for comparing our
$\zeta N$ production with experiment. Therefore, the $\gamma N \ra
2\pi N$ reaction is calculated in the model and included in the
rescattering summation, but not compared to experimental data; see
also Sec. \ref{secgdh}.
\par

\noindent $\mathbf{\gamma N \ra \gamma N}$:  \\
In addition to the data used in Ref. \cite{feusti99}, the differential
cross sections and beam-polarization data of Ref. \cite{datagg} are
implemented. Since the spin-$\ffh$ resonances $D_{15}(1675)$ and
$F_{15}(1680)$ are known to have large photon couplings \cite{pdg}, it 
is certain, that for higher energies their contributions will be
important. Therefore, we continue to compare Compton scattering only 
up to a maximum energy of $1.6$ GeV. 
\par

\noindent $\mathbf{\gamma N \ra \eta N}$:  \\
We have added the differential and total cross sections, beam- and
target polarizations from Ref. \cite{datage}. All of the published cross
section data concentrate almost exclusively on the energy region below
$1.7$ GeV. Only recently, the CLAS collaboration \cite{dugger} also
accessed the energy region above $1.7$ GeV. Therefore, the preliminary
CLAS data, more than 100 data points of which are directly included in 
the fitting procedure, are also important to obtain a handle on the
higher energy region of $\eta$ photoproduction and are consequently
included.
\par

\noindent $\mathbf{\gamma N \ra K\Lambda}$: \\
The recent cross section and $\Lambda$ polarization measurements of
the SAPHIR Collaboration \cite{datagl} have been added. 
\par

\noindent $\mathbf{\gamma N \ra K\Sigma}$: \\
For this reaction, experimental data on cross sections and the
$\Sigma$ polarization for $\gamma p \ra K^+ \Sigma^0/K^0 \Sigma^+$ are
included \cite{datags}. 
\par

\noindent $\mathbf{\gamma N \ra \omega N}$: \\
For this reaction, only the cross section measurements of the ABBHHM 
Collaboration \cite{erbeom} and of Crouch {\it et al.} \cite{crouch} are
published up to now. Using only these data, even in combination with
the pion-induced data, it is difficult to extract the $\omega N$
couplings reliably. Thus, in addition we have also considered the very
precise preliminary differential cross section data of the SAPHIR
Collaboration \cite{barthom}, more than 140 data points of which are
directly included in the fitting procedure. \par

Altogether, more than 4400 photoproduction (plus 2400 pion-induced)
data points are included in the fitting strategy, which are binned
into $96$ energy intervals; for each angle differential observable we
allow for up to $10-15$ data points per energy bin.

\section{Results on Photon-Induced Reactions}
\label{secphotoresults}

The details of the calculations to extract the resonance couplings and
masses by comparison with experimental data are discussed in
PMI \cite{pm1}. Here we only shortly review the properties of the global
calculations, where the photoproduction data are also considered
for the determination of the parameters. For these calculations, we
have extended the four best hadronic fits C-p-$\pi \pm$ and C-t-$\pi
\pm$. Here the first letter C denotes that the conventional
spin-$\fth$ couplings with spin-$\foh$ off-shell contributions are
used (see PMI \cite{pm1}), the next letter ``p'' or ``t'' denotes whether the
form factor $F_p$ [Eq. \refe{formfacp}] or $F_t$ [Eq. \refe{formfact}]
is used in the $t$-channel diagrams, $\pi$ stands for using only 
pion-induced data, and the last letter denotes the {\it a priori} unknown
sign of the coupling $g_{\omega \rho \pi}$. Note that this coupling
gives rise to the important $t$-channel $\rho$ exchange in $\pi N \ra 
\omega N$. Correspondingly, the four global calculations are labeled
C-p-$\gamma \pm$ and C-t-$\gamma \pm$. 

Similar to Feuster and Mosel \cite{feusti99}, our first attempt for
the inclusion of the photoproduction data in the calculation has been
to keep all hadronic parameters fixed to their values obtained in the
fit to the pion-induced reactions. In contrast to the findings of Ref.
\cite{feusti99}, no satisfactory description of the photoproduction
reactions has been achieved with these hadronic parameters. As a
consequence of the smaller data base used in Ref. \cite{feusti99} at most
three photoproduction reactions ($\gamma N \ra \gamma N$, $\gamma N
\ra \pi N$, $\gamma N \ra \eta N$) had to be fitted
simultaneously. Above 1.6 GeV, no data were available on $\eta$
photoproduction.

The extended model space and data base now 
constrains all production mechanisms more strongly, especially for
energies above 1.7 GeV, where precise photoproduction data on all
reactions (besides Compton scattering) are used. Due to the lack of 
precise data in the high energy region for pion-induced $\eta N$
and $\omega N$ production, these production mechanisms have not been
correctly decomposed in the purely hadronic calculations, thus leading
to contradictions in the photoproduction reactions when the hadronic
parameters are kept fixed. Moreover, as pointed out in PMI \cite{pm1},
the Born couplings in  the associated strangeness production
only play a minor role in the pion-induced reactions while, as a
result of the gauging procedure, these contributions are enhanced in
photoproduction thus allowing for a more reliable determination of the
corresponding couplings. Consequently, the $K\Lambda/\Sigma$ 
photoproduction also turns out to be hardly describable when the
hadronic parameters are kept fixed. Only when also these
parameters are allowed to vary a simultaneous description of all pion-
and photon-induced reactions is possible. 

The resulting $\chi^2$ values for the calculations C-p-$\gamma \pm$
are presented in Table \ref{tabchisquares};
\begin{table}
  \begin{center}
    \begin{tabular}
      {l|r|r|r|r|r|r|r}
      \hhline{========}
       Fit & Total $\pi$ & $\chi^2_{\pi \pi}$ & $\chi^2_{\pi 2\pi}$ & 
       $\chi^2_{\pi \eta}$ & $\chi^2_{\pi \Lambda}$ & $\chi^2_{\pi \Sigma}$
       & $\chi^2_{\pi \omega}$ \\ 
       \hline
       C-p-$\gamma +$ & 3.78 & 4.23 & 7.58 & 3.08 & 3.62 & 2.97 & 1.55 \\
       C-p-$\gamma -$ & 4.17 & 4.09 & 8.52 & 3.04 & 3.87 & 3.94 & 3.73 \\
       SM95-pt-3 & 6.09 & 5.26 & 18.35 & 2.96 & 4.33 & --- & --- \\
      \hhline{========}
      Fit & Total$^a$ & $\chi^2_{\gamma \gamma}$ & $\chi^2_{\gamma \pi}$ & 
      $\chi^2_{\gamma \eta}$ & $\chi^2_{\gamma \Lambda}$ & 
      $\chi^2_{\gamma \Sigma}$ & $\chi^2_{\gamma \omega}$ \\
      \hline
       C-p-$\gamma +$ & 6.57 & 5.30 & 10.50 & 2.45 & 3.95 & 2.74 & 6.25 \\
       C-p-$\gamma -$ & 6.66 & 5.15 & 10.54 & 2.37 & 2.85 & 2.27 & 6.40 \\
       SM95-pt-3 & 24.40 & 16.45 & 42.07 & 8.01 & 4.64 & --- & --- \\
      \hhline{========}
    \end{tabular}
  \end{center}
  \caption{Resulting $\chi^2$ of the various
    fits. For comparison, we have also applied the preferred
    parameter set SM95-pt-3 of Ref. \cite{feusti99} to our
    extended and modified data base for energies up to $1.9$
    GeV. $^a$: This value includes all pion- and photon-induced data
    points. 
    \label{tabchisquares}}
\end{table}
for the results of the calculations C-t-$\gamma\pm$ see below. 
Note that, in contrast to the previous analysis \cite{feusti99}, in
the present calculation we have included all experimental data up to
the upper end of the energy range, in particular also for all partial-
wave and multipole data up to $J=\fth$. At first sight it seems that
the total $\chi^2$ is only fair; however, one has to note that the main
part of this value stems from the pion-photoproduction multipoles
\cite{SP01}, which have very small error bars but also scatter a lot
(cf. Figs. \ref{figgpip} -- \ref{figgpi32} in Sec.
\ref{secresgp} below). Note, that in this channel, there are 40\% of
all data points. Taking this channel out, the total $\chi^2$ per data
point is reduced from $6.56$ to $3.87$ for the preferred global
fit. Thus a very good simultaneous description of all reactions is
possible, which shows that the measured data for all reactions are
compatible with each other, concerning the partial-wave decomposition
and unitarity effects. As a guideline for the quality of the present
calculation, we have also included a comparison with the preferred  
parameter set SM95-pt-3 of Ref. \cite{feusti99} applied to our extended
and modified data base. It is interesting to note that although this
comparison has only taken into account data up to 1.9 GeV for the
final states $\gamma N$, $\pi N$, $2\pi N$, $\eta N$, and $K \Lambda$,
the present best global calculation C-p-$\gamma +$ results in a better 
description in almost all channels; only for $\pi N \ra \eta N$ the
$\chi^2$ of Ref. \cite{feusti99} is slightly better. This is a consequence
of the fact that, for example, for the understanding of $K\Lambda$
production, the coupled-channel effects due to the final states
$K\Sigma$ and $\omega N$ have to be included. This is discussed in
Sec. \ref{secresgl} below. 

Moreover, while in Ref. \cite{feusti98} similar
results were found using either one of the form factors $F_t$ and
$F_p$ for the $t$-channel meson exchanges, and in Ref. \cite{feusti99} only
$F_t$ was applied, the extended data base and model space shows a
clear preference of using the form factor $F_p$ for all vertices,
i.e., also for the $t$-channel meson exchange. We have also tried to
perform global fitting calculations using $F_t$ in the $t$-channel
exchange processes (C-t-$\gamma \pm$), but have not found any 
satisfactory parameter set for a global description in this case. Even
when the fitting procedure has been reduced to the five most important
final states --- $\gamma N$, $\pi N$, $2\pi N$, $\eta N$, and $\omega
N$ --- we have found for $\gamma /\pi N \ra \eta N$ $\chi^2$'s of only 
$\approx 5$ and for $\gamma N \ra \omega N$ ($\pi N \ra \omega N$)
$\chi^2$'s of $\approx 30$ ($\approx 7$), while pion production and Compton
scattering have been only slightly worse as compared to C-p-$\gamma
\pm$. The much worse description using $F_t$ in the global fits can be
explained by the fact that, for the photon-induced reactions, the
$NN\omega$ coupling now not only appears as a final state coupling,
but also contributes in the production of $\pi N$ and $\eta
N$. Conversely the $\pi NN$ coupling constant is now also of great
importance in $\omega$ photoproduction. Thereby, the validity of the
form factors is tested in a wide kinematical region, since, in our
model, many of the $t$-channel meson couplings contribute to several
reactions and also as final state couplings (cf. Table
\ref{tabparticleprops} above). We conclude that $F_p$ is applicable 
to a much wider kinematic region (especially to higher energies) than
$F_t$. This comes about because of the quite different $q^2$-dependent
behavior of the two form factors $F_p$ and $F_t$ below the pole mass
and in the low $|t|=|q^2|$ region. 
To find satisfactory results with the form factor $F_t$ in the
present model, it would be necessary to lift the restriction of using
only one cutoff value $\Lambda_t$ for all $t$-channel diagrams. 

In the following sections, the photoproduction results of the two
global calculations C-p-$\gamma +$ and C-p-$\gamma -$ are discussed in
detail.

\subsection{Compton scattering}
\label{secresgg}

\begin{figure}
  \begin{center}
    \parbox{16cm}{\includegraphics[width=16cm]{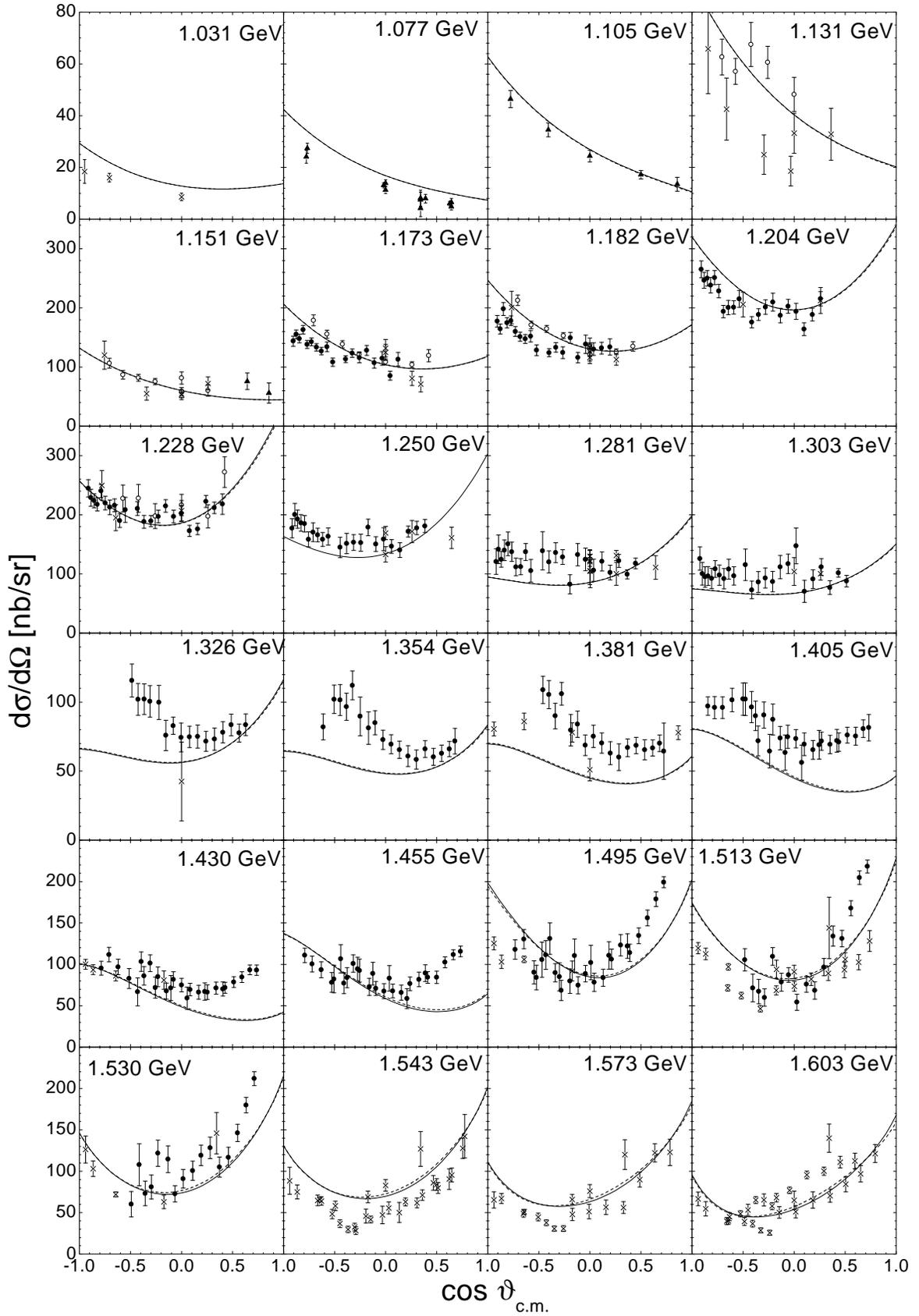}}
    \caption{$\gamma p \ra \gamma p$ differential cross
      section for different $\sqrt s$ as indicated in the
      figure. Calculation C-p$\gamma +$: solid line; C-p$\gamma -$: 
      dashed line. Data are from Ref. \cite{datagg}.
      \label{figggdif}}
  \end{center}
\end{figure}

A simultaneous description of Compton scattering together with the
inelastic channels is essential because this process is dominated by
the electromagnetic coupling and may thus impose more stringent
requirements on those. As a consequence of the new data from Ref.
\cite{datagg}
we have doubled the Compton scattering data base from 266 to 538 data
points as compared to Ref. \cite{feusti99}. This means that the description 
of Compton scattering becomes more difficult, resulting in larger
$\chi^2$ values than in Ref. \cite{feusti99}. However, as Fig. \ref{figggdif} 
shows, 
our calculations are able to describe the differential cross section
in the considered energy region up to $\sqrt s = 1.6$ GeV. Only in the 
intermediate energy region between 1.3 and 1.5 GeV are there 
indications of contributions missing in the present model. These
missing contributions are due to the lack of $2\pi N$ rescattering
contributions, since in the present model only resonant $2\pi N$
photoproduction mechanisms are included; see Sec.
\ref{expdata}. This leads to the lack of background contributions in
the low energy two-pion photoproduction; see also the discussion in
Sec. \ref{secgdh} below.

\begin{figure}
  \begin{center}
    \parbox{16cm}{\includegraphics[width=16cm]{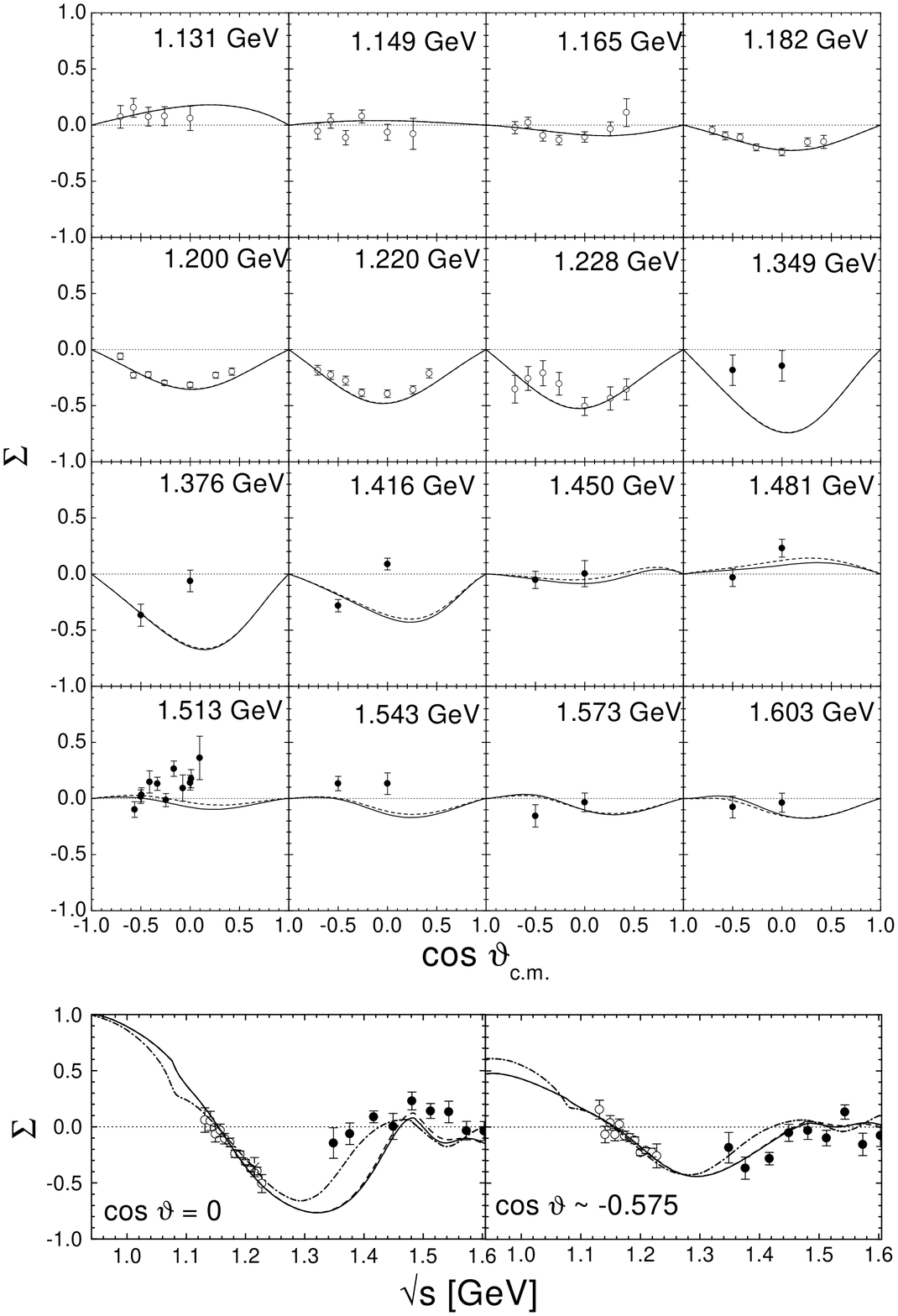}}
    \caption{$\gamma p \ra \gamma p$ differential beam
      polarization. 
      Line code and data as in Fig. \ref{figggdif}. In addition, the
      results of the analysis of L'vov {\it et al.} \cite{lvov} are given by 
      the dash-dotted line.
      \label{figggsig}}
  \end{center}
\end{figure}

The same discrepancy in this energy region can also be observed in the
$90^\circ$ region of the beam polarization (see Fig. \ref{figggsig}),
which is well described for energies below 1.3 and above 1.45 GeV and
also other angles. For comparison, we also display the results on
the beam polarization of the dispersion theoretical analysis of L'vov
{\it et al.} \cite{lvov}. In the model of Ref. \cite{lvov},
analyticity constraints are taken 
into account by saturating $s$-channel dispersion relations with use
of the VPI pion-photoproduction multipole analysis and resonance
photocouplings. In addition, two-pion photoproduction background
contributions are also taken into account. These authors' description 
of the beam asymmetry is rather close to our description, with the
exception of the above mentioned energy region and the $\pi N$
threshold region. This asserts the findings of Pearce and Jennings
\cite{pearce}, that, due to the extracted soft form factor, the off-shell
rescattering contributions of the intermediate two-particle
propagator in the scattering equation, which are neglected in the
$K$-matrix approximation, 
have to be damped by a very soft form factor in $\pi N$ elastic
scattering; see also PMI \cite{pm1}. Thus the effects of the
off-shell rescattering part only become visible very close to the
$\pi N$ threshold, in line with the above comparison of the present
model with the dispersion theoretical analysis of 
L'vov {\it et al.} \cite{lvov}. The cusp in the beam polarization at the
$\pi N$ threshold is due to the $\mct^{EE}_{1-}$ multipole amplitude
[cf. Eq. \refe{comptontmulti}], which has also been found by
Kondratyuk and Scholten \cite{kondratyuk}.

As expected, the two global fits C-p-$\gamma \pm$ lead to practically
identical results since Compton scattering is only considered up to
1.6 GeV, which is still far below the $\omega N$ threshold. The
dominant contributions stem from the nucleon, the $P_{33}(1232)$
resonance, and the $D_{13}(1520)$, while the $P_{11}(1440)$ and
$S_{11}(1535)$ only make small contributions.

\subsection{Pion photoproduction}
\label{secresgp}

Pion photoproduction is most precisely measured of all the channels
considered in the present work. This has also led to the development
of a large amount of models on this reaction (see references in Ref.
\cite{feusti99}), most of them concentrating on the low-energy
[$P_{33}(1232)$] region. The Mainz MAID isobar model of Drechsel {\it et
al.} \cite{maid} covers a similar energy region as the present
analysis. In MAID, the Born and vector meson background contributions
were $K$-matrix unitarized with the help of the VPI $\pi N \ra \pi N$
partial waves \cite{SM00}. Instead of using a form factor for the $\pi
NN$ vertex, a pseudovector (PV)- pseudoscalar (PS) mixture scheme is
introduced to regularize the nucleon contributions at higher
energies. Since the resonance contributions are generated by
unitarized Breit-Wigner descriptions, the resonances do not 
create additional background by $u$-channel diagrams. The advantage of 
this procedure is that the inclusion of spin-$\ffh$ resonances is
straightforward and, consequently, the $F_{15}(1680)$ is also taken
into account. The free parameters (e.g., the vector meson couplings)
are adjusted to the VPI multipoles \cite{SP01} and a very good
description is achieved. As a consequence of the Breit-Wigner
description and the restriction on pion photoproduction, the extracted
electromagnetic helicity amplitudes of the resonances are very close
to the Particle Data Group (PDG) values \cite{pdg}, while in our
analysis all resonance contributions are also constrained by Compton
scattering, $\eta N$, $K\Lambda /\Sigma$, and $\omega N$
photoproduction data.

\begin{figure}
  \begin{center}
    \parbox{16cm}{\includegraphics[width=16cm]{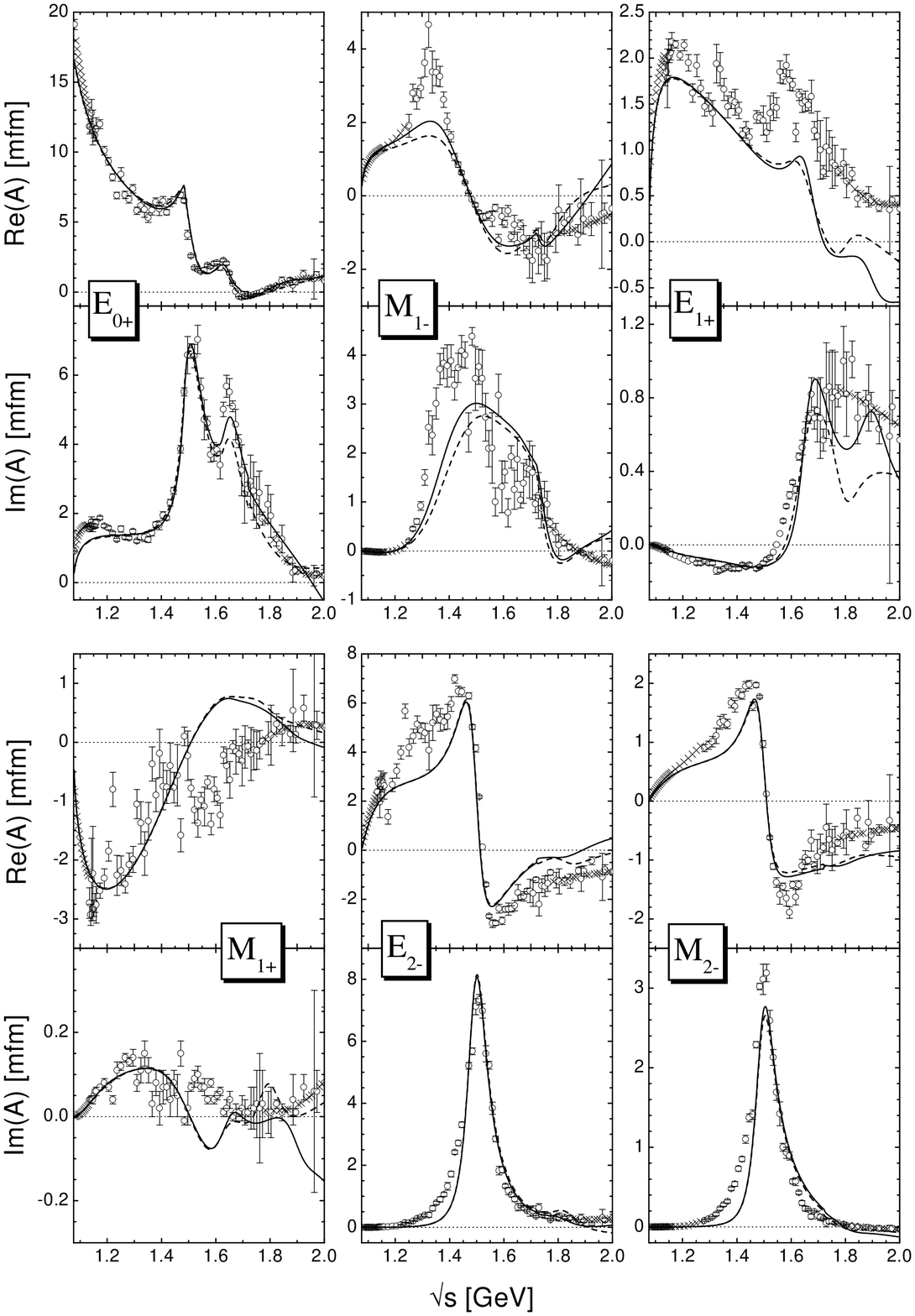}}
    \caption{$\gamma N \ra \pi N$ proton [see Eq. \refe{isogampi} in
      Appendix \ref{appisogampi}] multipoles. Line code as in 
      Fig. \ref{figggdif}. Data are from the VPI \cite{SP01}
      single-energy ($\circ$) and energy-dependent ($\times$)
      solution.
      \label{figgpip}}
  \end{center}
\end{figure}

\begin{figure}
  \begin{center}
    \parbox{16cm}{\includegraphics[width=16cm]{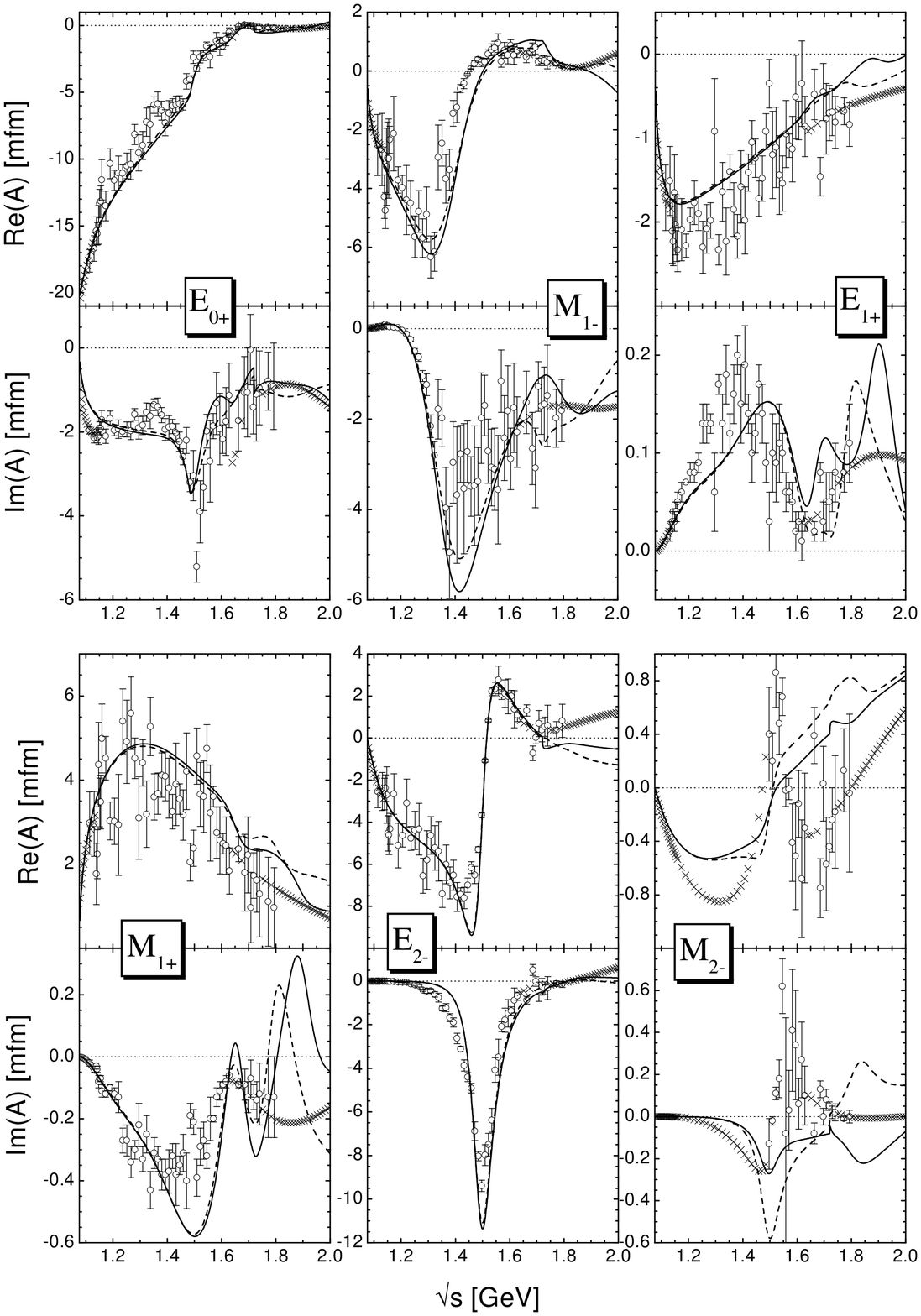}}
    \caption{$\gamma N \ra \pi N$ neutron [see Eq. \refe{isogampi} in
      Appendix \ref{appisogampi}] multipoles. Line code as in 
      Fig. \ref{figggdif}, data as in Fig. \ref{figgpip}.
      \label{figgpin}}
  \end{center}
\end{figure}

\begin{figure}
  \begin{center}
    \parbox{16cm}{\includegraphics[width=16cm]{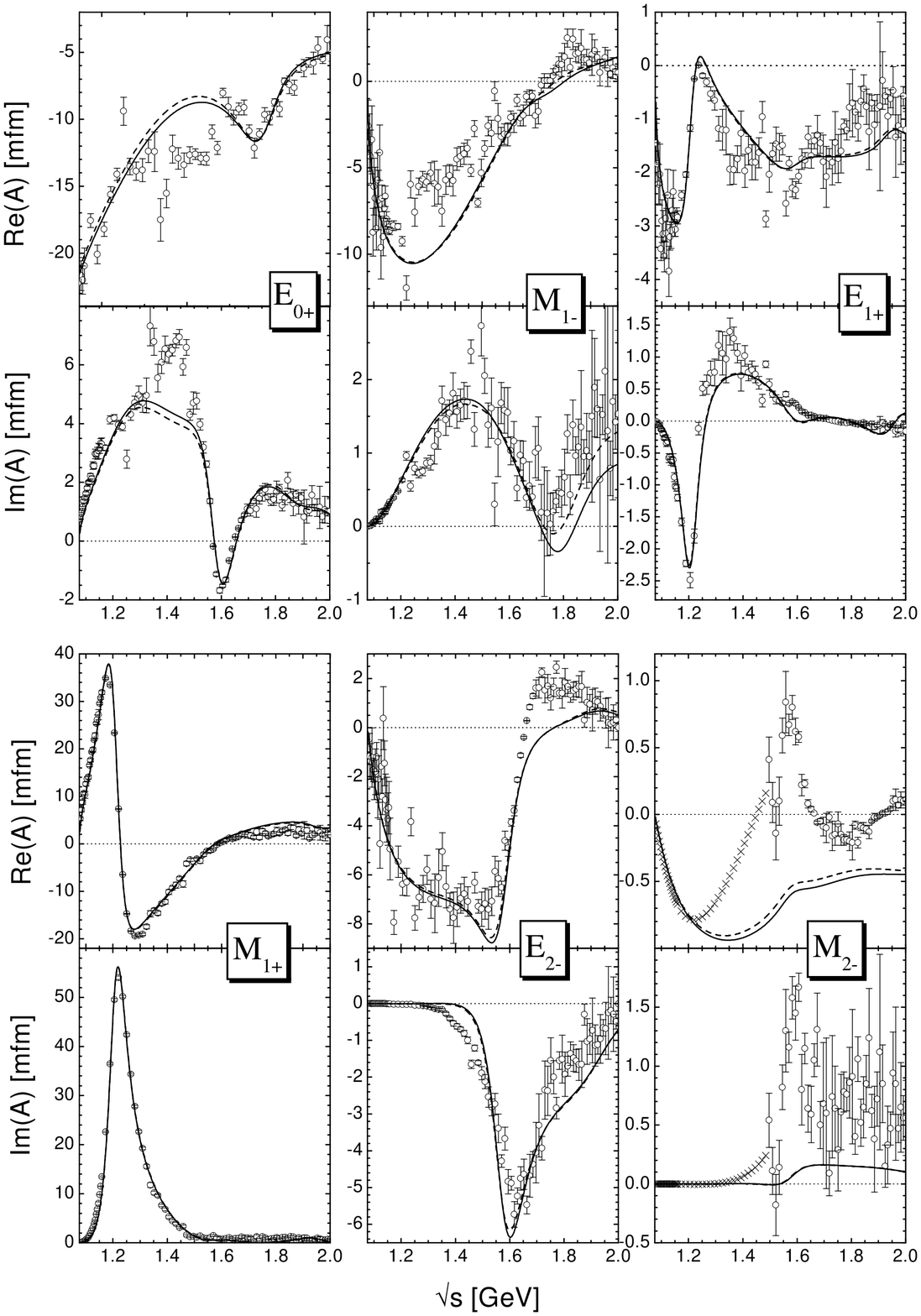}}
    \caption{$\gamma N \ra \pi N$ $I=\fth$ multipoles. Line code as in 
      Fig. \ref{figggdif}, data as in Fig. \ref{figgpip}.
      \label{figgpi32}}
  \end{center}
\end{figure}

As a consequence of the precise experimental data, the 
pion-photoproduction channel is of great importance in our data base
and contains about 40\% of all data points, 
many of which have very small error bars. Thus this channels strongly
influences the photon and pion couplings and also the masses of the
resonances. For example, the masses of the $S_{11}(1535)$,
$S_{31}(1620)$, $P_{31}(1750)$, and $D_{33}(1700)$ are influenced 
by the pion-photoproduction multipoles; see Figs. \ref{figgpip} --
\ref{figgpi32} and PMI \cite{pm1}.
Although the resulting $\chi^2$ seems to be rather high ($\sim 10$),
Figs. \ref{figgpip} $-$ \ref{figgpi32} reveal, that the properties of
almost all multipoles up to $J=\fth$ are well described in the present 
model. 

The largest contributions to the total $\chi^2$ stem from the real
parts of the $E_{1+}^p$, $E_{2-}^p$, $M_{1+}^\fth$, and $E_{2-}^\fth$
multipoles. In the latter three cases, this is a consequence of the
fact, that around the resonances $D_{13}(1520)$, $P_{33}(1232)$, and
$D_{33}(1700)$ the multipoles are known with very high accuracy, and
thus even very small deviations in the calculation lead to a large
$\chi^2$. For the $D_{13}(1520)$ multipoles $E_{2-}^{p/n}$ and
$M_{2-}^p$, but also for the $D_{33}(1700)$ multipole $E_{2-}^\fth$,
in the imaginary parts we observe the same problem of the increasing
behavior below the resonance position as in the corresponding $\pi N$
partial waves (see PMI \cite{pm1}), which is probably due to
deficiencies in the present model concerning the $2\pi N$ final state
description. 
In the case of the $E_{1+}^p$ multipole the deviation is due 
to the lack of some background contribution, which might be related to 
the problem in the description of the $\pi N\ra \pi N$ 
$P_{13}$ partial wave described in PMI \cite{pm1} due to a 
missing ($3\pi N$) inelastic channel. It is interesting to note that
the discrepancy between the calculation and the VPI data points in the
$E_{1+}^p$ multipole starts around 1.6 GeV, which is the same
energy, where the problems in the $P_{13}$ $\pi N\ra \pi N$ wave arise, 
and also where a sudden increase in the total cross section of $\gamma 
p \ra p \pi^+ \pi^- \pi^0 / n \pi^+ \pi^+ \pi^-$ was observed in
experiments \cite{erbeom,wisskirchen}. For the neutron 
multipoles, there are only data of the energy-dependent solution
available at energies above 1.8 GeV. Since these data are
model dependent, they only enter with enlarged error bars in the
present calculation, and the high-energy tails of the neutron multipoles
are not well fixed. This explains the pronounced resonant structure in
the imaginary part of the $E_{1+}^n$ and $M_{1+}^n$ multipoles, not
observed in the VPI multipole data \cite{SP01}.

As can be seen in Figs. \ref{figgpip} $-$ \ref{figgpi32}, the 
differences between the two global calculations C-p-$\gamma +$ and
C-p-$\gamma -$ can be mainly found in the $J^P=\fth^+$ proton and
neutron multipoles
above the $\omega N$ threshold. This is a consequence of the fact
that these multipoles give important contributions to the $\omega N$
production mechanism (see Sec. \ref{secresgo} below and also the
results on $\pi N \ra \omega N$ in PMI \cite{pm1}) and are thus very
sensitive to the change of sign of the $t$-channel background
contribution in $\pi N \ra \omega N$.

\begin{figure}
  \begin{center}
    \parbox{16cm}{\includegraphics[width=16cm]{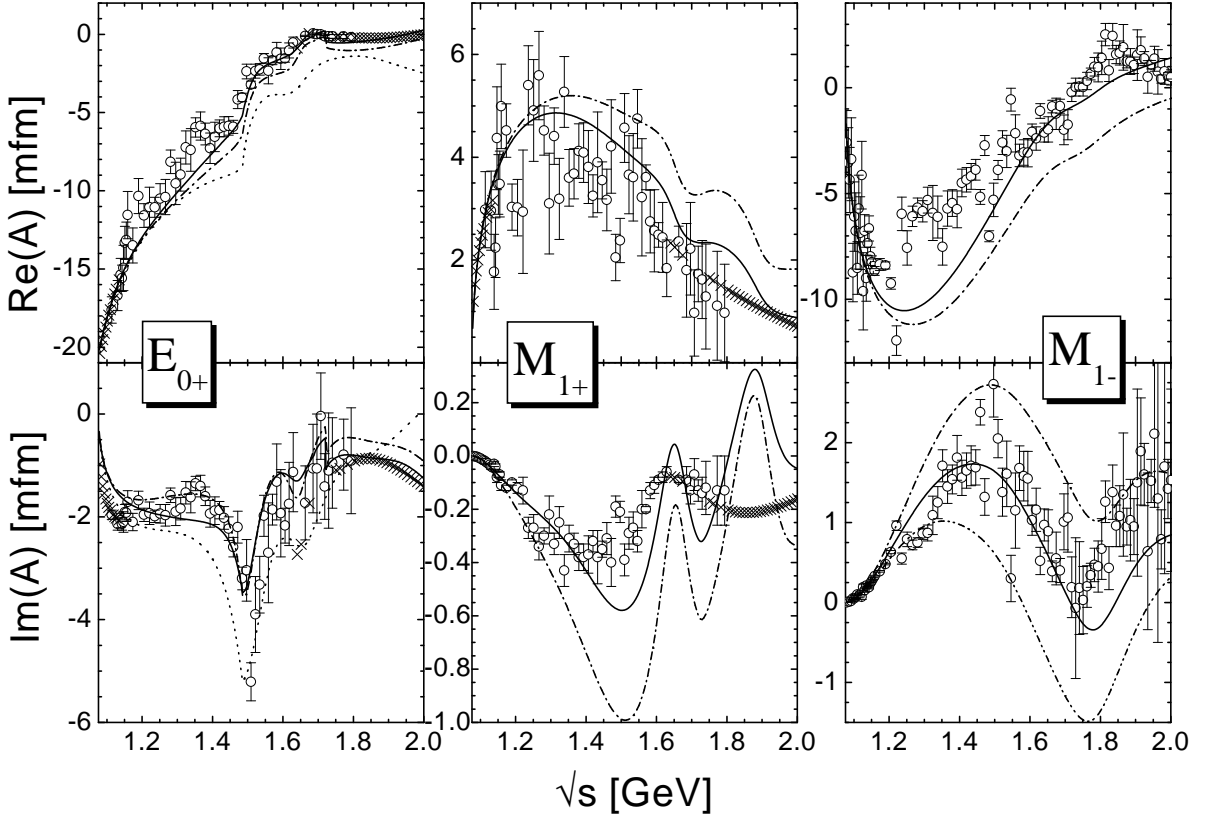}}
    \caption{Examples for the influence of the nucleon cutoff value 
      $\Lambda_N$ on the pion-photoproduction multipoles: 
      neutron $E^n_{0+}$ (\textit{left}), neutron $M^n_{1+}$
      (\textit{middle}), $I=\fth$ $M^\fth_{1-}$
      (\textit{right}). C-p-$\gamma 
      +$ with $\Lambda_N = 0.96$ GeV: solid line; C-p-$\gamma
      +$ with $\Lambda_N = 1.16$ GeV: dash dotted line. For
      $E^n_{0+}$, also the calculation of Ref. \cite{feusti99} is 
      displayed (dotted line). For the imaginary part of
      $M^\fth_{1-}$, the calculation C-p-$\gamma +$ using the
      Haberzettl gauging procedure is also shown (dash-double dotted line).
      \label{figgpe0pnuc}}
  \end{center}
\end{figure}

Apart from the $E_{1+}^p$ multipole discussed above, we find
indications for missing background only in the $M_{2-}^n$ and
$M_{2-}^\fth$ multipoles, while in all other multipoles the background
contributions seem to be in line with the VPI \cite{SP01}
analysis. Since the background is mainly generated by the Born terms,
the multipoles strongly influence the nucleon cutoff value
$\Lambda_N$. In Fig. \ref{figgpe0pnuc} 
we show the sensitivity of the $E^n_{0+}$, $M^n_{1+}$, and
$M^\fth_{1-}$ multipoles to the cutoff value $\Lambda_N$, which is
used in the $\pi NN$ form factor. As we have pointed out in PMI
\cite{pm1}, the $S_{11}$ and $P_{11}$ $\pi N \ra \pi N$ partial 
waves are more poorly described once the pion-photoproduction data is
included. This effect can be traced back to the necessity of reducing
the value of $\Lambda_N = 1.16$ GeV of the hadronic calculation to
$\Lambda_N = 0.96$ GeV in the global calculation. Using the latter
value, the background contributions in the multipoles are in line with
the VPI analysis \cite{SP01}, while with the former value the
incorrect background description leads to largely increased $\chi^2$
values. The price one has to pay for the improvement in the mentioned
multipoles is the deterioration in the low-energy $S_{11}$ and
$P_{11}$ $\pi N$ elastic partial waves leading also to an increase of
the $P_{11}(1440)$ mass and width. Since the Born terms are very
sensitive to the gauging procedure, the resulting good description of
most of the background features also indicates, that the
Davidson-Workman gauging procedure [Eq. \refe{davidsongauge}] is supported
by the pion-photoproduction data. As an example, we show the effect of
switching to the Haberzettl gauging procedure [Eq. \refe{habergauge}]
in the imaginary part of the $M_{1-}^\fth$ multipole in
Fig. \ref{figgpe0pnuc}. Similar observations are also made in other
multipoles. This is also related to the large $\chi^2$ improvement of
the present calculation as compared to Ref. \cite{feusti99}, where the
Haberzettl gauging procedure has been used. The largest differences as
compared to Feuster and Mosel \cite{feusti99} can be observed in the
real part of the $I=\foh$ $E_{0+}$ multipoles; see, e.g., $E^n_{0+}$ in
Fig. \ref{figgpe0pnuc}. Note that it was already speculated in Ref.
\cite{feusti99}, that modifying the gauging procedure might improve
the description in these multipoles. 

In the $M_{2-}^\fth$ multipole, in addition to the missing background
mentioned above, also a too small resonance contribution is 
extracted in the present model. However, this contribution is also
strongly constrained by the spin-$\foh$ off-shell contributions of the
$D_{33}(1700)$ to the $E_{0+}^\fth$ and $M_{1-}^\fth$
multipoles. Since these multipoles are more precisely known than the
$M_{2-}^\fth$ multipole, the fitting procedure is dominated by the
background contributions of the $D_{33}(1700)$ in the spin-$\foh$
multipoles, resulting in photon couplings which deteriorate the
$M_{2-}^\fth$ description.

\subsection{$\eta$ Photoproduction}
\label{secresge}

\begin{figure}
  \begin{center}
    \parbox{16cm}{
      \parbox{75mm}{\includegraphics[width=75mm]{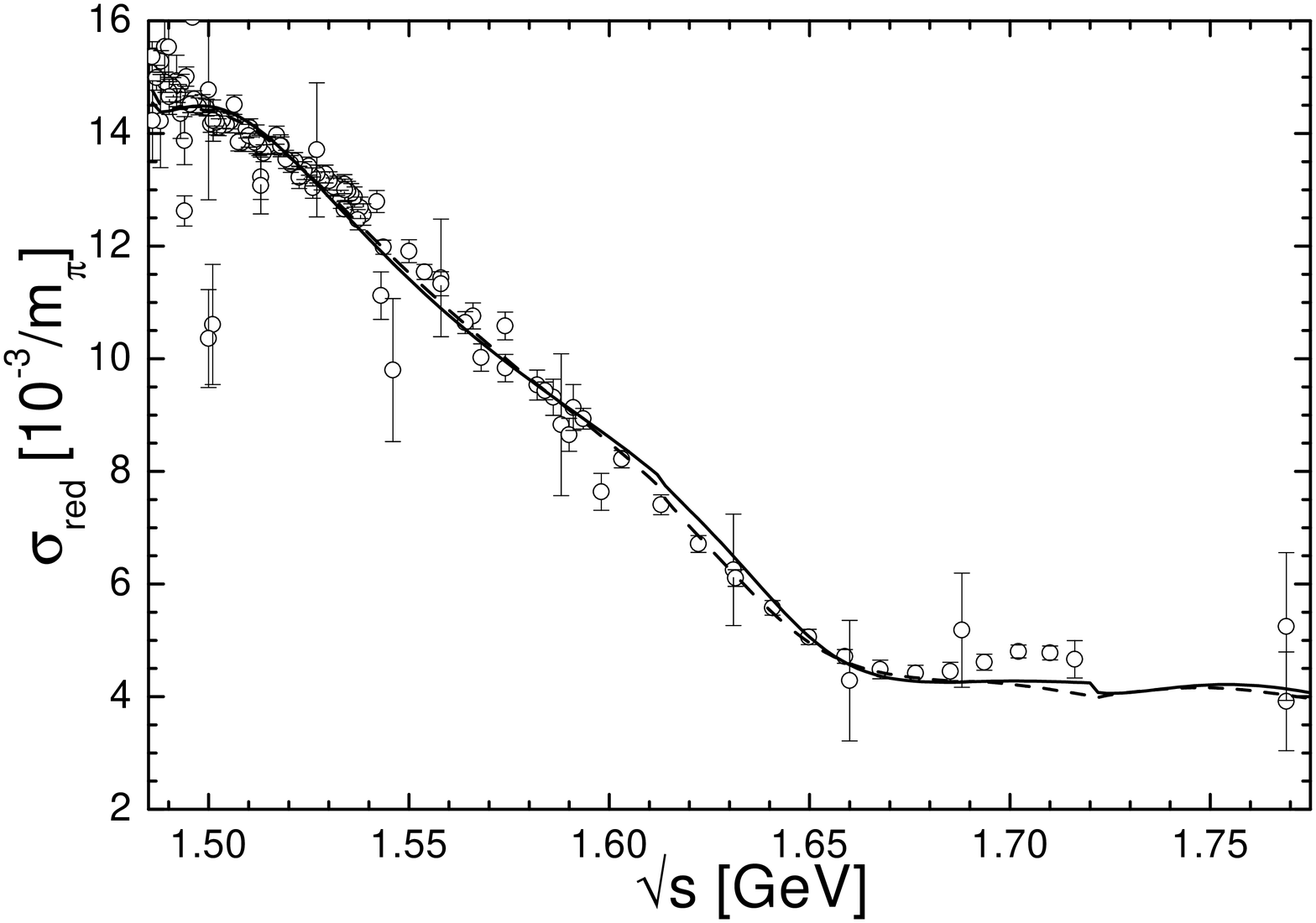}}
      \parbox{75mm}{\includegraphics[width=75mm]{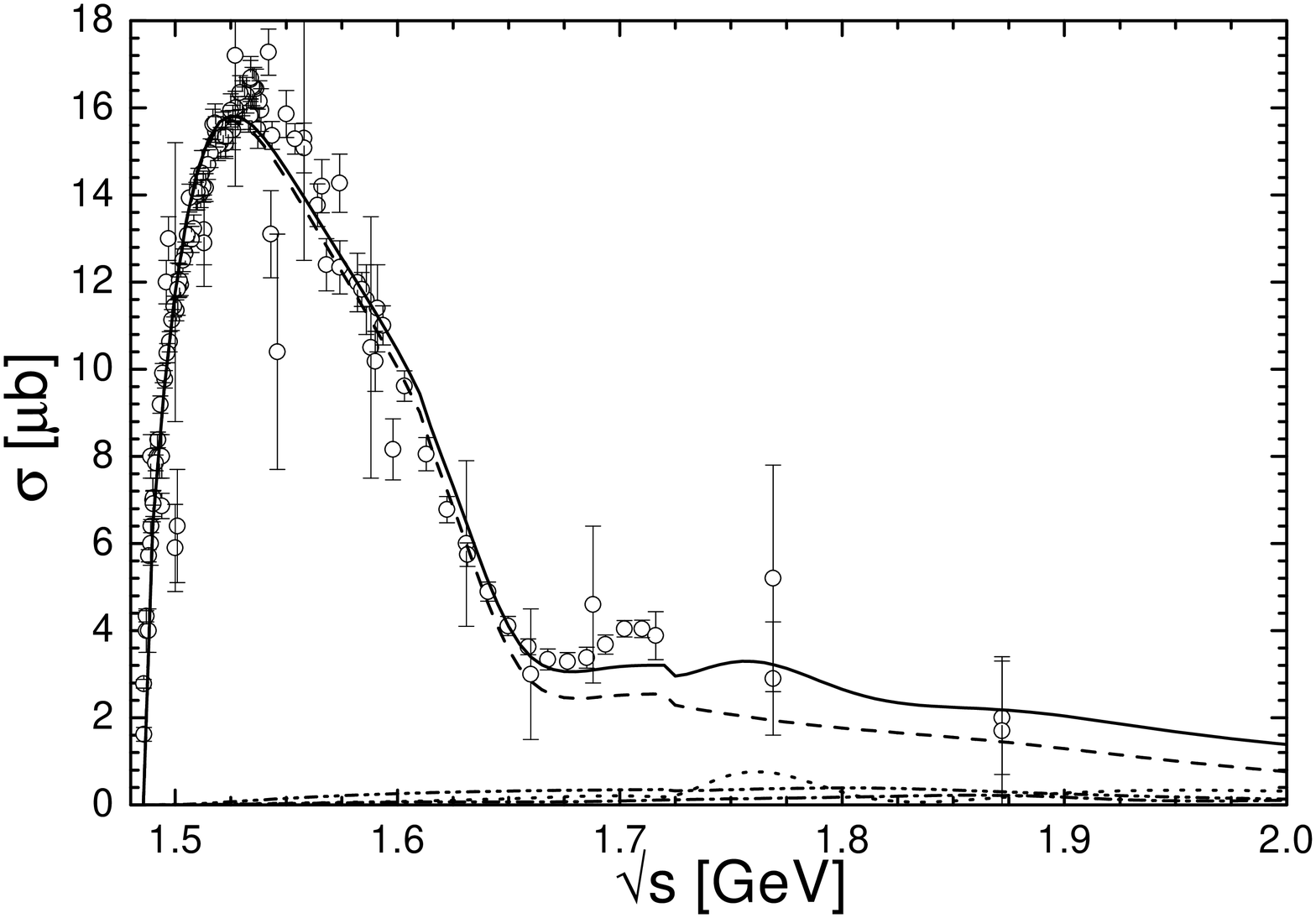}}
      }
    \caption{$\gamma p \ra \eta p$. Data as given in Sec. \ref{expdata}.
      \textit{Left:} Reduced cross section. Line code as in
      Fig. \ref{figggdif}. \textit{Right:} Partial-wave decomposition
      of the total cross section, $J^P=\foh^-$: dashed; $\foh^+$:
      dotted; $\fth^+$: dash dotted; $\fth^-$: dash-double dotted.
      \label{figgetot}}
  \end{center}
\end{figure}

Several investigations \cite{ben95,feusti99,sauermann} showed,
that the $\eta N$ photoproduction is dominated by a $J^P = \foh^-$
production mechanism, in particular at threshold. While we find in the
pion-induced reaction still important $\foh^+$ and $\fth^+$ 
cross-section contributions, only a small contribution of the
$P_{11}(1710)$ 
is visible in the photon-induced reaction, and the $\foh^-$
contribution is by far dominant up to 2 GeV, see Fig. \ref{figgetot}.
Here we have also displayed the so-called reduced cross section,
which takes out effects caused by phase space and is given by
$\sigma_{red} = \sqrt{\sigma_{tot} \avec k / (4\pi \avec k')}$ 
(cf. Appendix \ref{appobs}), and allows for more conclusive
investigations close to threshold. As can be clearly seen in
Fig. \ref{figgetot}, the production mechanism is well under control in 
the present model down to the very threshold. Thus the energy
dependence of the $\eta N$ total cross section is correctly described, 
although the inclusion of the pion photoproduction $E_{0+}^p$
multipole data requires a reduction of the $S_{11}(1535)$ mass from
$\approx 1.544$ to $\approx 1.526$ GeV; see also PMI
\cite{pm1}. Note that our calculations do not follow the increase of
the GRAAL total cross section 
\cite{datage}
around 1.7 GeV, which is not observed in the estimated total cross
section from the CLAS collaboration \cite{dugger} either.

\begin{figure}
  \begin{center}
    \parbox{16cm}{\includegraphics[width=16cm]{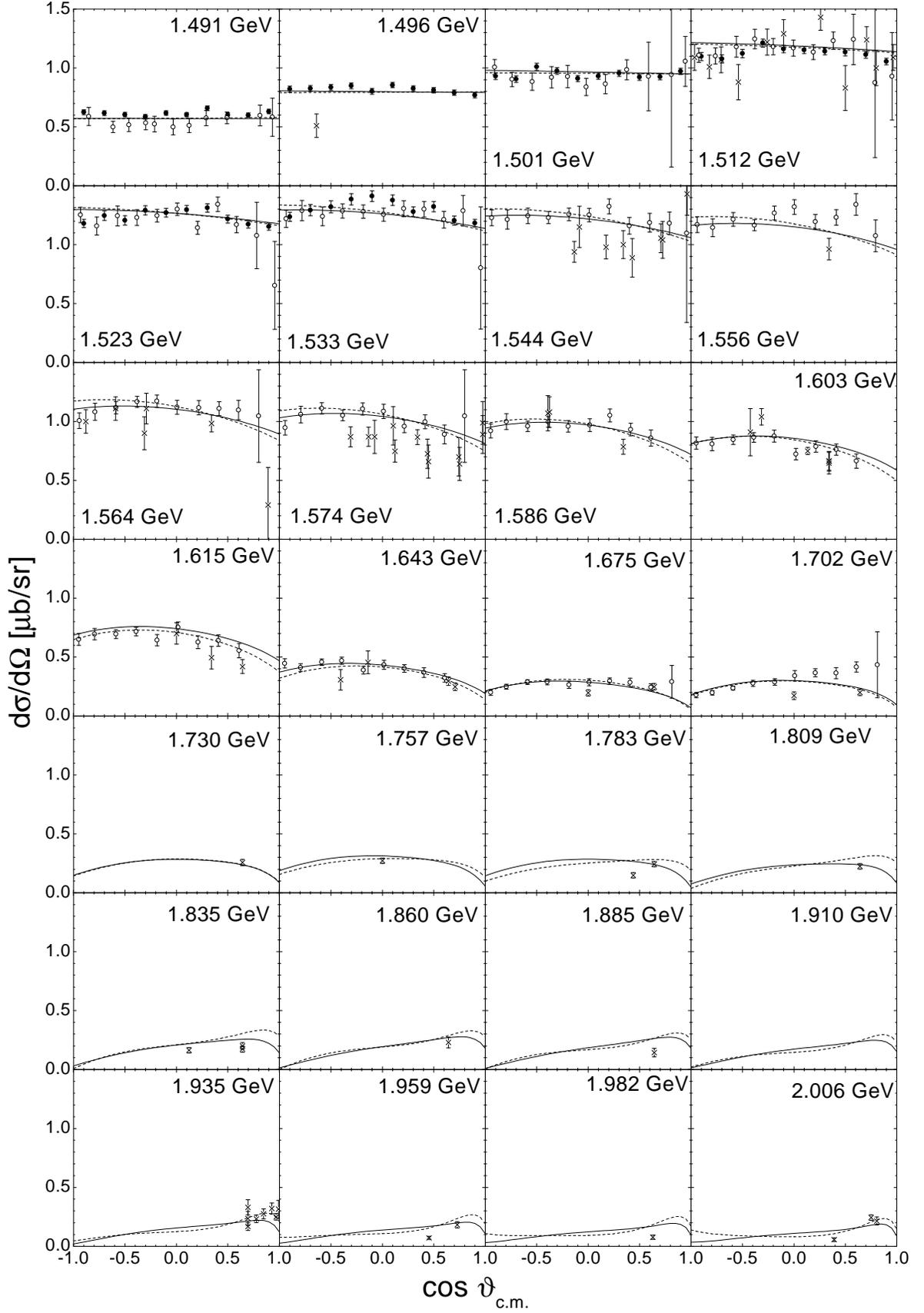}}
    \caption{$\gamma p \ra \eta p$ differential cross section. Line
      code as in Fig. \ref{figggdif}. Data are 
      as given in Sec. \ref{expdata}. The data from Ref. \cite{dugger}
      are not shown.
      \label{figgedif}}
  \end{center}
\end{figure}
In the first coupled-channel model on photon- and pion-induced $\eta
N$ production up to $\sqrt s = 1.75$ GeV by Sauermann {\it et
al.} \cite{sauermann}, it has been found, that an important production
mechanism is due to the vector meson ($\rho$ and $\omega$)
exchanges. In line with these authors' findings, it also turns out in
the present model that these exchanges give important contributions
in all partial waves and the neglect would lead to total cross
sections below the experimental data already at 1.55 GeV. Note 
that in the present calculation the forward peaking behavior of the
differential cross section at higher energies is less pronounced as
compared to Ref. \cite{feusti99} (see Fig. \ref{figgedif}), 
\begin{figure}
  \begin{center}
    \parbox{16cm}{\includegraphics[width=16cm]{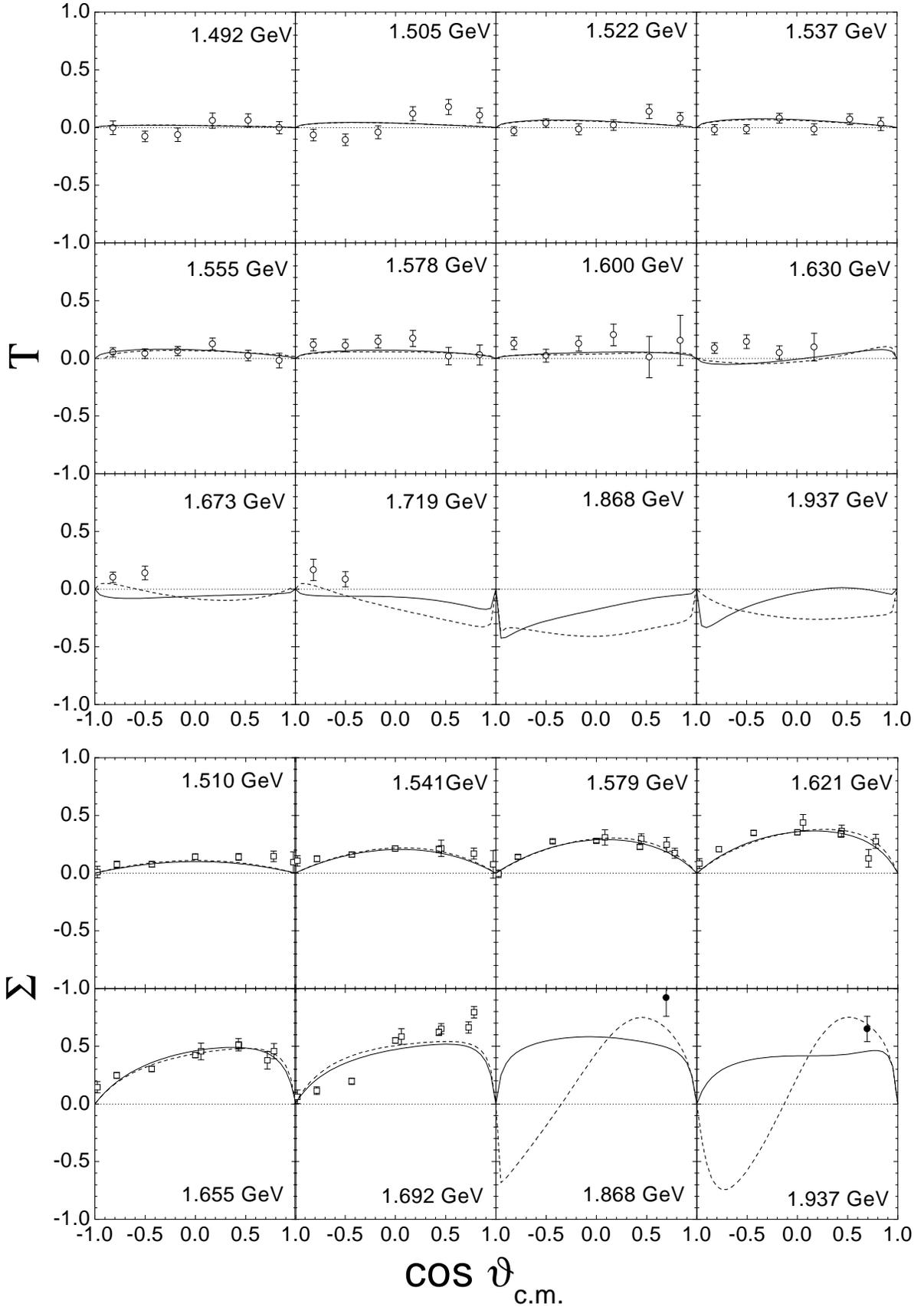}}
    \caption{$\gamma p \ra \eta p$ target- (\textit{upper panel}) and
      beam- (\textit{lower panel}) polarization measurements. Line code
      as in Fig. \ref{figggdif}. Data are
      as given in Sec. \ref{expdata}.
      \label{figgepol}}
  \end{center}
\end{figure}
which is in line with the preliminary CLAS \cite{dugger} and the older
experimental data\footnote{Although we have included the preliminary
  CLAS data \cite{dugger} in our data base and the displayed energy
  bins for the differential cross section are chosen
  accordingly, we do not reproduce these data here, because they have
  not yet been published by the CLAS Collaboration.}.

The resulting decomposition of the $\eta N$ photoproduction describes
the differential cross sections and polarization measurements very
well in the complete considered energy region; see
Figs. \ref{figgedif} and \ref{figgepol}. As pointed out in Sec.
\ref{expdata}
prior to the differential cross section measurements of the CLAS
Collaboration \cite{dugger}, there were hardly any measurements
taken above 1.7 GeV. Consequently, the preliminary CLAS data give
strong constraints on the reaction mechanism in the upper energy
region, which would otherwise be mainly determined by the 
pion-induced $\eta N$ data being of poor quality at higher energies;
see PMI \cite{pm1}.

It is interesting to note that we find a considerably smaller
$D_{13}(1520)$ $\eta N$ width than, e.g., Batini\'c {\it et
al.} \cite{batinic}. However, since the $D_{13}(1520)$ basically gives
the only contribution to the low energy behavior of the beam
polarization $\Sigma$ \cite{feusti99}, our value of around 20 KeV (as
compared to 140 KeV) is strongly corroborated by the measurements of
the GRAAL collaboration \cite{datage}, since these data are very well
described in the complete measured region; see Fig. \ref{figgepol}. 
Note also that Tiator {\it et al.} \cite{tiator99} deduced from the
GRAAL beam asymmetry data a $D_{13}(1520)$ $\eta N$ branching ratio of 
$0.8\pm 0.1$\permil, which is about half of our value. This is related 
to the fact, that in these authors' analysis, the PDG \cite{pdg}
electromagnetic helicity amplitudes have been used, which are larger
than the ones deduced from our analysis; see Table \ref{tabhelii12}
below. In Ref. \cite{tiator99} it was also shown, that the
forward-backward asymmetry of the beam polarization $\Sigma$ between
1.65 and 1.7 GeV (see Fig. \ref{figgepol}) can only be explained by
contributions with spin $J\geq \ffh$. Since in the present model no
$J\geq \ffh$ resonances are included, the asymmetric behavior
is generated by the vector meson exchanges. Since
the GRAAL data cannot be completely described at 1.69 GeV, this might
be an indication that spin-$\ffh$ resonances indeed play a role in
$\eta N$ photoproduction. 
At higher energies ($\sqrt s > 1.8$ GeV), an opposite behavior of
the beam asymmetry for our two calculations at backward angles is
observed. Since there are no data points, only the behavior at forward 
angles is fixed. The difference in the two calculations can be
explained by the opposite photon helicity amplitudes of the
$D_{13}(1950)$ (see Table \ref{tabhelii12} in Sec. \ref{secresheli}
below) and the different $\eta N$ strength (5.4\% for C-p-$\gamma +$
and 8.6\% C-p-$\gamma -$). Thus 
beam-asymmetry measurements at energies above 1.7 GeV for $\eta N$
photoproduction would be a great tool to study the properties of this
``missing'' resonance and also the necessity for the inclusion of a
spin-$\ffh$ resonance in more detail.

For the target polarization, we find small values in the complete
energy region; see Fig. \ref{figgepol}. 
Only in the lowest energy bins, the experimental data seem to indicate 
a nodal structure. Tiator {\it et al.} \cite{tiator99} showed, that this 
behavior can only be explained by a strong energy dependence of the
relative phase between the $S_{11}(1535)$ and $D_{13}(1520)$
contributions, which is not found in the present calculation. 
For the region above 1.6 GeV, our calculations change from positive to
negative values, which seems not to be supported by the Mainz data 
\cite{datage}
at backward angles. It turns out that the target polarization is
dominated in our calculation by the $P_{11}(1710)$ resonance
properties, and, hence, more experimental data on the target
polarization at higher energies would also help to clarify whether
this resonance plays such an important role in $\eta N$
photoproduction as found in the present analysis.

\subsection{$K\Lambda$ photoproduction}
\label{secresgl}

\begin{figure}
  \begin{center}
    \parbox{16cm}{
      \parbox{75mm}{\includegraphics[width=75mm]{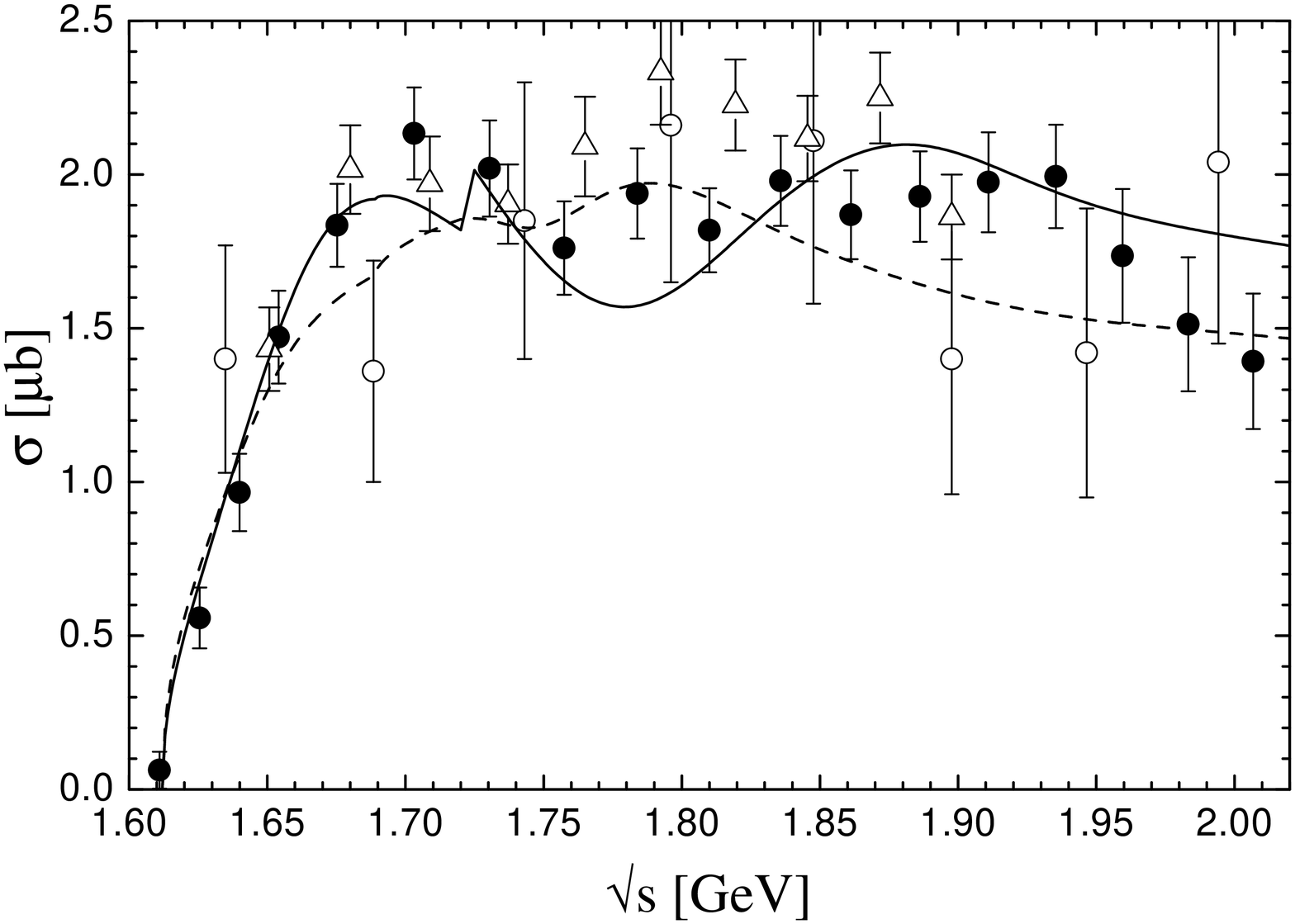}}
      \parbox{75mm}{\includegraphics[width=75mm]{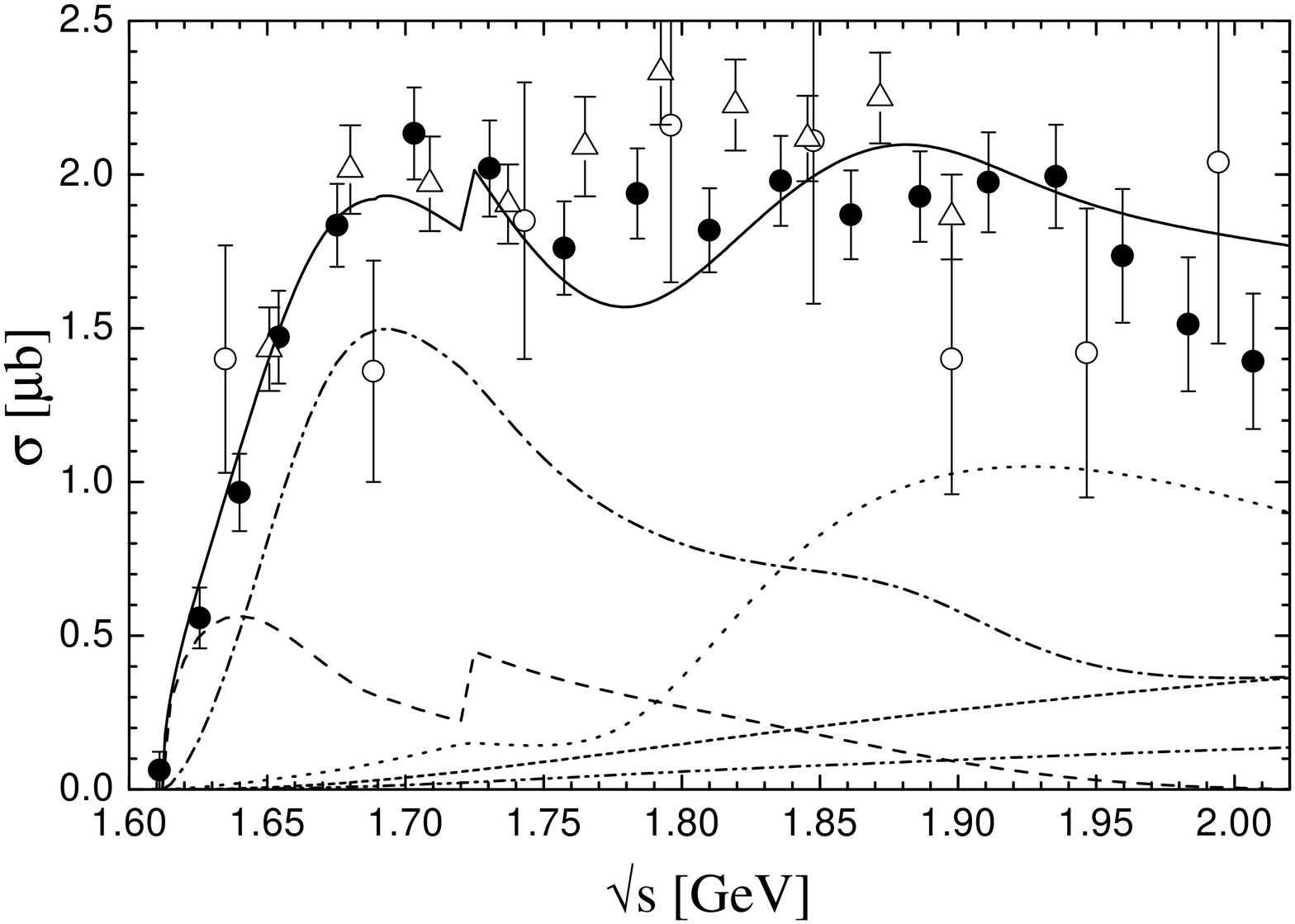}}
      }
    \caption{$\gamma p \ra K^+ \Lambda$ total cross section. Data are
      as given in Sec. \ref{expdata}.
      \textit{Left:} Line code as in
      Fig. \ref{figggdif}. \textit{Right:} Partial-wave
      decomposition. Notation as in Fig. \ref{figgetot}. In addition,
      the contribution of higher partial waves ($J\geq \ffh$) is
      indicated by the short-dashed line.
      \label{figgltot}}
  \end{center}
\end{figure}

The decomposition of the $K\Lambda$ photoproduction channel turns out
to be very similar to the pion-induced reaction. In contrast to
Feuster and Mosel \cite{feusti99}, where the $S_{11}(1650)$ and the
$P_{11}(1710)$ dominated this reaction, in the present calculation the 
former one turns out to be important only very close to threshold,
while the latter one hardly gives any sizeable contribution at all;
see Fig. \ref{figgltot}. 
\begin{figure}
  \begin{center}
    \parbox{16cm}{
    \parbox{16cm}{\includegraphics[width=16cm]{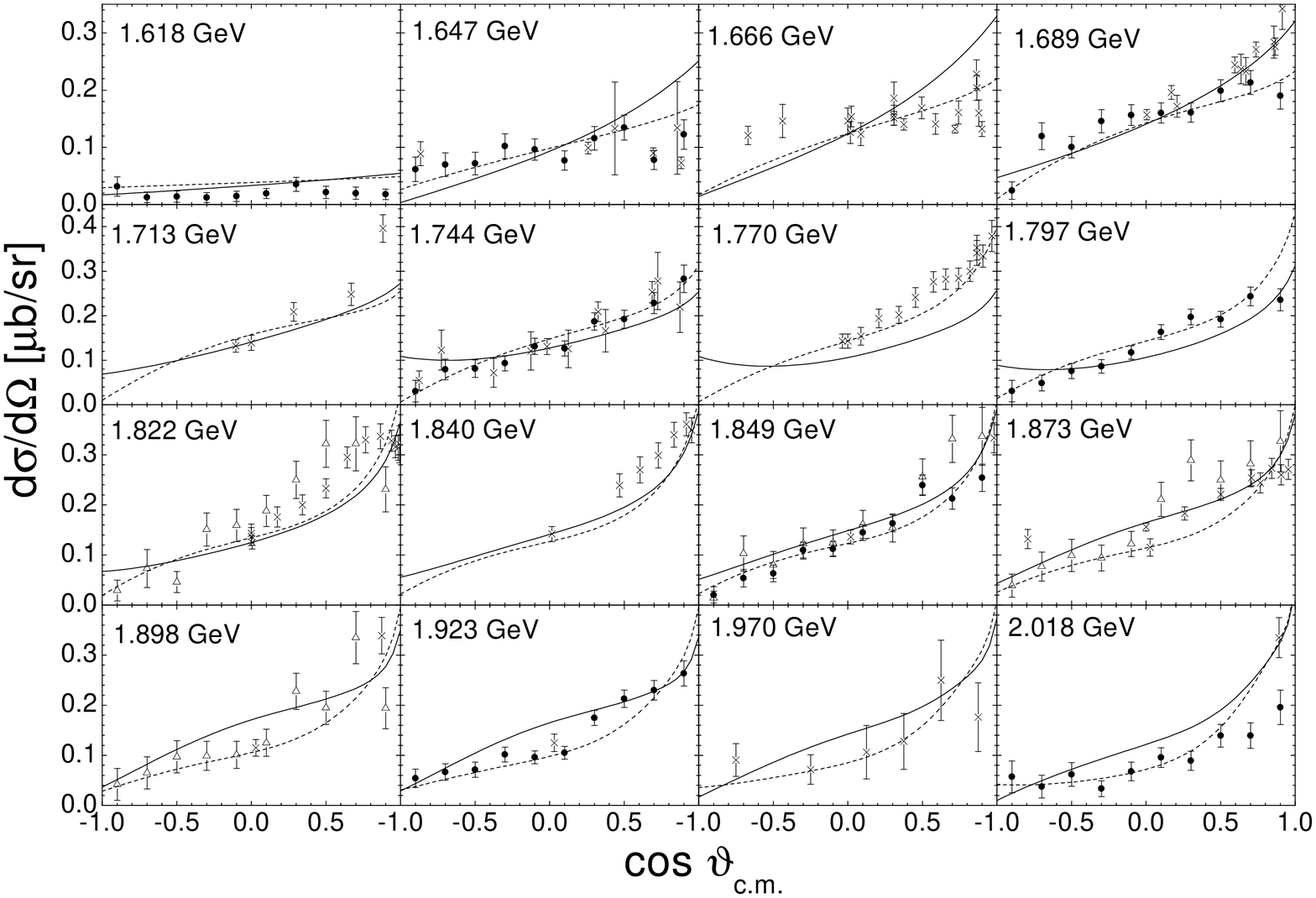}}
    \parbox{16cm}{\includegraphics[width=16cm]{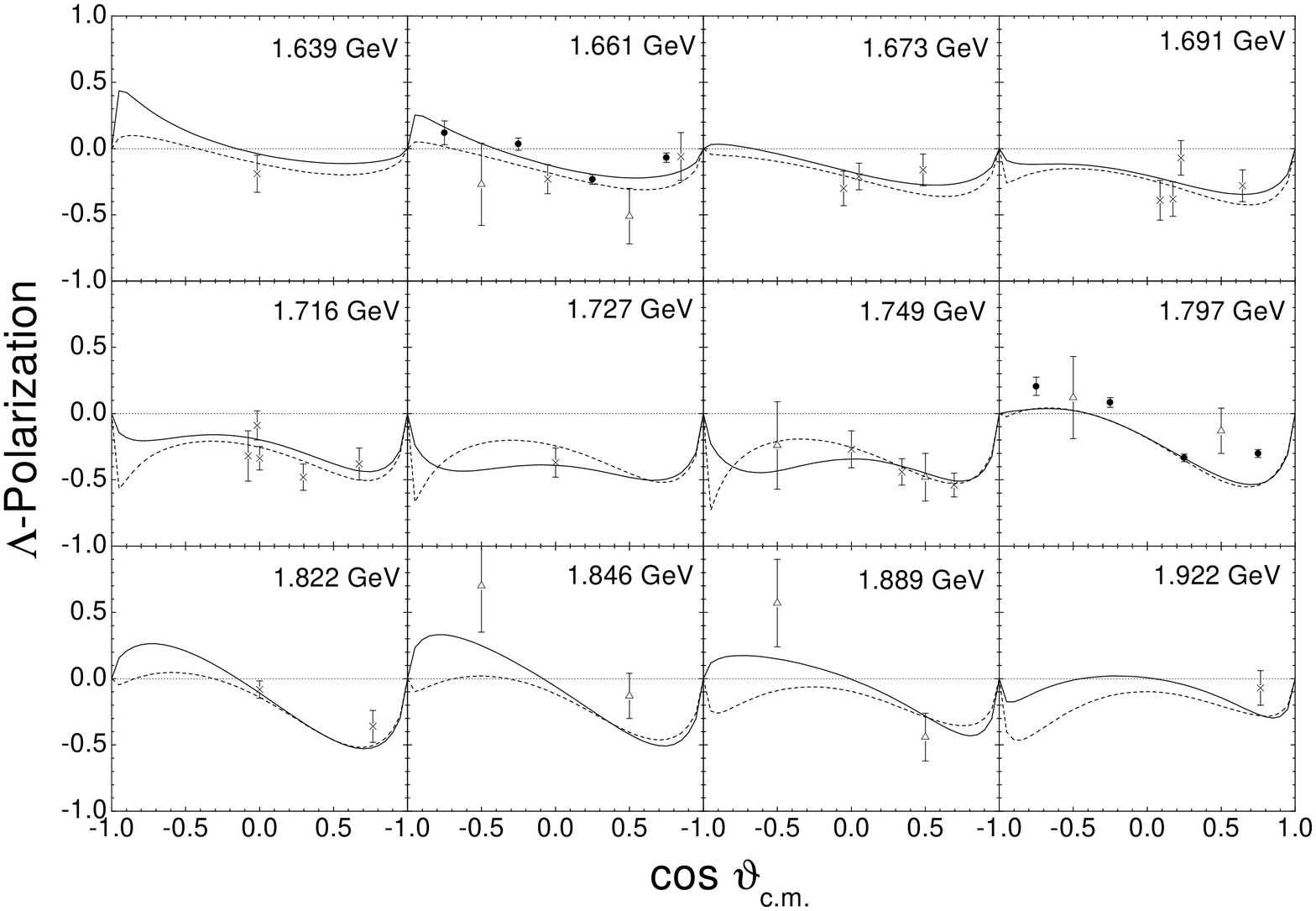}}
    }
    \caption{$\gamma p \ra K^+ \Lambda$ differential cross
      section (\textit{upper panel}) and $\Lambda$-recoil polarization
      (\textit{lower panel}). Line code as in Fig. \ref{figggdif}. Data
      are as given in Sec. \ref{expdata}.
      \label{figgldifpol}}
  \end{center}
\end{figure}
At low energies, the $P_{13}(1720)$
($J^P=\fth^+$) resonance is dominating, causing a resonant structure
around 1.7 GeV. At higher energies, the $P_{13}(1900)$ still makes
important contributions due to rescattering in spite of its small
$K\Lambda$ width. The strong $\foh^-$ contribution very close to
threshold, which is caused by the $S_{11}(1650)$, 
is strongly influenced by the $\omega N$ threshold leading to a
sudden increase in the total cross section. Note, that the finite
width of the $\omega$ meson of 8 MeV, which is not taken into account
in the present model, smears out this threshold effect. A similar 
observation of the feeding of $K\Lambda$ (and also $K\Sigma$, see
Sec. \ref{secresgs} below) photoproduction through threshold
effects has also been made in the coupled-channel model of Lutz {\it et
al.} \cite{lutz}. As a consequence 
of the inclusion of the $K^*$ and $K_1$ meson exchanges, we also find
important contributions to the total cross section by partial waves
with $J\geq \ffh$, cf. Fig. \ref{figgltot}. 

A striking difference to the pion-induced $K\Lambda$ production
mechanism is observed in the $\foh^+$ wave, which exhibits a structure 
resonating around 1.9 GeV, where a second peak is also visible in the
SAPHIR total cross section data \cite{datagl}.
However, there is no $P_{11}$ resonance included in the present model
around this energy. It turns out that the $\foh^+$ behavior is caused 
by the interference of the nucleon and $K^*$ contributions. Switching
these two contributions off leads to a $\foh^+$ wave, which is
practically zero for energies higher than the $P_{11}(1710)$
peak. This is in contrast to the findings of the single-channel
model of Mart and Bennhold \cite{martbenn}, where the peaking behavior
in the SAPHIR total cross section \cite{datagl}
was explained by the same $D_{13}(1950)$ resonance, which was
found by Feuster and Mosel \cite{feusti98,feusti99} around 1.9
GeV. This example emphasizes the importance of coupled-channel
analyses for the correct identification of missing
resonances. Although the $D_{13}(1950)$ is included in
the present calculation, in the simultaneous analysis of all channels
it turns out to be of negligible importance for $K\Lambda$
photoproduction. Similar results were already found by
Janssen {\it et al.} \cite{janssen01}. Using a field-theoretic model, these
authors deduced that the present $K\Lambda$-photoproduction data
alone are insufficient to identify the exact properties of a missing
resonance in a single-channel analysis on $K\Lambda$
photoproduction. Moreover, these properties also depend on the
background contributions. Since in the present model the background
is uniformly generated for the various reaction channels, and pion-
and photon-induced data are analyzed simultaneously,
the extracted background and resonance contributions are more strongly 
constrained than in Ref. \cite{martbenn}, and more reliable
conclusions can be drawn.

The recoil polarization (see Fig. \ref{figgldifpol}) is equally
well described in the two global calculations C-p-$\gamma +$ and
C-p-$\gamma -$, although the difference in the $g_{\omega \rho \pi}$
sign leads to changes in the $P$-wave resonance couplings. However,
since the differential cross section displayed in
Fig. \ref{figgldifpol} is $P$-wave dominated, slight changes in the
forward peaking and backward decrease can be seen in this
observable. This different behavior is the reason for the better
$\chi^2$ value of C-p-$\gamma -$ as compared to C-p-$\gamma +$, and
again shows, that $K\Lambda$ production reacts very sensitive on 
rescattering effects due to $\omega N$.

As a consequence of the inclusion of the photoproduction data, the
$NK\Lambda$ coupling is only reduced from $-18.8$ to $-12.2$ from the
best hadronic (C-p-$\pi +$; see PMI \cite{pm1}) to the best global
(C-p-$\gamma +$) fit (see also Table \ref{tabborncplgs} in Sec.
\ref{secresbackborn} below). Thus, in 
contrast to other models on $K\Lambda$ photoproduction, the resulting
agreement of the present calculation with experimental data is neither
achieved with a very low $NK\Lambda$ coupling far off SU(3)
predictions, nor with a very soft nucleon form factor; see Table
\ref{tabcutoff} in Sec. \ref{secresbackborn}. Note that the same
cutoff value $\Lambda_N = 0.96$ GeV is used in all nucleon $s$- and
$u$-channel diagrams.

\subsection{$K\Sigma$ Photoproduction}
\label{secresgs}

\begin{figure}
  \begin{center}
    \parbox{16cm}{
    \parbox{75mm}{\includegraphics[width=75mm]{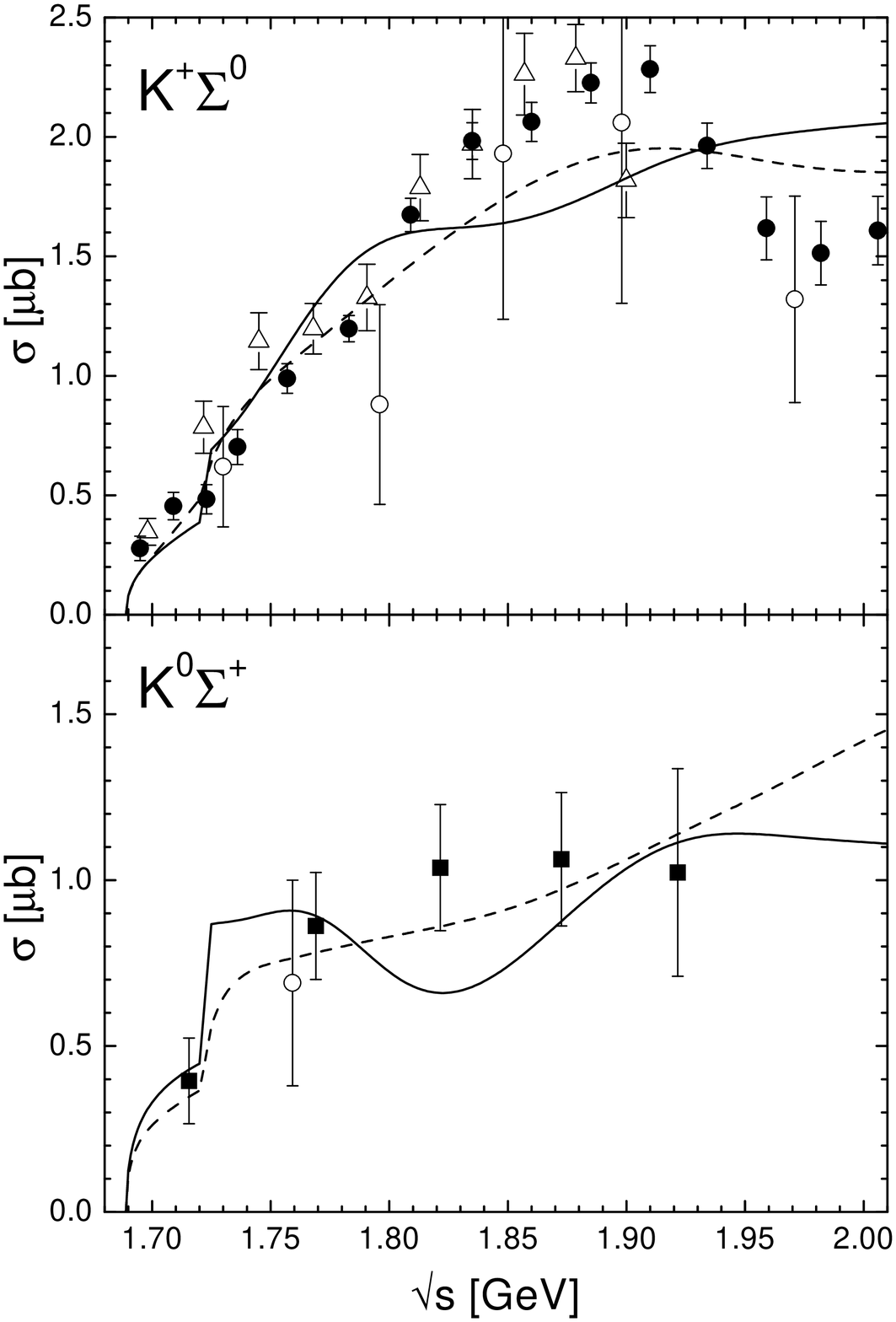}}
    \parbox{75mm}{\includegraphics[width=75mm]{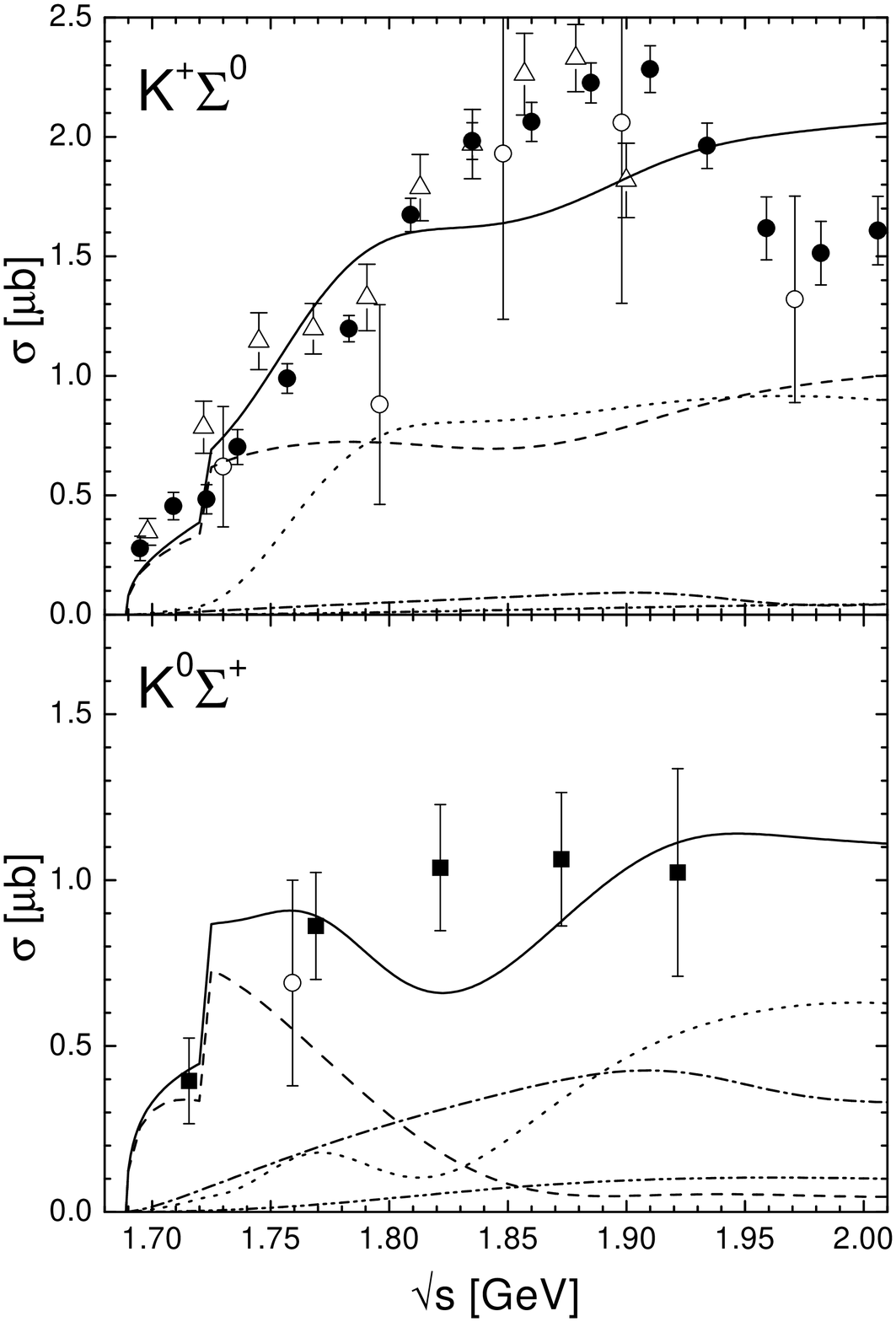}}
    }
    \caption{$\gamma p \ra K\Sigma$ total cross sections. Data are
      as given in Sec. \ref{expdata}.
      \textit{Left:} Line code as in
      Fig. \ref{figggdif}. \textit{Right:} Partial-wave
      decomposition. Line code as in Fig. \ref{figgetot}.
      \label{figgstot}}
  \end{center}
\end{figure}

As it turns out in the present model, it is also possible to
simultaneously describe both measured $\gamma p \ra K\Sigma$ charge
reactions (see Table \ref{tabchisquaregs} and Fig. \ref{figgstot}),
while still being in line with all three pion-induced $K\Sigma$ charge 
channels (see Table \ref{tabchisquares} and PMI \cite{pm1}). Similarly
to $K\Lambda$ photoproduction, the 
$K\Sigma$ mechanism also proves to be very sensitive to rescattering
effects via $\omega N$. The $IJ^P=\foh \foh^-$ $K\Sigma$ wave is fed
by the $\omega N$ channel, leading to a sudden increase in the
$K^+\Sigma^0$ and $K^0\Sigma^+$ total cross sections. As pointed out
in Sec. \ref{secresgl}, such an effect has also been observed in
the coupled-channel model of Lutz {\it et al.} \cite{lutz}. Note that the
finite width of the $\omega$ meson of 8 MeV, which is not taken into
account in the present model, smears out this threshold effect.
\begin{table}
  \begin{center}
    \begin{tabular}
      {l|c|c|c}
      \hhline{====}
       Fit & Total $\chi^2_{\gamma \Sigma}$ & 
       $\chi^2 (\gamma p \ra K^+ \Sigma^0)$ &
       $\chi^2 (\gamma p \ra K^0 \Sigma^+)$ \\
       \hline
       C-p-$\gamma +$ & 2.74 & 2.81 & 2.38 \\ 
       C-p-$\gamma -$ & 2.27 & 2.28 & 2.17 \\ 
      \hhline{====}
    \end{tabular}
  \end{center}
  \caption{Resulting $\chi^2$ of the two global fits for the two
    different charge reactions in $\gamma p \ra 
    K\Sigma$.\label{tabchisquaregs}}
\end{table}

The total cross section of $\gamma p \ra K^+\Sigma^0$ is dominantly
composed of $J^P=\foh^-$ and $\foh^+$ contributions, where the latter
is generated by the $P_{31}(1750)$ and $K^*$ exchange
contributions. The higher partial waves, especially those with $J\geq
\ffh$, hardly play any role. In the $\gamma p \ra K^0\Sigma^+$
reaction, the situation is changed in such a way that the
contribution of the $P_{11}(1710)$ becomes more pronounced, and the
the $J^P=\fth^+$ contribution due to the $P_{33}(1920)$ and in
particular the $P_{13}(1900)$ is emphasized. The $J^P=\fth^-$ and
higher partial-wave contributions remain negligible. A similar
decomposition of the $K\Sigma$-photoproduction mechanism was
found by Janssen {\it et al.} \cite{janssen02}. By applying a tree-level
isobar model, these authors were able to exclude any relevance
of the $D_{13}$ wave and to identify important
contributions from the $P_{11}(1710)$ and $S_{11}(1650)$ as in our
model. Also $P_{13}$, $S_{31}$, and $P_{31}$ contributions have been
identified, however, those have been attributed to the $P_{13}(1720)$,
$S_{31}(1900)$, and $P_{31}(1910)$ resonances instead of
$P_{13}(1900)$, $S_{31}(1620)$, and $P_{31}(1750)$ in the present
model. Note, that we have checked for the importance of $S_{31}(1900)$
and $P_{31}(1910)$ contributions within the present model (see PMI 
\cite{pm1}), but have not found any sizable contributions. 

\begin{figure}
  \begin{center}
    \parbox{16cm}{
    \parbox{16cm}{\includegraphics[width=16cm]{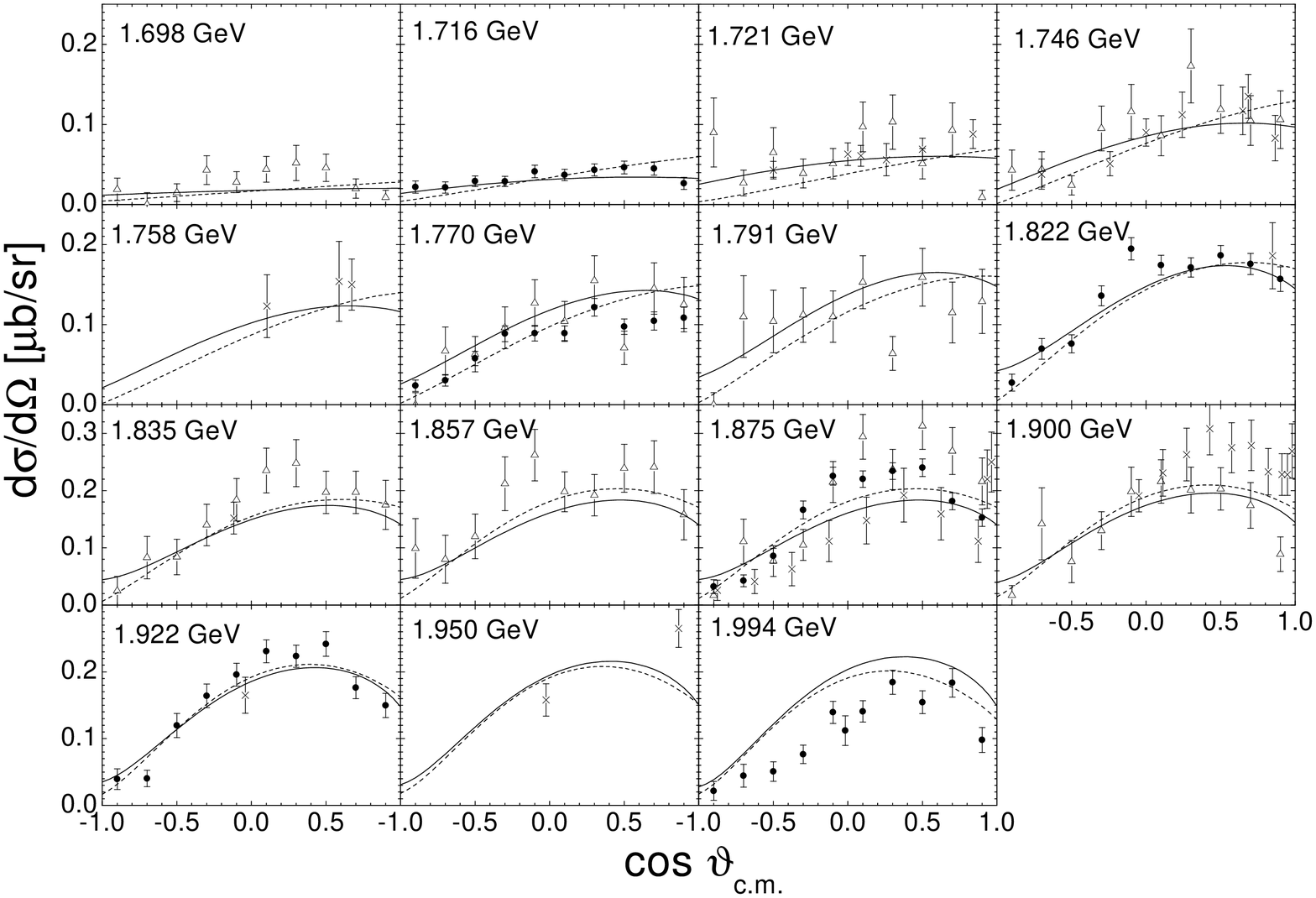}}
    \parbox{16cm}{\includegraphics[width=16cm]{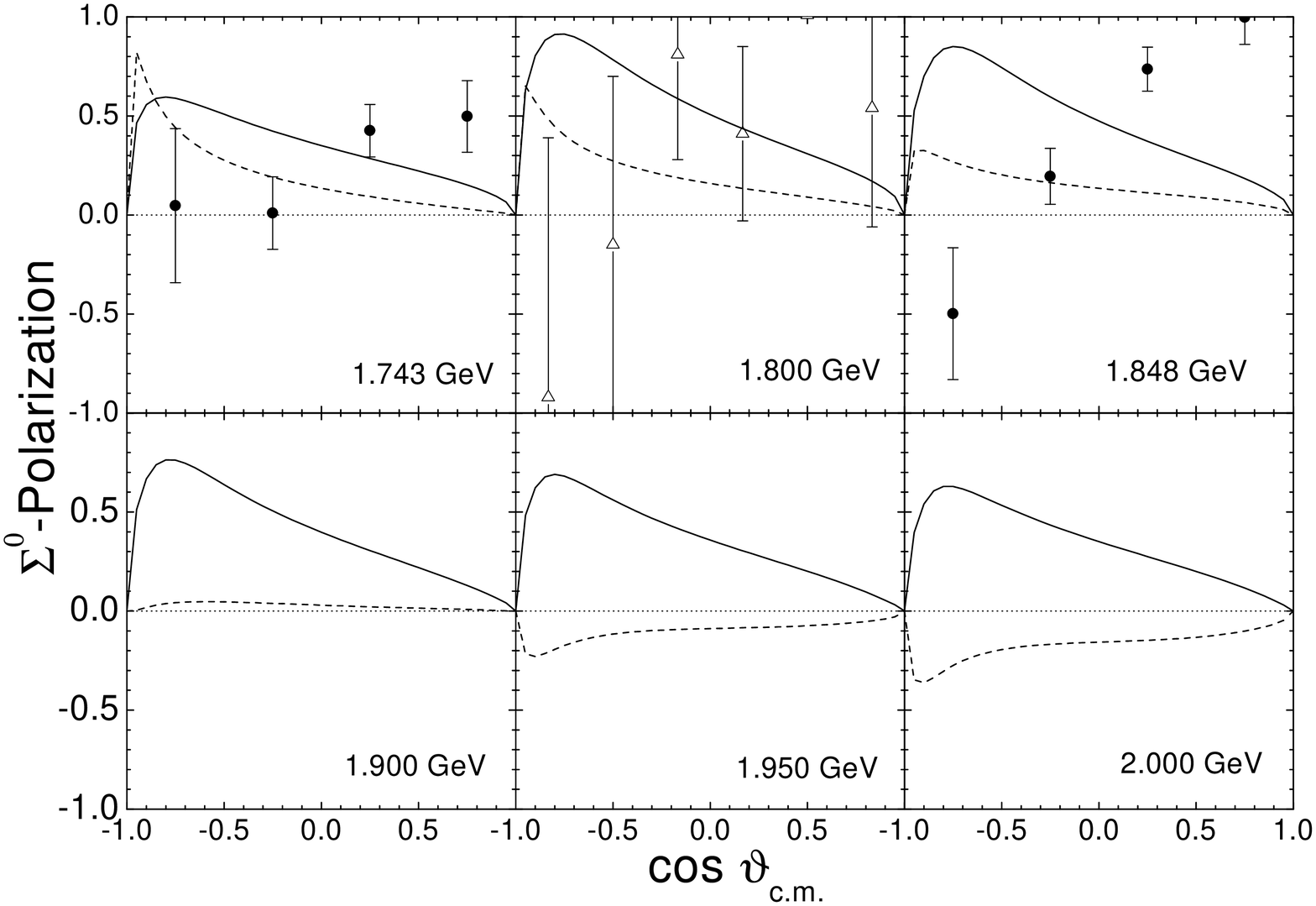}}
    }
    \caption{$\gamma p \ra K^+\Sigma^0$. \textit{Upper panel:}
      differential cross section, \textit{lower panel:} 
      $\Sigma^0$-recoil polarization. Line code as in
      Fig. \ref{figggdif}. Data are from Ref. \cite{datags}.
      \label{figgs0difpol}}
  \end{center}
\end{figure}

The differential cross section behavior of $\gamma p \ra K^+
\Sigma^0$, shown in Fig. \ref{figgs0difpol}, 
\begin{figure}
  \begin{center}
    \parbox{16cm}{\includegraphics[width=16cm]{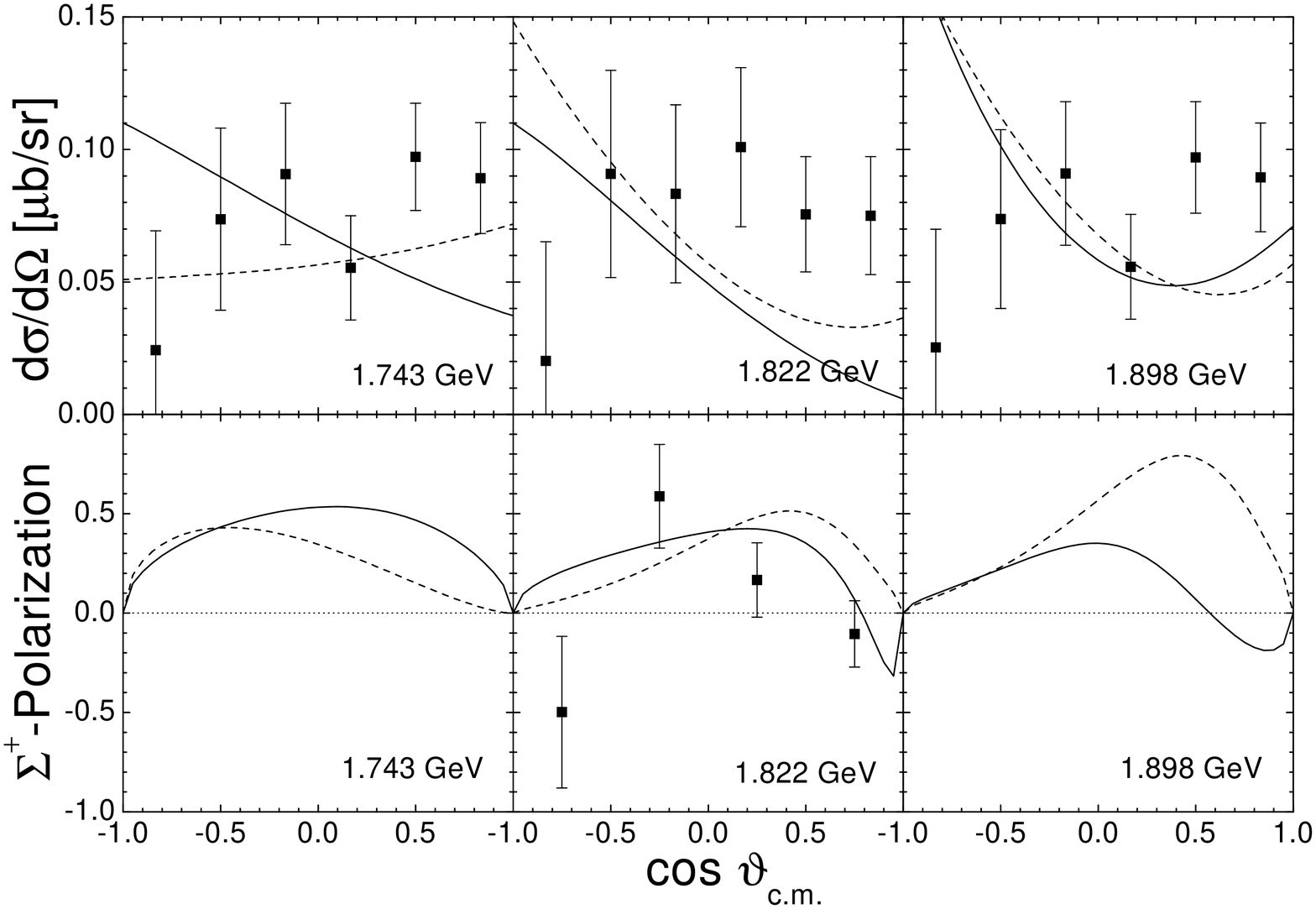}}
    \caption{$\gamma p \ra K^0\Sigma^+$ differential cross section and 
      $\Sigma^+$ recoil polarization. Line code as in
      Fig. \ref{figggdif}. Data are from Ref. \cite{datags}.
      \label{figgspdifpol}}
  \end{center}
\end{figure}
is very similar for the two global calculations C-p-$\gamma \pm$ with
different coupling signs of $g_{\omega \rho \pi}$. Both
describe the angular structure of the cross sections very well and show
a tendency to decrease at forward angles for higher energies, which is
caused by the $K^*$ exchange. In the $\Sigma^0$ recoil polarization of
$\gamma  p \ra K^+ \Sigma^0$, the two calculations C-p-$\gamma +$ and 
C-p-$\gamma -$ show behaviors opposite in sign for energies above 1.9 
GeV. This difference can
be traced back to the different $P_{11}(1440)$, $P_{11}(1710)$, and
$D_{13}(1950)$ contributions in the two calculations. Thus more
precise experimental data in the higher-energy region on the
$\Sigma^0$ polarization would certainly help to clarify the exact
decomposition. 

We also observe a very similar behavior of the two calculations for
the $\gamma p \ra K^0 \Sigma^+$ (see Fig. \ref{figgspdifpol})
differential cross section and  $\Sigma^+$
polarization. Unfortunately, the few SAPHIR data points \cite{datags} 
are not precise enough to judge the quality of the
description. 

As a result of the inclusion of the photoproduction data, the
$NK\Sigma$ coupling is reduced from 15.4 to 2.5 from the best hadronic 
(C-p-$\pi +$) to the best global (C-p-$\gamma +$) fit. As pointed
out in Sec. \ref{secresbackborn} and PMI \cite{pm1}, the pion-
induced reactions are only slightly influenced by the exact $NK\Sigma$
coupling value and are thus still be well described in the global
calculation. The final value for the $NK\Sigma$ coupling is close to
SU(3) expectations; see Sec. \ref{secresbackborn}.

\subsection{$\omega$ photoproduction}
\label{secresgo}

\begin{figure}
  \begin{center}
    \parbox{16cm}{
      \parbox{75mm}{\includegraphics[width=75mm]{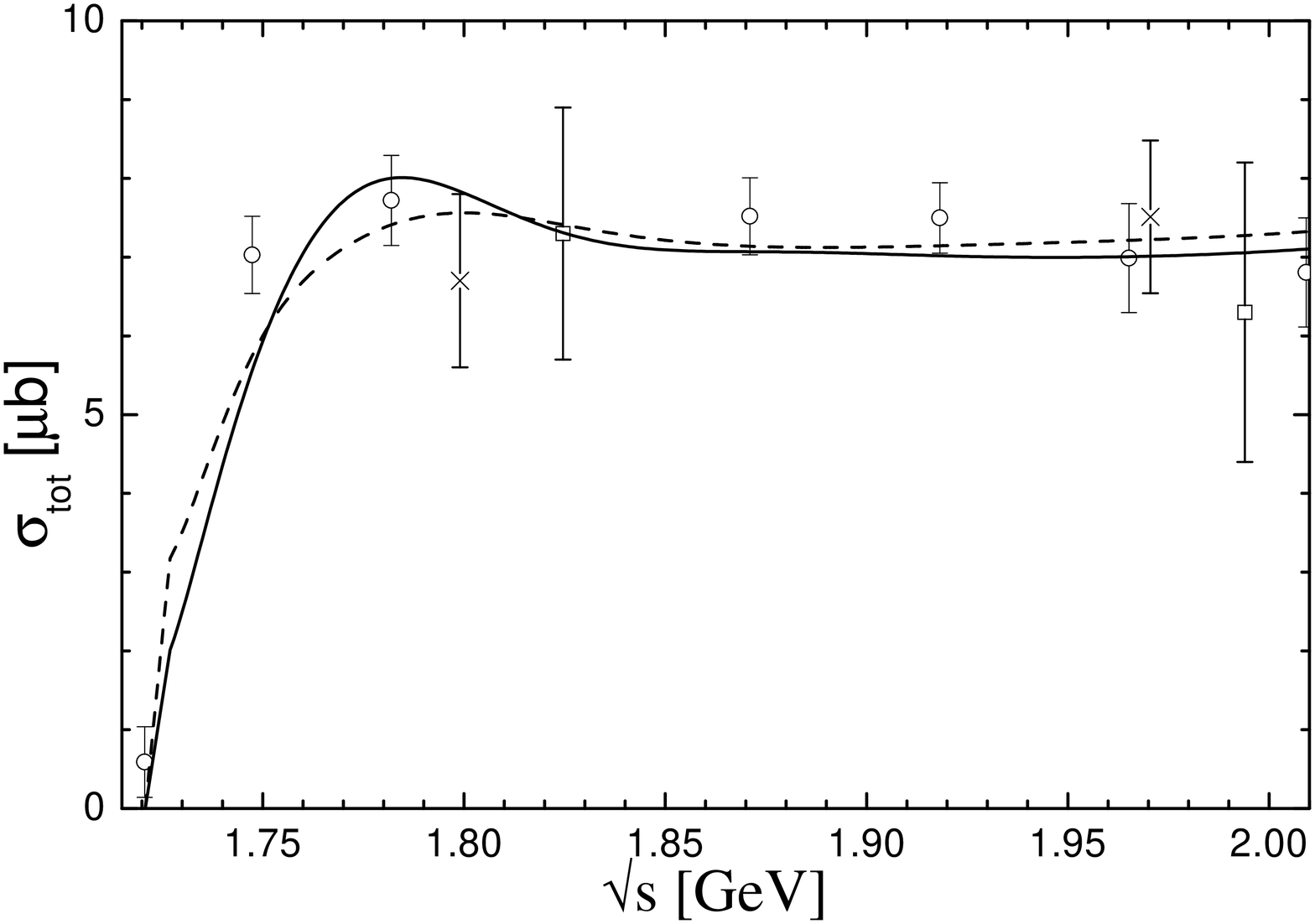}}
      \parbox{75mm}{\includegraphics[width=75mm]{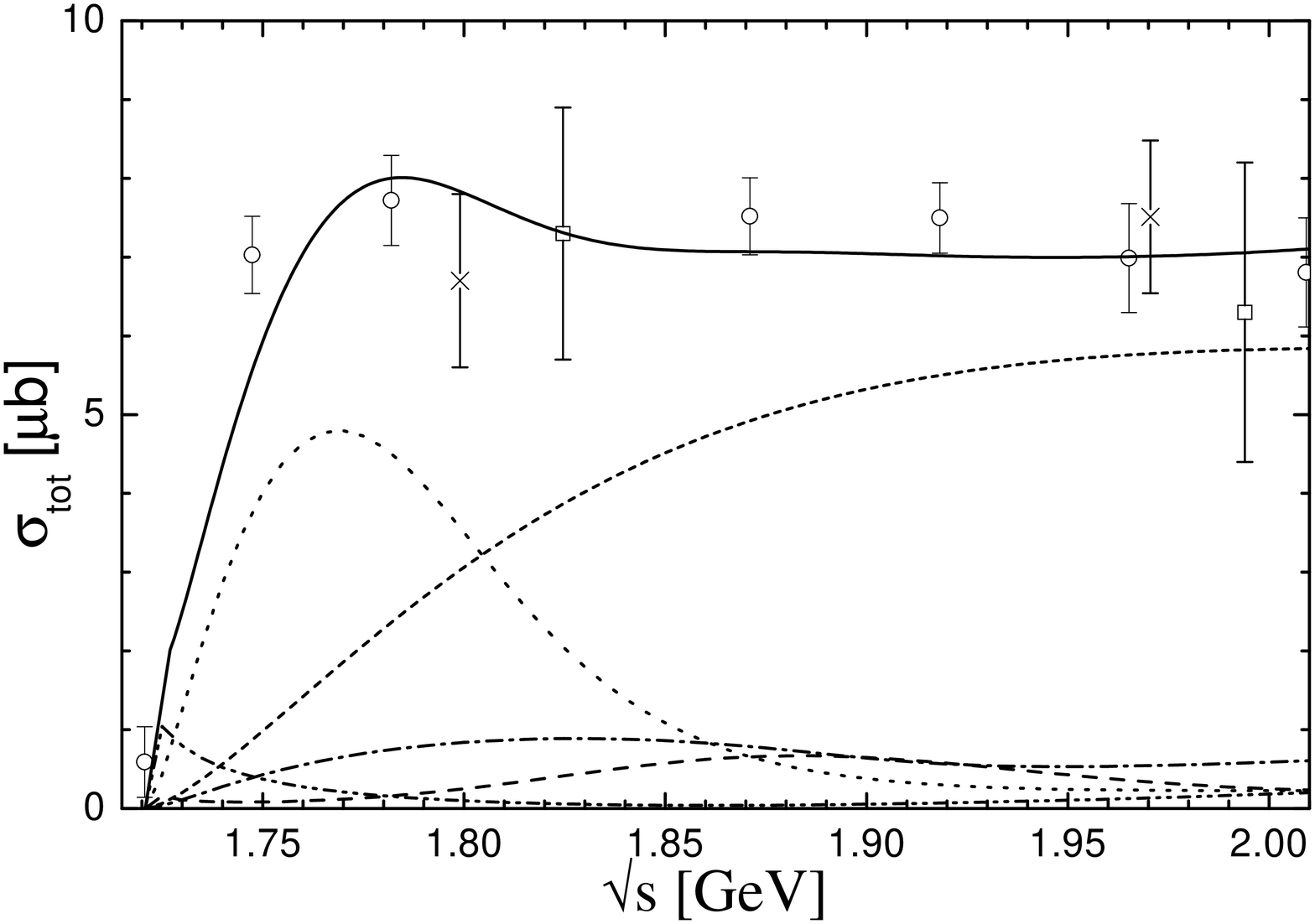}}
      }
    \caption{$\gamma p \ra \omega p$ total cross section. Data are from 
      $\circ$ \cite{kleingo}, $\times$ \cite{erbeom}, $\Box$
      \cite{crouch}. The data from Ref. \cite{barthom} are not shown.
      \textit{Left:} Line code as in
      Fig. \ref{figggdif}. \textit{Right:} Partial-wave
      decomposition. Notation as in Fig. \ref{figgltot}.
      \label{figgotot}}
  \end{center}
\end{figure}

The literature on $\omega$ photoproduction does not offer a clear
picture of the importance of individual resonance mechanisms in this
channel, which is due to the fact that basically all models are only
single-channel analyses. Hence rescattering effects and the impact of
the drawn conclusions on other channels are neglected. While Titov 
and Lee \cite{titov02} recently found important contributions of
the sub-threshold $D_{13}(1520)$ and $F_{15}(1680)$ resonances, Oh {\it et
al.} \cite{oh01} extracted dominant contributions from a $P_{13}(1900)$
and a $D_{13}(1960)$ resonance. Furthermore, in the model of Zhao
\cite{zhao01} the $P_{13}(1720)$ and $F_{15}(1680)$ were shown to give
dominant contributions, but the low lying $S_{11}(1535)$ and
$D_{13}(1520)$ were also important. All models agree, however, on the
importance of the $\pi^0$ exchange, which has already been considered
in one of the first models on $\omega$ photoproduction by Friman and
Soyeur \cite{frimansoyeur}. The higher partial-wave contributions of
the $\pi^0$ mechanism also dominate the cross
section behavior above $\sqrt s \approx 1.82$ GeV in the present
model; see Fig. \ref{figgotot}.
The clear dominating threshold contribution stems from the
$P_{11}(1710)$, just as in the pion-induced case (see PMI
\cite{pm1}). The importance of the other resonances, however, is  
reduced, and only the $J^P=\fth^+$ contributions of the $P_{13}(1720)$
and $P_{13}(1900)$ remain non-negligible. 

\begin{figure}
  \begin{center}
    \parbox{16cm}{
      \parbox{16cm}{\includegraphics[width=16cm]{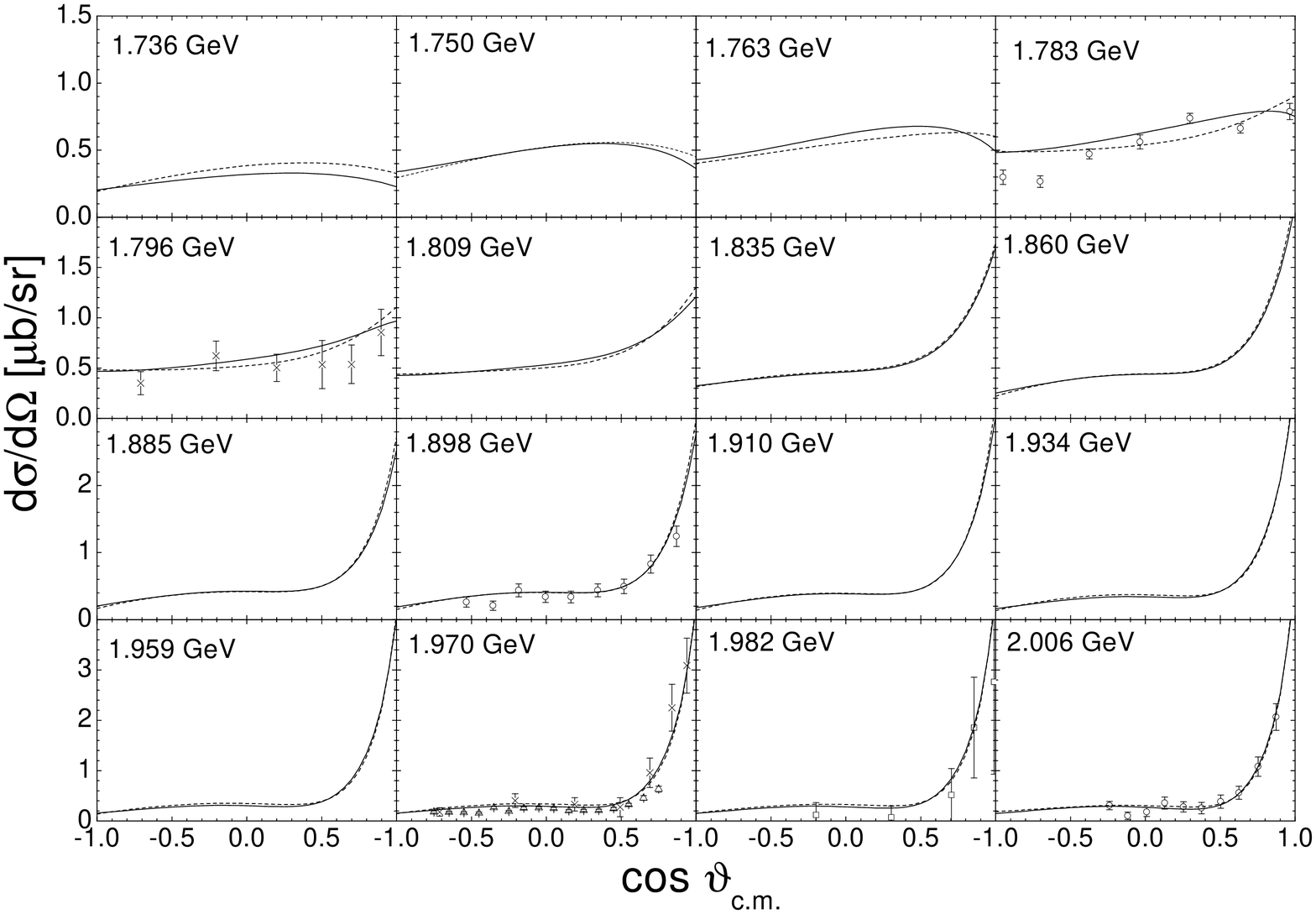}}
      \parbox{16cm}{\includegraphics[width=16cm]{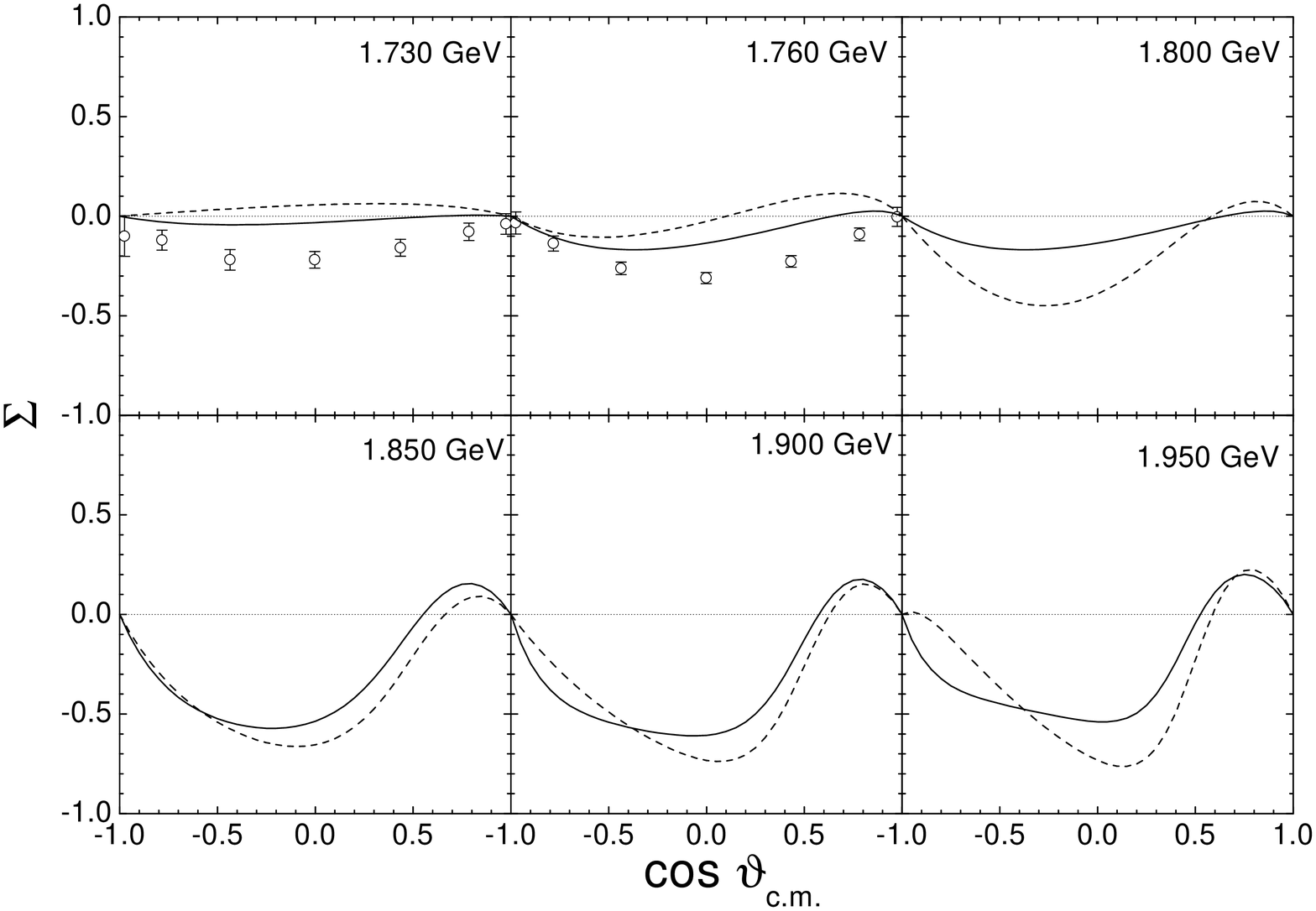}}
      }
    \caption{$\gamma p \ra \omega p$. Line code as in
      Fig. \ref{figggdif}. \textit{Upper panel:} differential cross 
      section. Data are as in Fig. \ref{figgotot}. The preliminary
      CLAS data \cite{manak} 
      ($\triangle$) have not been used in the fitting procedure.
      \textit{Lower panel:} Beam asymmetry $\Sigma$. Preliminary data
      are from Ref. \cite{ajakago}.
      \label{figgodifsig}}
  \end{center}
\end{figure}

\begin{figure}
  \begin{center}
    \parbox{16cm}{
      \parbox{75mm}{\includegraphics[width=75mm]{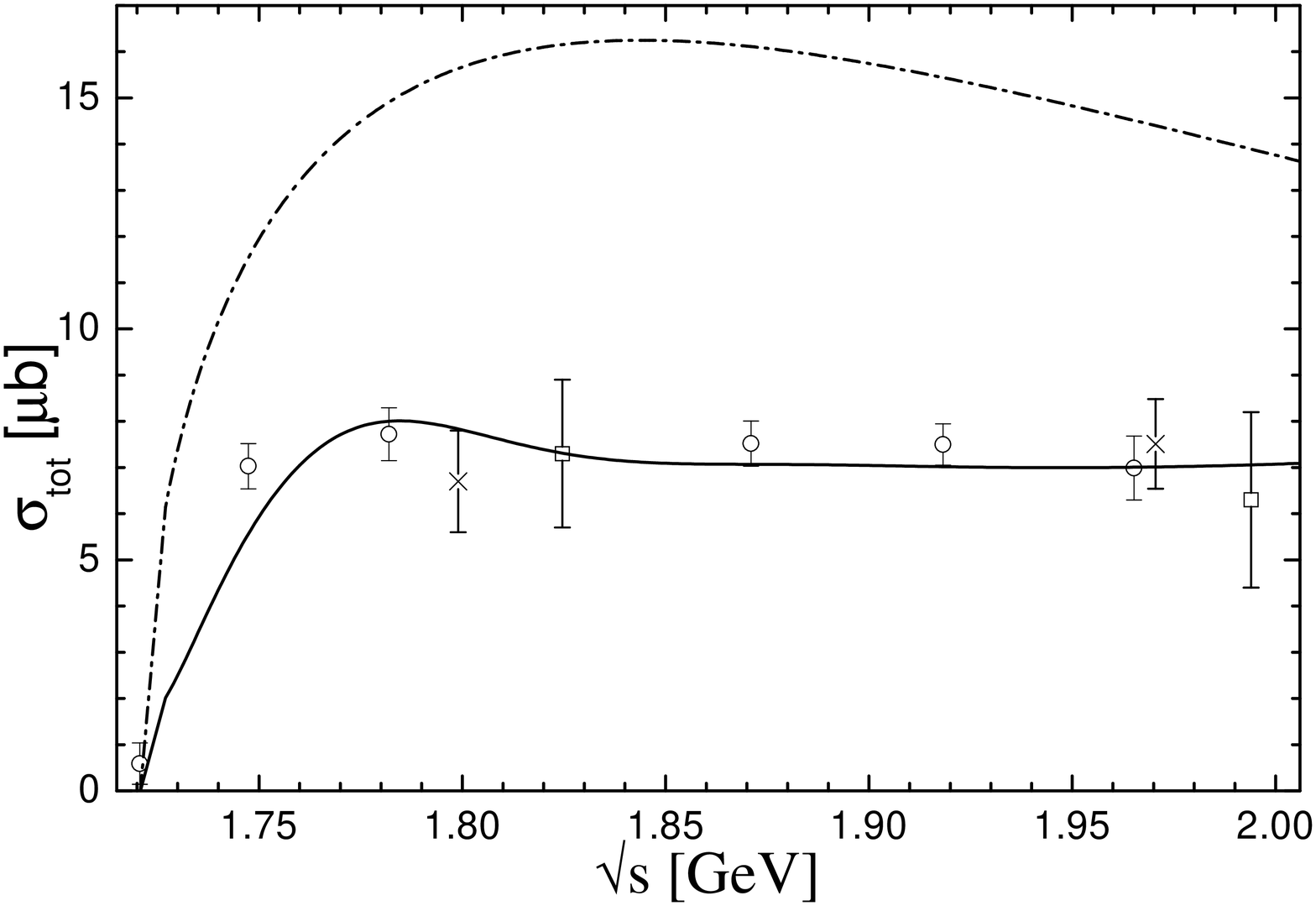}}
      \parbox{75mm}{\includegraphics[width=75mm]{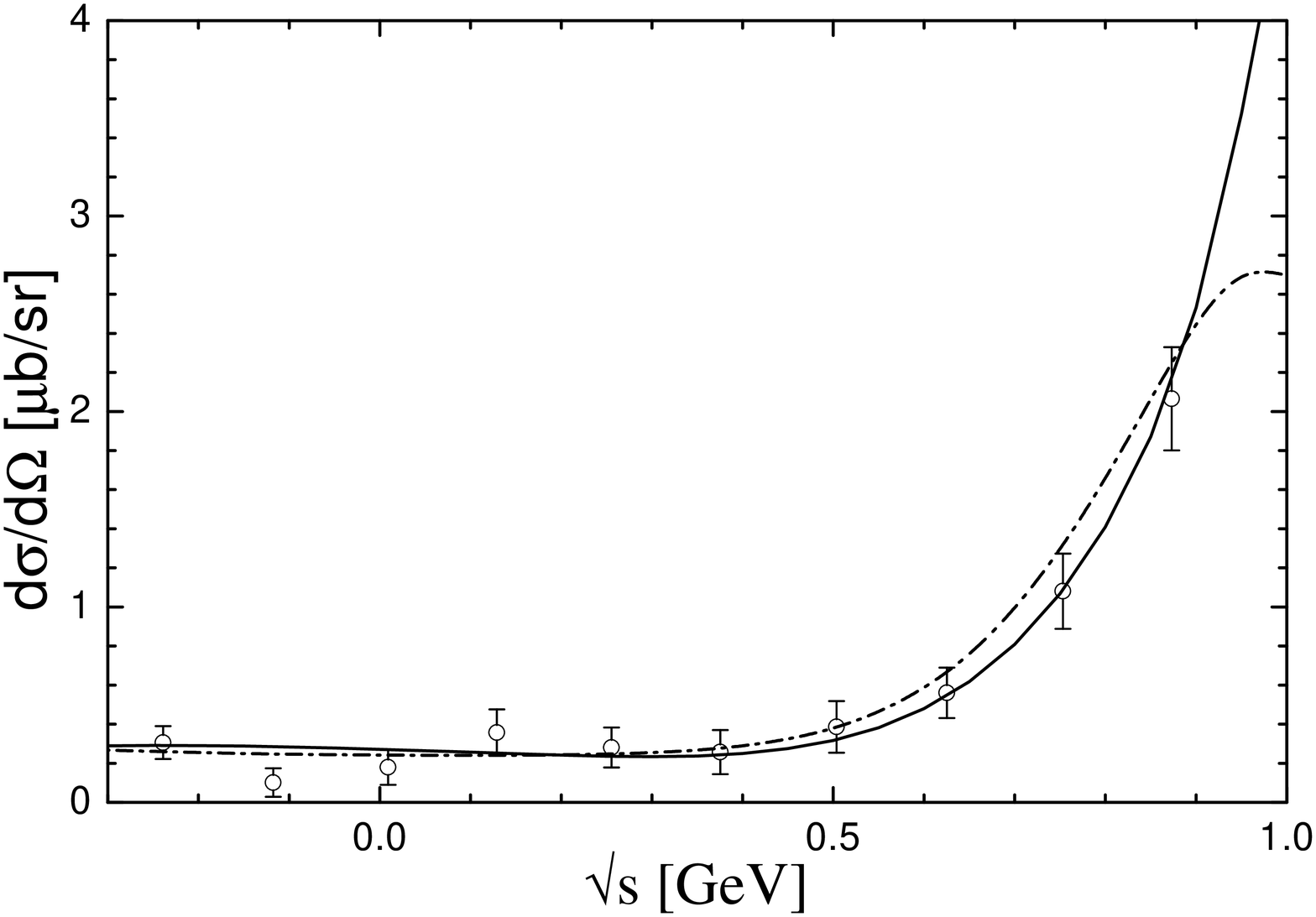}}
      }
    \caption{$\gamma p \ra \omega p$. Data as in
      Fig. \ref{figgotot}. \textit{Left:} total cross 
      section. Solid line: full calculation C-p-$\gamma
      +$. Dash-dotted line: C-p-$\gamma +$ with resonance
      contributions switched off. \textit{Right:} Solid line: full
      calculation C-p-$\gamma +$. Dash-dotted line: C-p-$\gamma +$
      with $J_{max}=\frac{11}{2}$; see the text.
      \label{figgochecks}}
  \end{center}
\end{figure}

The dominance of the $\pi^0$ exchange mechanism becomes even more
obvious in the differential cross section; see Fig. \ref{figgodifsig}.
However, in particular in the middle- and backward-angle region the
resonance contributions destructively interfering with the pion
exchange are mandatory to describe the precise preliminary SAPHIR
data \cite{barthom}, which cover the complete angular
range\footnote{Although we have included the preliminary 
  SAPHIR data \cite{barthom} in our data base and the displayed energy
  bins for the differential cross section are chosen
  accordingly, we do not reproduce these data here, because they have
  not yet been published by the SAPHIR Collaboration.}. When
these resonance contributions are neglected, the total cross section
behavior is strongly altered and the calculation largely
overestimates the total cross section; see Fig. \ref{figgochecks}.

The upper limit of the partial-wave decomposition $J_{max}$ turns out
to be essential for the $\omega$ photoproduction channel because of
the importance of the pseudoscalar $\pi^0$ exchange. Performing the
decomposition only up to $J_{max}=\frac{11}{2}$ as in Refs.
\cite{feusti98,feusti99}, the full upward bending behavior at forward
angles is not reproduced. This is displayed in
Fig. \ref{figgochecks}. We have checked for $J_{max}$ providing good
convergence in the angular structure and found a satisfying behavior
for $J_{max} \approx \frac{27}{2}$, which is consequently used in the
partial-wave decomposition for the present calculation. The necessity
of the consideration of higher partial waves when pseudoscalar
exchange mechanisms are included was also pointed out recently by 
Davidson and Workman \cite{davidson01b}. These authors demonstrated
striking differences in the forward peaking behavior for a
pion-photoproduction calculation at 1.66 GeV using the VPI multipoles
only up to $\ell_\pi=5$ ($\Leftrightarrow J_{max}=\frac{11}{2}$) or
additionally taking into account the full angular structure of the
Born terms, in particular the pion-Bremsstrahlung contribution.

Although the inclusion of the precise SAPHIR photoproduction data
\cite{barthom} allows for a better disentangling of the importance
of different resonances, the various resonance (helicity) couplings to
$\omega N$ cannot be fixed with certainty; see PMI \cite{pm1}. To
clarify the situation, there is an urgent need 
for data on polarization observables of $\omega N$ photoproduction,
as, e.g., currently extracted at GRAAL. For comparison, we give our
results on the beam asymmetry $\Sigma$ in
Fig. \ref{figgodifsig}. Note that the preliminary GRAAL data
\cite{ajakago} have not yet been included in the fit.

\subsection{Photoabsorption on the nucleon}
\label{secgdh}

\begin{figure}
  \begin{center}
    \parbox{8cm}{\includegraphics[width=8cm]{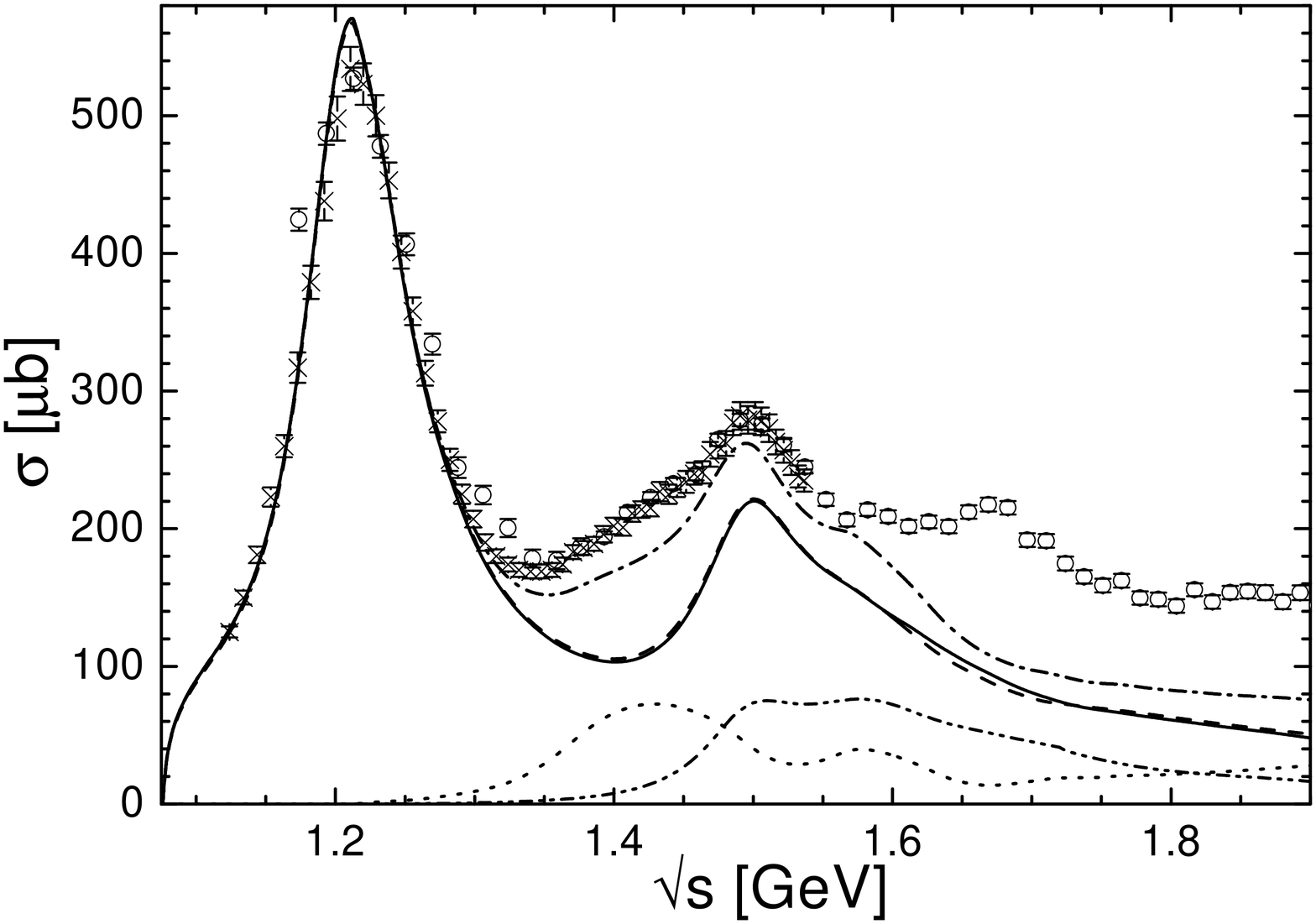}}
    \caption{Total photoabsorption cross section on the proton $\gamma
      p \ra X$. Calculation C-p-$\gamma +$: solid line; C-p-$\gamma
      -$: dashed line; cross section $\Delta \sigma$ of Ref. \cite{lvov}
      (see the text): dotted line; sum of calculation C-p-$\gamma +$ and 
      $\Delta \sigma$: dash-dotted line; $\gamma p \ra 2\pi N$ of
      C-p-$\gamma +$: dash-double-dotted line (see the text). Data are
      from $\times$ \cite{maccormick} and $\circ$ \cite{armstrong}.
      \label{figggabs}}
  \end{center}
\end{figure}

In the present model, we have included all important inelastic $\pi N$
channels below $\sqrt s = 2$ GeV, and, hence, we can also compare the
resulting total photoabsorption cross section $\sigma_{abs}^T = \foh
(\sigma^\foh_{abs} + \sigma^\fth_{abs})$ on the proton with
experimental data \cite{armstrong,maccormick}. As can be seen from
Fig. \ref{figggabs}
our model is in line with experiment all through the $\Delta(1232)$
region, but we cannot describe the total photoabsorption cross 
section $\sigma_{abs}^T$ above the $2\pi N$ threshold. 
This is not unexpected: the photoproduction of $2\pi N$ cannot be
described within our model as well as the pion-induced $2\pi N$
production, since in the photon-induced reaction, e.g., $\rho N$ 
or $\pi \Delta$ contact (Kroll-Rudermann like) interactions are also
known to be important \cite{hirata,murphy,oset}. 

In the dispersion theoretical analysis of Compton scattering by L'vov
{\it et al.} \cite{lvov}, exactly this part of the total photoabsorption
cross section has been determined. By subtracting from the
experimental total photoabsorption cross section $\sigma_{abs}^T =
\foh (\sigma^\foh_{abs} + \sigma^\fth_{abs})$ on the proton 
\cite{armstrong,maccormick} the 
single-pion photoproduction cross section, determined via the VPI
multipoles, and their $2\pi N$ cross section simulated via 
nucleon resonances, they extracted a remaining cross section $\Delta
\sigma$ supposed to be due to the aforementionned background
interactions. Ignoring interference effects (see Ref. \cite{lvov}),
one can just add $\Delta \sigma$ to our total photoabsorption cross
section. The resulting sum is remarkably close to the experimental
photoabsorption cross section \cite{armstrong,maccormick} 
up to about 1.6 GeV (see Fig. \ref{figggabs}), 
above which important contributions of spin-$\ffh$ resonances can be
expected, which are so far missing in our analysis. Thus it seems
that the resonance contributions to the $2\pi N$ photoproduction,
displayed in Fig. \ref{figggabs} by the dash-double-dotted line, 
are rather well described within the present model. This
provides an additional cross check that at least up to 1.6 GeV all
important channels are correctly described in our model. Above 1.6
GeV the data of the ABBHHM Collaboration on $3\pi N$ photoproduction
\cite{erbeom} indicate that this channel contributes $\approx
30-40$ $\mu$b to $\sigma^T_{abs}$, less than 10 $\mu$b of which are
due to $\omega N$.

Realizing the limitations of the present model, we can nevertheless
give estimates on the contributions of the various final states to
the Gerasimov-Drell-Hearn (GDH) sum rule \cite{gdh} (see also Ref.
\cite{drechsel} and references therein), which allows one to relate
the static property of the anomalous magnetic 
moment of the nucleon to the photoabsorption cross section
difference $\sigma^\foh_{abs} - \sigma^\fth_{abs}$ via
dispersion relations. The contributions of the individual reactions on 
the proton target up to $\sqrt s = 2$ GeV are given in Table
\ref{tabgdh}.
\begin{table}[hbt]
  \begin{center}
    \begin{tabular}
      {l|l|l|l|l|l}
      \hhline{======}
       $\;\; \pi N$ & $2\pi N$ & $\;\eta N$ & $\;K\Lambda$ &
       $\;K\Sigma$ & $\;\omega N$ \\
       \hline
       $-157.5$ & $-21.2$ & $\;\: +9.2$ & $+1.1$   & $+1.6$   & $+0.8$ \\ 
       $-162.7$ & $-20.7$ & $\;\: +8.6$ & $+0.9$   & $+1.8$   & $+0.1$ \\ 
       $-171^a$ & $-45^a$ &     $+15^a$  & $+1.7^b$ & $+2.4^b$ & $-2.0^c$ \\ 
      \hhline{======}
    \end{tabular}
  \end{center}
  \caption{Contributions (in $\mu$b) of the individual final states to
    the GDH sum rule up to $\sqrt s = 2$ GeV on the proton
    target. First line: Calculation C-p-$\gamma +$; second line:
    C-p-$\gamma -$; third line: Values are from Refs. $^a$: 
    \cite{tiatorgdh}, $^b$: \cite{sumowidagdo}, and $^c$: \cite{zhao02}.
    \label{tabgdh}}
\end{table}
As is clear from the above discussion, our estimates for $\pi N$ and
$2\pi N$ deviate from the rather well known values for reasons well
understood. For all other final states ($\eta N$, $K\Lambda$,
$K\Sigma$, and $\omega N$) our model is compared to all available
experimental observables and thus allows for reasonable estimates of
the contributions to the GDH sum rule. It is interesting to note (see
Table \ref{tabgdh}) that our values for the contributions from $\eta
N$, $K \Lambda$, $K \Sigma$, and in particular $\omega N$ deviate from
the values of Refs. \cite{tiatorgdh,sumowidagdo,zhao02}, all of which
have been extracted in single-channel analyses.

\subsection{Born and background couplings}
\label{secresbackborn}

The values of all Born and background couplings of our two global fits
and the extracted cutoff values $\Lambda$ are summarized in Tables
\ref{tabborncplgs} and \ref{tabcutoff}.
\begin{table}[t]
  \begin{center}
    \begin{tabular}
      {l|r|l|r|l|r|l|r}
      \hhline{========}
      $g$ & Value & $g$ & Value & $g$ & Value & $g$ & Value \\
      \hhline{========}
      $g_{NN\pi}$      &  12.85 & $g_{NN\sigma} \cdot g_{\sigma \pi \pi}$ & $ 11.46$ & $g_{NN\rho}$       &   4.53 & $\kappa_{NN\rho}$     &   1.47 \\
      &  12.75 & & $ 12.57$ & &   4.40 & &   1.41 \\
      \hline
      $g_{NN\eta}$     &   0.10 & $g_{NNa_0}$ & $-35.30$ & $g_{NN\omega}$     &   3.94 & $\kappa_{NN\omega}$        &  $-0.94$ \\
      &   0.12 & & $-22.91$ & &   3.87 & &   0.17 \\
      \hline
      $g_{N\Lambda K}$ & $-12.20$ & $g_{N\Lambda K_0^*}$ &  26.27 & $g_{N\Lambda K^*}$ & $-27.61$ & $\kappa_{N\Lambda K^*}$ &  $-0.50$ \\
      & $-12.88$ & & $  1.16$ & & $-28.29$ & &  $-0.55$ \\
      \hline
      $g_{N\Sigma K}$  &   2.48 & $g_{N\Sigma K_0^*}$ & $-26.15$ & $g_{N\Sigma K^*}$  &   4.33 & $\kappa_{N\Sigma K^*}$  &  $-0.86$ \\
      &   1.56 & & $-27.22$ & &    3.88  & &  $-0.98$ \\
      \hline
      $g_{N\Lambda K_1}$ & $-19.20$ & $\kappa_{N\Lambda K_1}$ & $-1.83$ & $g_{N\Sigma K_1}$ & $22.80$ & $\kappa_{N\Sigma K_1}$  &  $2.40$ \\
      & $-24.35$ & &  $-1.99$ & &  23.29 & &   2.06 \\
      \hhline{========}
    \end{tabular}
  \end{center}
  \caption{Nucleon and $t$-channel couplings. First line: C-p-$\gamma
    +$; second line: C-p-$\gamma -$. 
    \label{tabborncplgs}}
\end{table}
\begin{table}[t]
  \begin{center}
    \begin{tabular}
      {r|r|r|r|r|r}
      \hhline{======}
      $\Lambda_N$ [GeV] & 
      $\Lambda^h_\foh$ [GeV] & $\Lambda^\gamma_\foh$ [GeV] & 
      $\Lambda^h_\fth$ [GeV] & $\Lambda^\gamma_\fth$ [GeV] & 
      $\Lambda^h_t$ [GeV] \\
      \hhline{======}
      0.96 & 4.00 & 1.69 & 0.97 & 4.30 & 0.70 \\
      0.96 & 4.30 & 1.59 & 0.96 & 4.30 & 0.70 \\
      \hhline{======}
    \end{tabular}
  \end{center}
  \caption{Cutoff values for the form factors. First line: C-p-$\gamma
    +$; second line: C-p-$\gamma -$. The upper
    index $h$ or $\gamma$ denotes whether the value is applied to a
    hadronic or electromagnetic vertex, while the lower one denotes
    the particle going off-shell, i.e., $N$: nucleon; $\foh$:
    spin-$\foh$ resonance; $\fth$: spin-$\fth$ resonance; $t$:
    $t$-channel meson.
    \label{tabcutoff}}
\end{table}
Since these values were already discussed in PMI \cite{pm1}, we
only outline the main properties.

When realizing that the values of $g_{\pi NN}$ are lower than the
values extracted by other groups, for example the value of $g_{\pi NN}
= 13.13$ from the VPI group \cite{SM00}, one has to keep in
mind that the present calculation considers a large energy region
using only one $\pi NN$ coupling constant, and that the $\pi
NN$coupling is especially influenced by the $t$-channel 
pion exchange mechanism of $\omega N$ photoproduction. Remember that
only one cutoff value $\Lambda_t=0.7$ GeV (see Table \ref{tabcutoff})
is used for all $t$-channel 
diagrams. As a result of gauge invariance, the importance of the Born
diagrams is enhanced in the photoproduction reactions and
consequently, the other Born couplings can also be more reliably
extracted in the global calculations than when just the pion-induced
data is considered. As found in previous analyses
\cite{sauermann,feusti98,feusti99} the $\eta NN$ coupling turns out to 
be very small and the precise value thus hardly influences the
$\chi^2$ of the $\eta N$ production. The $K\Lambda$ and $K\Sigma$
couplings turn out to be larger than extracted in other
calculations. Thus the resulting relations between the Born couplings
for the pseudoscalar mesons of our best global fit are actually close
to SU(3) relations with $\alpha_{FD} = F/(F+D) \in [0.25;0.41]$ (see, e.g., 
Ref. \cite{dumbrajs}), which is around the value o $\alpha_{FD} \approx
0.35$ predicted by the Cabibbo-theory of weak interactions and the
Goldberger-Treiman relation \cite{dumbrajs}. Furthermore, the $\omega
NN$ coupling constants are also larger than extracted in other
calculations, which is only possible since rescattering effects are
properly taken into account in the present model. Note, that our value
for the nucleon cutoff $\Lambda_N=0.96$ GeV (see Table \ref{tabcutoff}) 
is the same for all final states.

The only $t$-channel meson which exclusively contributes to
photoproduction reactions in the present model, is the $K_1$
meson. Although the couplings are almost identical in both
calculations, we find that it only plays a minor role in $K\Lambda$
and $K\Sigma$ photoproduction; far more important are the
contributions from $K^*$ exchange (see also Secs. \ref{secresgl}
and \ref{secresgs}).

\subsection{Resonance electromagnetic helicity amplitudes}
\label{secresheli}

In Tables \ref{tabhelii12}, \ref{tabhelii32}, and
\ref{tabresagamma} the extracted electromagnetic properties of the 
resonances are summarized in comparison with the values of the PDG
\cite{pdg}, Feuster and Mosel \cite{feusti99}, and the pion
photoproduction analysis of Arndt {\it et al.} \cite{arndt02}.
\begin{table}[t]
  \begin{center}
    \begin{tabular}
      {l|ll|ll}
      \hhline{=====}
      $L_{2I,2S}$ & $A_\foh^p$ & $A_\foh^n$ & $A_\fth^p$ & $A_\fth^n$ \\
      \hhline{=====}
      $S_{11}(1535)$ & $\;\:90/93$ &  $-24/-34$ & \multicolumn{2}{c}{---} \\
                     & $\;\:90(30)$ & $-46(29)$ & \multicolumn{2}{c}{---} \\
                     & 106 & $-63$ & \multicolumn{2}{c}{---} \\
                     & \multicolumn{2}{c|}{NG} & \multicolumn{2}{c}{---} \\
      \hline 
      $S_{11}(1650)$ & 49/47 & $-11/-13$ & \multicolumn{2}{c}{---} \\
                     & 53(16) & $-15(21)$ & \multicolumn{2}{c}{---} \\
                     & 45 & $-26$ & \multicolumn{2}{c}{---} \\
                     & 74(1) & $-28(4)$ & \multicolumn{2}{c}{---} \\
      \hhline{=====}
      $P_{11}(1440)$ & $-87/-81$ & 121/112 & \multicolumn{2}{c}{---} \\
                     & $-64(4)$ & $\;\: 40(10)$ & \multicolumn{2}{c}{---} \\
                     & $-84$ & $\;\: 47$ & \multicolumn{2}{c}{---} \\
                     & $-67(2)$ & $\;\: 47(5)$ & \multicolumn{2}{c}{---} \\
      \hline 
      $P_{11}(1710)$ &   44/28 &  $-24/41$ & \multicolumn{2}{c}{---} \\
                     & $\;\: 9(22)$ & $\;\: -2(14)$ & \multicolumn{2}{c}{---} \\
                     & 19 & $-19$ & \multicolumn{2}{c}{---} \\
                     & \multicolumn{2}{c|}{NG} & \multicolumn{2}{c}{---} \\
      \hhline{=====}
      $P_{13}(1720)$ & $-53/-65$ & $-4/3$ & $\;\;\: 27/34$ &  $\;\;\;\;\: 3/2$ \\
                     & $\;\;\: 18(30)$ & $\;\;\: 1(15)$ & $-19(20)$ & $-29(61)$ \\
                     & $\;\;\: 23$ & $\;\;\: 2$ & $\;\;\: 75$ & $-17$ \\
                     & \multicolumn{2}{c|}{NG} & & \\
      \hline 
      $P_{13}(1900)$ &  $-17/-18$ &  $-16/-21$ &   31/8 &   $-2/-28$ \\
                     & & & & \\
                     & \multicolumn{2}{c|}{NC} & & \\
                     & & & & \\
      \hhline{=====}
      $D_{13}(1520)$ &  $\;\: -3/1$ &  $-84/-74$ &  151/153 & $-159/-161$ \\
                     & $-24(9)$ & $-59(9)$ & 166(5) & $-139(11)$ \\
                     & $\;\;\;\;\: 3$ & $-47$ & 136 & $\;\:-98$ \\
                     & $-24(2)$ & $-67(4)$ & 135(2) & $-112(3)$ \\
      \hline 
      $D_{13}(1950)$ &  $12/-1$ &   $23/-15$ &  $-10/-22$ &   $\;\:-9/22$ \\
                     & & & & \\
                     & $\;\: 5$ & 47 & $\;\;\: 41$ & $-55$ \\
                     & & & & \\
      \hhline{=====}
    \end{tabular}
  \end{center}
  \caption{Electromagnetic helicity amplitudes (in $10^{-3}$
    GeV$^{-\foh}$) of $I=\foh$ resonances
    considered in the calculation. First line: C-p-$\gamma 
    +$/C-p-$\gamma -$; second line PDG \cite{pdg}; third line Feuster and
    Mosel \cite{feusti99}; fourth line: Arndt {\it et al.} \cite{arndt02}. In
    brackets, the estimated errors are given. ``NF'': not
    found. ``NG'': not given. ``NC'': not considered (energy range
    ended at 1.9 GeV).
    \label{tabhelii12}}
\end{table}
\begin{table}[t]
  \begin{center}
    \begin{tabular}
      {l|l|l||l|l|l}
      \hhline{===:t:===}
      $L_{2I,2S}$ & $A_\foh$ & $A_\fth$ & 
      $L_{2I,2S}$ & $A_\foh$ & $A_\fth$ \\
      \hhline{===||===}
      $S_{31}(1620)$ &  $-50/-53$ & ---      & $P_{33}(1232)$ & $-128/-129$ & $-247/-248$ \\
                     & $\;\;\: 27(11)$ & --- &                & $-135(6)$ & $-255(8)$ \\
                     & $\;\: -4$ & ---       &                & $-126$ & $-233$ \\
                     & $-13(3)$ & ---        &                & $-129(1)$ & $-243(1)$ \\
      \hhline{===||-|-|-}
      $P_{31}(1750)$ &   53/30 & --- & $P_{33}(1600)$ &  $\;\;\;\;\: 0/0$ &  $-24/-24$ \\
                     & &             &                & $-23(20)$ & $\;\:-9(21)$ \\
                     & NC & ---      &                & $-26$ & $-52$ \\
                     & &             &                & & \\
      \hhline{===||-|-|-}
      $D_{33}(1700)$ &  $\;\: 96/96$ & $154/153$    & $P_{33}(1920)$ &   $-7/-9$ &   $-1/-2$ \\
                     & 104(15) & $\;\: 85(22)$      &                & $\:40(14)$ & $\:23(17)$ \\
                     & $\;\: 75$ & $\;\: 98$        &                & NC & \\
                     & $\;\: 89(10)$ & $\;\: 92(7)$ &                & & \\
      \hhline{===:b:===}
    \end{tabular}
  \end{center}
  \caption{Electromagnetic helicity amplitudes of $I=\fth$
    resonances. Notation as in Table \ref{tabhelii12}.
    \label{tabhelii32}}
\end{table}
One has to note that in the present model the helicity amplitudes of 
the resonances are not only determined by one specific reaction alone, 
but by a simultaneous consideration of all included photoproduction
reaction channels, largely reducing the freedom of the choice of these 
values. This in particular holds true for the proton helicity
amplitudes. For the neutron, these values can be determined only from 
pion photoproduction data on the deuteron; such data for other final
states are very scarce. Moreover, in some
neutron pion photoproduction multipoles ($M_{2-}^n$) the data
situation is not very good, and above 1.8 GeV only the
energy-dependent VPI \cite{SP01} solution (see Secs.
\ref{expdata} and \ref{secresgp}) is available, hindering a
reliable extraction of the neutron helicity amplitudes of the
corresponding resonances (see also below). This problem can only be
overcome, once data for more final states are available on the
deuteron target.

In the following, the helicity amplitudes of the resonances are
discussed in detail. A guideline for their uncertainty within the
present model is given by the variation between the two calculations; 
cf. Tables \ref{tabhelii12} and \ref{tabhelii32}.

\subsubsection{Isospin-$\foh$ resonances}

\noindent $\mathbf{S_{11}:}$\\
In contrast to Arndt {\it et al.} \cite{arndt02} the properties and in
particular the helicity amplitudes of the $S_{11}(1535)$ can be well
fixed in the present calculation, which is a result of the inclusion
of the $\eta$-photoproduction data. The extracted lower value for 
$A_\foh^p$ as compared to Feuster and Mosel \cite{feusti99} is caused
by the different gauging procedure and the fact, that a lower mass is
extracted in the present calculation. The differences in the neutron
value, however, can be explained by the improved data base underlying
the pion-photoproduction neutron multipoles; see
Fig. \ref{figgpin}.\\
The helicity coupling of the $S_{11}(1650)$ is also influenced by
$K\Lambda$ photoproduction in our analysis, but the extracted value
agrees well with the PDG \cite{pdg} value. However, the most recent
VPI photoproduction single-energy analysis presented in Ref. \cite{arndt02} 
indicates, that the structure of this resonance is enlarged as
compared to the analysis \cite{SP01} used in the present
calculation, which leads to the larger values found by Arndt {\it et al.}.
\par

\noindent $\mathbf{P_{11}:}$\\
The $P_{11}(1440)$ values are extremely sensitive to the damping of
the nucleon contributions and consequently the gauging procedure. This
leads to large differences in the neutron amplitude as compared to
Feuster and Mosel, the PDG, and Arndt {\it et al.}. However, the error
bars in the neutron multipole allow for a large range of resonance
contributions (see Fig. \ref{figgpin}). As a consequence of the large
$P_{11}(1440)$ mass and width (see Sec. \ref{secresgp} and PMI
\cite{pm1}) the resonant behavior of the $M_{1-}^p$ pion
photoproduction multipole between 1.25 and 1.4 GeV cannot be
completely described; see Fig. \ref{figgpip}. The $P_{11}(1440)$
proton helicity amplitude is mostly constrained by the small error
bars in the real part of $M_{1-}^p$ between 1.4 and 1.5 GeV. A
different helicity amplitude would largely deteriorate the overall
description of this multipole. Summarizing, as a consequence of the
large $P_{11}(1440)$ width necessary in the present model, the
$P_{11}(1440)$ helicity amplitudes cannot be reliably fixed. Possible
reasons for this problem are the lack of analyticity in the present
model leading to shortcomings close to the $2\pi N$ threshold, and the 
missing background contributions in the $2\pi N$ photoproduction (see
Secs. \ref{expdata}, \ref{secresgg}, and \ref{secgdh}).\\
Similarly to Arndt {\it et al.}, the electromagnetic properties of the
second $P_{11}$ cannot be completely fixed in the present
calculation. While in the proton case, the $P_{11}(1710)$ photon
coupling is roughly identical for both global calculations, the lack
of precise neutron target pion-photoproduction data especially above
1.8 GeV (see Fig. \ref{figgpin}) does not allow one to pin down the
$P_{11}(1710)$ neutron coupling.
\par

\noindent $\mathbf{P_{13}:}$\\
Since both $P_{13}$ resonances considered in the present calculation 
not only give important contributions to pion photoproduction, but
also to $K\Lambda$ and $\omega$ photoproduction, the resulting proton
couplings are rather well determined, although the structure in the
$E_{1+}^p$ pion photoproduction multipole cannot be completely
described (see Sec. \ref{secresgp}). This is in contrast to Arndt
{\it et al.} \cite{arndt02}, where the values of the $P_{13}(1720)$ are not
given. Note that our 
coupling signs for the $P_{13}(1720)$ are opposite to the PDG values,
but in line with the ones of Arndt, Strakovsky, and Workman
\cite{arndt96}: $A_\foh^p = -15(15)$ and $A_\fth^p = 7(10)$ (in brackets,
the estimated errors are given). The newly
included $P_{13}(1900)$ also influences the $P_{13}(1720)$ properties, 
thus explaining the differences in the couplings of the latter to
Feuster and Mosel \cite{feusti99}. As pointed out in Sec.
\ref{secresgp}, the lack of neutron data for the pion-photoproduction
multipoles above 1.8 GeV leaves the $P_{13}(1900)$ neutron photon
couplings essentially undetermined.
\par

\noindent $\mathbf{D_{13}:}$\\
As shown in Ref. \cite{feusti99}, the $D_{13}(1520)$ photon couplings are
extremely sensitive to Compton scattering. Therefore and due to the
enlarged Compton data base, the differences to the values of Arndt {\it et
al.} \cite{arndt02} and Feuster and Mosel \cite{feusti99} can be
understood. Furthermore, as pointed out in Sec. \ref{secresgp}, the 
$D_{13}(1520)$ neutron photon couplings are also influenced by the 
lack of precise $M_{2-}^n$ multipole data, thus fixing the
$D_{13}(1520)$ neutron photon couplings partially by its influences on 
the $J=\foh$ multipoles. The 
$D_{13}(1950)$ photon couplings always result in small values, since
neither in pion photoproduction nor in the other photoproduction
channels such a resonant structure is found. However, more
polarization measurements on the nonpion photoproduction data would
allow for a more closer determination of the electromagnetic
properties of this resonance.
\par

\subsubsection{Isospin-$\fth$ resonances}

\noindent $\mathbf{S_{31}:}$\\
Similarly to the $IJ=\foh \foh$ channels, the $E_{0+}^\fth$
multipole is also very sensitive to background contributions. Thus,
although in our calculation and in the analysis of Arndt {\it et
al.} \cite{arndt02} the resonance peak of the $S_{31}(1620)$ is nicely
described, the extracted helicity amplitude differs by a factor of
4. Feuster and Mosel \cite{feusti99} also found a smaller
helicity value, which, however, can be explained by the fact that, in
the older multipole analysis used in Ref. \cite{feusti99}, this
resonance's peak was less pronounced.
\par

\noindent $\mathbf{P_{31}:}$\\
As a consequence of the large error bars in the $M_{1-}^\fth$
multipole, the photon coupling of the $P_{31}(1750)$ differs in the
two global calculations. However, the extracted values describe the
tendency in the data correctly and are also in line with the influence 
of the $P_{31}(1750)$ on $K\Sigma$ photoproduction.
\par

\noindent $\mathbf{P_{33}:}$\\
Although Compton scattering is simultaneously analysed in the present
model, our helicity coupling nicely agrees with the recent analysis of
Arndt {\it et al.}, corroborating the compatibility of the
Compton and pion-photoproduction experimental data. The ratio of
electric and magnetic transition strength for the 
$\Delta$ [$P_{33}(1232)$] resonance is of special interest, because it
vanishes for a zero quadrupol deformation of this excited nucleon
state. Combining Eqs. \refe{photomultiheli} and 
\refe{gammaheli32} and using the normalization entering
Eq. \refe{multipoles}, we find
\bea
R^\Delta_{E/M} = 
\frac{A^\Delta_\foh - A^\Delta_\fth / \sqth}
{A^\Delta_\foh + \sqth A^\Delta_\fth}
= - \frac{g^\Delta_1 - g^\Delta_2\frac{m_\Delta}{2m_N}}
{g^\Delta_1 \frac{3m_\Delta+m_N}{m_\Delta-m_N} - g^\Delta_2\frac{m_\Delta}{2m_N}}
\label{emratiodelta} \; .
\eea
Our value of $-2.6$\% ($-2.5$\%) of calculation C-p-$\gamma +$
(C-p-$\gamma -$) is also identical with the PDG \cite{pdg} value of
$-2.5\pm 0.5$\% and the one of Tiator {\it et al.} \cite{tiator01} $-2.5\pm
0.1$, even though the $E_{1+}^\fth$ multipole is very sensitive to
rescattering \cite{feusti99}.\\
For the two higher lying $P_{33}$ resonances, we find small
electromagnetic contributions resulting in hardly any visible
structure in the $M_{1+}^\fth$ and $E_{1+}^\fth$ pion
multipoles. However, since these resonances also influence $K\Sigma$
photoproduction, both global calculations result in basically
identical values.
\par

\noindent $\mathbf{D_{33}:}$\\
As pointed out in Sec. \ref{secresgp}, we observe problems in the
description of the  $M_{2-}^\fth$ multipole due to the lack of a
background contribution in this multipole and the helicity amplitudes
are difficult to extract. Moreover, since $K\Sigma$ photoproduction
also proves to be sensitive to the $D_{33}(1700)$ helicity amplitudes, 
our $A_\fth$ values differ from those of the other references. Note
that in Ref. \cite{feusti99} similar observations were also made and
the extracted $A_\fth$ strength ranged from $98$ to $172$. 
\par

\subsubsection{Electromagnetic off-shell parameters}

\begin{table}[t]
  \begin{center}
    \begin{tabular}
      {l|r|r||l|r|r}
      \hhline{===:t:===}
      $L_{2I,2S}$ & $a_{\gamma_1}$ & $a_{\gamma_2}$ &
      $L_{2I,2S}$ & $a_{\gamma_1}$ & $a_{\gamma_2}$ \\
      \hhline{===||===}
      $P_{13}(1720)$ & $-1.324$ &   0.266 & $P_{33}(1232)$ &  0.471 &   0.932 \\
                     & $0.148$  &   0.429 &                &  0.538 &   0.809 \\
                     & $-0.352$ & 1.586   &                & $0.233$ & $-0.158$ \\
      \hhline{---|---}
      $P_{13}(1900)$ & $-3.599$ &  0.488 & $P_{33}(1600)$ & $-2.006$ &   2.650 \\
                     & $2.893$  &  0.149 &                & $-3.281$ &   3.000 \\
                     & NC &              &                & 3.282 & $-3.979$ \\
      \hhline{===||-|-|-}
      $D_{13}(1520)$ &  0.075 &  $-0.571$ & $P_{33}(1920)$ &   4.000 &  $-0.579$ \\
                     &  0.002 &  $-0.873$ &                &   4.000 &  $-2.123$ \\
                     & $0.235$ & $0.025$  &                & NC & \\
      \hhline{-|-|-||===}
      $D_{13}(1950)$ & $0.035$ & $1.101$   & $D_{33}(1700)$ & $-3.999$ &  $-1.580$ \\
                     & $-2.114$ & $-3.944$ &                & $-3.993$ &  $-1.666$ \\
                     & $-0.671$ & $-1.822$ &                & $0.962$ & $-0.362$ \\
      \hhline{===:b:===}
    \end{tabular}
  \end{center}
  \caption{Electromagnetic off-shell parameters $a_\gamma$ of
    spin-$\fth$ resonances. First line: C-p-$\gamma +$; second line:
    C-p-$\gamma -$; third line: SM95-pt-3 of Ref. \cite{feusti99}. ``NC'':
    not considered (energy range ended at 1.9 GeV).
    \label{tabresagamma}}
\end{table}
The electromagnetic off-shell parameters $a_\gamma$ (see Sec.
\ref{secelmgpot}) turn out to be
mostly well fixed in the two global calculations; see Table
\ref{tabresagamma}. Exceptions are the $a_{\gamma_1}$ values of the
$P_{13}$ resonances, which can, however, be explained by the fact that 
the corresponding couplings $g_{\gamma_1}$ are very small and thus the 
off-shell parameters are very sensitive to any changes. In the
$D_{13}(1950)$ case, the differences between the two calculations are
related to the fact that the helicity amplitudes can also not be well
fixed; see Table \ref{tabhelii12}. Since the off-shell parameters
determine the background contributions in the $J=\foh$ waves, it is
also quite clear that these parameters are very sensitive to the
gauging procedure, which has already been found by Feuster and Mosel
\cite{feusti99}. This explains, why even in the case of the 
$P_{33}(1232)$ resonance, our values differ from those extracted in
Ref. \cite{feusti99}, where the Haberzettl gauging procedure
[Eq. \refe{habergauge}] was used instead of the Davidson-Workman
procedure [Eq. \refe{davidsongauge}] (note that the values of Ref. 
\cite{feusti99} for the hadronic off-shell parameters are mostly 
similar to ours; see PMI \cite{pm1}).

\section{Summary and Outlook}

The presented model provides a tool for nucleon resonance
analysis below energies of $\sqrt s = 2$ GeV. Unitarity effects are
correctly taken into account, since all important final states,
i.e., $\pi N$, $2\pi N$, $\eta N$, $K\Lambda$, $K\Sigma$, and $\omega
N$, are included. Since the driving potential is built up by the use
of effective Lagrangians for Born-, $t$-channel, spin-$\foh$, and
spin-$\fth$ resonance contributions, the background contributions
are also generated consistently and the number of parameters is greatly
reduced. The dependence on different descriptions for the spin-$\fth$
resonance vertices has been investigated for the pion-induced
reactions and similar results have been found. 

The simultaneous consideration of the $\gamma N$ 
final state guarantess access to a much larger and more precise data
base allowing for strong tests on all resonance contributions. It has
turned out that the inclusion of photoproduction data is inevitable to
extract the resonance masses and widths reliably. A side effect is
that within such a model the consistency of the experimental data for
the various reactions can be checked, and no discrepancies are found.

A simultaneous description of all pion and photon-induced reactions on 
these final states is possible with one parameter set. Although we
have largely extended our data base on pion photoproduction and
Compton scattering, both channels (and $\eta N$ photoproduction) are
still well described in the energy region
below $\sqrt s = 1.6$ GeV. The extracted
electromagnetic properties of the $P_{33}(1232)$ resonance perfectly
agree with other analyses. In general, the agreement with the previous 
analysis of Feuster and Mosel \cite{feusti99} is quite good. The main
differences are found for resonances in those partial waves, where
additional higher lying states have been added, and in the
electromagnetic off-shell parameters $a$ of the spin-$\fth$
resonances, which is a consequence of the different applied gauging
procedures.

No global fit has been possible when the form factor $F_t$
\refe{formfact} is used for the $t$-channel exchange diagrams. Even
when using $F_p$ a readjustment of the parameters obtained from purely 
hadronic reactions is necessary, 
since, especially in the $\eta N$ and $\omega N$ channels, the
resonance contributions cannot be well fixed using the pion-induced
data alone. In addition, in the associated strangeness channels the
Born couplings have to be readjusted, since the corresponding
contributions are largely enhanced as a consequence of the gauging
procedure. The resulting Born couplings of the global parameter set
are close to SU(3) predictions. The background in pion
photoproduction is very sensitive on nucleon contribution, and in
particular on the gauging procedure. Although this background is fixed
by only a few parameters, it is well described in most multipoles,
thus giving confidence in the applied Davidson-Workman gauging
procedure [Eq. \refe{davidsongauge}].

In the $K\Lambda$, $K^0\Sigma^+$, and $\omega N$ channels we find
a strong need for contributions of a $P_{13}(1900)$ resonance between
1.9 and 2 GeV, similar to the pion-induced reactions. The inclusion of
this resonane also leads to changes in the properties of the
$P_{13}(1720)$ as compared to previous analyses. In particular, we
find that the role of the $P_{13}(1720)$ is largely enhanced in
$K\Lambda$ photoproduction. However, for a clear disentanglement
of the resonant contributions in the energy region above 1.7 GeV more 
polarization measurements in particular on $\omega N$ and $\eta N$ are
needed to completely determine the $P_{13}(1900)$ and
also the $D_{13}(1950)$ resonance properties. 

The associated strangeness photoproduction
channels prove to be very sensitive to the $\omega N$ threshold and
interference effects. This leads to the explanation of a
resonancelike structure in the $K\Lambda$ total cross section by an 
interference of $K^*$ and nucleon contributions, instead of a
resonance. The $\omega N$ production is mostly dominated by the
$\pi^0$ exchange mechanism, but large interference effects due to the
implemented resonances are necessary to find a satisfactory description
of the preliminary SAPHIR data \cite{barthom}. The pseudoscalar nature 
of the $\pi^0$ exchange mechanism requires the inclusion of partial
waves up to $J_{max} = \frac{27}{2}$ in the PWD. The threshold
behavior of this reaction is mostly explained by a large
$P_{11}(1710)$ contribution, in contrast to all other models on
$\omega N$ photoproduction. 

The good description of all photoproduction channels enables us
to evaluate the GDH sum rule contributions of the various
final states. We find small values for the contributions of $\eta N$,
$K\Lambda$, $K\Sigma$, and $\omega N$, which are remarkably different
from those extracted in single-channel analyses. 

Deficiencies of the present model
concerning the $2\pi N$ production are visible in Compton
scattering, where a background contribution in the energy region
between the $P_{33}(1232)$ and $D_{13}(1520)$ resonance is
missing. We have nevertheless shown that the resonance contributions
to $2\pi N$ photoproduction are well under control in the present
model. Moreover, similar to $\pi N$ elastic scattering, there are 
also evidences of the influence of a $3\pi N$ final state in the
$J^P=\fth^+$ multipole $E_{1+}^p$. As a consequence of the lack of
spin-$\ffh$ resonances, the analysis of Compton scattering is
restricted to energies below $\sqrt s = 1.6$ GeV. Since all data on
$\eta N$, $K\Lambda$, $K\Sigma$, and $\omega N$ are well described
without such resonances, they seem to be of minor importance in these 
reactions. This point is being investigated further at present
\cite{vitali}. 

Using the generalization of the partial-wave decomposition presented
here for the inclusion of the $\omega N$ final state a more realistic
description of the $2\pi N$ final state in terms of $\rho N$ and $\pi
\Delta$ is now possible. The inclusion of these final states allows
one to mimic the three particle phase space while still dealing with
two-body unitarity. Accounting for the spectral function of the $\rho$ 
meson and the $\Delta$ baryon would then allow for the complete
description of $2\pi N$ production within the present model 

While for larger energies threshold effects due to unitarity are of
main importance, at lower energies considerable effects are known to
be caused by analyticity. This has been demonstrated by the comparison
of the present analysis with models also taking analyticity into
account. Therefore, also work along analytic extensions of the
$K$-matrix ansatz, e.g., in the direction proposed by Kondratyuk and
Scholten \cite{kondratyuk}, should be pursued.

\begin{acknowledgments}
We are thankful to B. Ritchie and M. Dugger for providing us with the
preliminary CLAS data on $\eta$ photoproduction
\cite{dugger}. Furthermore, we would like to thank W. Schwille and J. Barth
for making the prelinary SAPHIR data on $\omega$ photoproduction
\cite{barthom} available to us. One of the authors (G.P.) is grateful
to C. Bennhold for the hospitality at the George Washington
University, Washington, D.C., in the early stages of this work.
This work was supported by DFG and GSI Darmstadt.
\end{acknowledgments}

\begin{appendix}

\section{Lagrangians, Couplings, and Helicity Amplitudes}
\label{applagr}

\subsection{Background}

The Born contributions are generated by
\bea
\mcl &=& - e \bar u_{B'} (p') \left[ 
\left( \hat e \gamma_\mu A^\mu + 
  \frac{\kappa}{2m_N} \sigma_{\mu \nu} F^{\mu \nu} \right) +
\frac{g_\varphi}{m_B+m_{B'}} \gamma_5 \gamma_\mu A^\mu
\right] u_B (p) 
\nonumber \\
&& -\mi \hat e e \varphi^* \left( 
\partial_\mu^\varphi - \partial_\mu^{(\varphi^*)} 
\right) \varphi A^\mu
-e \frac{g_\varphi}{m_B+m_{B'}}
\bar u_{B'} (p') \gamma_5 \gamma_\mu u_{B} (p) A^\mu \; ,
\eea
with the asymptotic baryons $B,B'=(N,\Lambda,\Sigma)$, the
pseudoscalar mesons $(\varphi,\varphi')=(\pi,K)$, and 
$F^{\mu \nu} = \partial^\mu A^\nu - \partial^\nu A^\mu$.
Accounting correctly for the masses entering the hyperon anomalous
magnetic moments, the $\kappa$ values that enter the Lagrangian above
can be extracted from the PDG \cite{pdg} values for the magnetic
moments:
\bea
\ba{lcrclcr}
\kappa_\Lambda &=& -0.613 \; , & & 
\kappa_{\Sigma^0 \ra \Lambda \gamma} &=& 1.610 \; , \\
\kappa_{\Sigma^+} &=& 1.671 \; , & & 
\kappa_{\Sigma^-} &=& -0.374 \; . \\
\ea
\eea

For the intermediate ($t$-channel) mesons the additional Lagrangian
\bea
\mcl =
- \mi g_{K_1} \left( \gamma_\mu K_1^\mu + 
  \frac{\kappa_{K_1}}{2m_N} \sigma_{\mu \nu} K_1^{\mu \nu} \right) \gamma_5
u_B (p)
- \frac{g}{4 m_\varphi} \ve_{\mu \nu \rho \sigma} V^{\mu \nu}
{V'}^{\rho \sigma} \varphi
+ e \frac{g_{K_1 K \gamma}}{2 m_K} K
F_{\mu \nu} K_1^{\mu \nu}
\eea
is taken. $V^{\mu \nu}$ and $K_1^{\mu \nu}$ are defined in analogy to
$F^{\mu \nu}$. Using the values for the decay widths from Refs. \cite{pdg}
and \cite{alavi} [$\Gamma (K^0_1(1270)\ra K^0\gamma) = 73$ keV], the
following couplings are extracted:
\bea
\ba{lcrclcr}
g_{\rho \pi \gamma}    &=&  0.105 \; , & & 
g_{\rho \eta \gamma}   &=& -0.805 \; , \\
g_{\omega \pi \gamma}  &=&  0.313 \; , & & 
g_{\omega \eta \gamma} &=& -0.291 \; , \\
g_{{K^*}^+ K^+ \gamma} &=& -0.414 \; , & & 
g_{{K^*}^0 K^0 \gamma} &=&  0.631 \; , \\
g_{K_1^+ K^+ \gamma}   &=&  0.217 \; , & & 
g_{K_1^0 K^0 \gamma}   &=&  0.217 \; . \\
g_{\pi \gamma \gamma}  &=&  0.037 \; , & & 
g_{\eta \gamma \gamma} &=&  0.142 \; , \\
\ea
\label{mesdeccons}
\eea
Note that an isospin averaged value for the $g_{\rho \pi \gamma}$
coupling is used; see Ref. \cite{gregiphd}. The ratio between the radiative
decay of the charged and the neutral $K_1(1270)$ meson has not yet
been measured; for simplicity, we use $g_{K_1^+ K^+ \gamma} = g_{K_1^0
  K^0 \gamma}$. For the relative sign between the charged and 
the neutral $K^*$ coupling, we follow the quark model prediction of
Singer and Miller \cite{singer}. 

A remark on the $\rho$ and $\omega$ radiative decays into
$\eta \gamma$ is in order. Unfortunately, the decay widths are known
only with large uncertainties; the values above represent the
estimated mean given in Ref. \cite{pdg}. Taking into account the given
errors, the ranges for these couplings are: 
\bea
\left| g_{\rho \eta \gamma} \right| \in \left[ 0.636,0.930 \right]
\; , \hspace{5mm} 
\left| g_{\omega \eta \gamma} \right| \in \left[ 0.268,0.313 \right]
\; .
\eea
Due to the uncertainties, these couplings are also allowed to vary
within the given ranges during the fitting procedure. However, in all 
calculations, larger values for both couplings are preferred and 
consequently, these couplings are set to $g_{\rho \eta \gamma} =
-0.930$ and $g_{\omega \eta \gamma} = -0.313$. Note that all other
meson decay constants are also kept fixed to the values given in Eq. 
\refe{mesdeccons}.

\subsection{Resonances}

The radiative decay of the spin-$\foh$ resonances is described by
\bea
\mcl_{\foh N\gamma} 
= - e \frac{g_1}{4 m_N} \bar u_R 
\left( \begin{array}{c} 1 \\ - \mi \gamma_5 \end{array} \right)
\sigma_{\mu \nu} u_N F^{\mu \nu} \; ,
\eea
and for the spin-$\fth$ resonances by
\bea
\mcl_{\fth N\gamma} = 
\bar u_R^\mu e
\left( \begin{array}{c} \mi \gamma_5 \\ 1 \end{array} \right)
\left( \frac{g_1}{2m_N} \gamma^\nu + \mi \frac{g_2}{4 m_N^2} 
  \partial^\nu_N \right) u_N F_{\mu \nu} \; .
\label{lagr32gamapp}
\eea
In both cases, the upper (lower) factor corresponds to positive-
(negative-) parity resonances. Note that, in the spin-$\fth$ case,
both couplings are also contracted by an off-shell projector
$\Theta_{\mu \nu} (a) = g_{\mu \nu} - a \gamma_\mu \gamma_\nu$, 
where $a$ is related to the commonly used off-shell parameter $z$
by $a = (z + \sfoh )$ (see PMI \cite{pm1} for more details).

In analogy to the the $\omega N$ helicity amplitudes (see PMI
\cite{pm1}) the electromagnetic helicity amplitudes, which are
normalized by an additional factor $(2 E_\gamma)^{-\foh}$
\cite{warns}, are extracted:
\bea
A^{\gamma N}_\foh &=& \frac{\xi_R}{\sqrt{2 E_\gamma}}
\langle u_R, \lambda_R = \sfoh | \Gamma_\mu | u, \lambda = \sfoh \rangle 
A^\mu 
\nonumber \\
&=& - e g \frac{\xi_R}{2 m_N}
\frac{\sqrt{m_R^2 - m_N^2}}{\sqrt{2 m_N}}
\label{gammaheli12}
\eea
for spin-$\foh$ resonances, and 
\bea
A^{\gamma N}_\foh &=& 
+ \frac{e \xi_R}{4 m_N}
\frac{\sqrt{m_R^2 - m_N^2}}{\sqrt{3 m_N}}
\left( \pm g_1 \frac{m_N}{m_R} - g_2 \frac{m_N \mp m_R}{4 m_N} \right) 
\; ,
\nonumber \\
A^{\gamma N}_\fth &=& 
\pm \frac{e \xi_R}{4 m_N}
\frac{\sqrt{m_R^2 - m_N^2}}{\sqrt{m_N}}
\left( g_1 + g_2 \frac{m_N \mp m_R}{4 m_N} \right)
\label{gammaheli32} 
\eea
for spin-$\fth$ resonances. Here, $\xi_R$ denotes the phase at
the $RN\pi$ vertex. The lower indices correspond to the $\gamma N$
helicities and are determined by the $\gamma$ and nucleon helicities:
$\foh$: $\lambda_\gamma - \lambda_N = 1 - \foh = \foh$ and $\fth$: $1
+ \foh = \fth$. Note the differences of Eq. \refe{gammaheli32} to the
formulas given in Ref. \cite{feusti99}, which are due to the different sign
choice for the $g_1$ coupling in Eq. \refe{lagr32gamapp} for negative
parity resonances.

\section{Calculation of Amplitudes}
\label{appampli}

The scattering amplitude $\mct^{fi}_{\lambda' \lambda} (\vt)$ and the
$K$ matrix amplitude $\mck^{fi}_{\lambda' \lambda} (\vt)$, which enter 
the partial-wave decomposed Bethe-Salpeter equation \refe{tkrelwint},
are defined by
\bea
\mct^{fi}_{\lambda' \lambda} &\equiv& 
- \frac{\sqrt{\avec p \avec p' m_{B'} m_B}}{(4\pi)^2 \sqrt s} 
\langle f | M | i \rangle \; ,
\label{tandm} \\
\mck^{fi}_{\lambda' \lambda} &\equiv&
- \frac{\sqrt{\avec p \avec p' m_{B'} m_B}}{(4\pi)^2 \sqrt s} 
\langle f | K | i \rangle \; ,
\label{kandv}
\eea
where $K = V$ in the $K$-matrix Born approximation and $\langle f|$ 
and $|i\rangle$ denote the final and initial two-particle
momentum states, respectively; see PMI \cite{pm1}.

The calculation of the amplitudes $\mcv^{fi} \equiv \langle f | V | i
\rangle$ which enter Eq. \refe{kandv} are extracted from the Feynman
diagrams via
\bea
\mcv_{\lambda' \lambda}^{fi} &=&
\bar u(p',\lambda_{B'}) \Gamma (s, u) u(p,\lambda_B) 
\nonumber \\
&=&
\frac{4 \pi \sqrt s }{\sqrt{m_B m_{B'}}} \chi^\dagger_{\lambda_{B'}}
\mcf(s, u) \chi_{\lambda_B} \; .
\label{basicdecomp}
\eea

\subsection{Photoproduction of (pseudo-) scalar mesons}

The calculation of the spin-dependent amplitudes $\mcv_{\lambda'
  \lambda}^{fi}$ is identical in this case to the reactions $\pi
/\zeta N \ra VN$ (see also PMI \cite{pm1}): 
Replacing the Dirac operator $\Gamma \ra \Gamma_\mu
\varepsilon^\mu_{\lambda_V}$ the general form of $\Gamma_\mu$ is 
\bea
\Gamma_\mu (s, u) = \Theta &\cdot& \left( 
A_p p_\mu + A_{p'} p'_\mu + (B_p p_\mu + B_{p'} p'_\mu) \slash k +
C \gamma_\mu + D \slash k \gamma_\mu \right) \; ,
\label{pivectorgam}
\eea
with $\Theta = \mi \gamma_5$ for pseudoscalar and $\Theta = \umat_4$
for scalar outgoing mesons. Applying gauge invariance considerations
($\Gamma_\mu k^\mu = 0$), $\Gamma_\mu$ can be recasted into the usual
form of pseudoscalar meson electroproduction \cite{gregiphd}. For
example for real photons ($k^2 = \slash k^2 = 0$), the relation to the 
standard set of four gauge invariant amplitudes (see, e.g., Ref.
\cite{feusti99})
\bea
\mcv_{fi} &=& \bar u(p',s') \sum_{j=1}^4 A_j M_j u(p,s) \; ,
\hspace{5mm} \mbox{with} \nonumber \\
M_1 &=& - \mi \gamma_5 \slash \varepsilon \slash k \; ,
\nonumber \\ 
M_2 &=& 2 \mi \gamma_5 ( \etp k \cdot p' - \etpp k \cdot p ) \; ,
\nonumber \\  
M_3 &=& \mi \gamma_5 ( \slash \varepsilon k \cdot p - \slash k \etp )
\; ,
\nonumber \\ 
M_4 &=& \mi \gamma_5 ( \slash \varepsilon k \cdot p' - \slash k \etp' )
\; ,
\label{piphogaugeset}
\eea
is given by
\bea
A_1 &=& D \; , \hspace{2cm} 
A_2 = - \frac{A_{p'}}{2 p \cdot k} = \frac{A_p}{2 p' \cdot k} \; ,
\nonumber \\ 
A_3 &=& - B_p \; , \hspace{16mm} 
A_4 = - B_{p'} \; .
\label{relphotonsets}
\eea
$\mcf$ of Eq. \refe{basicdecomp} is constructed in analogy to the
virtual photon case \cite{ben95},
\bea
\mcf &=& 
  \mi \vec \sigma \cdot \vec \varepsilon \mcf_1 + 
  \vec \sigma \cdot \vec{\hat{k'}} \vec \sigma \cdot (\vec{\hat{k}} 
  \times \vec \varepsilon) \mcf_2 +
  \mi \vec \sigma \cdot \vec{\hat{k}} 
  \vec \varepsilon \cdot \vec{\hat{k'}} \mcf_3 + 
  \mi \vec \sigma \cdot \vec{\hat{k'} }
  \vec \varepsilon \cdot \vec{\hat{k'}} \mcf_4 
  - \mi \varepsilon^0 
  (\vec \sigma \cdot \vec{\hat{k'}} \mcf_5 +
   \vec \sigma \cdot \vec{\hat{k}}  \mcf_6) \; ,
\label{fpivecdecomp}
\eea
with $\varepsilon_{\lambda_V}^\mu = (\varepsilon^0,
\vec{\varepsilon})$. Obviously, $\mcf_5$ and $\mcf_6$ only contribute 
for longitudinal polarizations. This $\mcf$ has to be replaced for
scalar meson production by $\mcf \ra -\mi \vec \sigma \cdot
\vec{\hat{k'}} \mcf$. The decompositions in Eqs. \refe{pivectorgam}
and \refe{fpivecdecomp} are related via 
\bea
\mcf_1 &=& \frac{1}{8 \pi \sqrt s} \sqrt{R'_\pm R_+}
  \left( C - S_- D \right) \; , \nonumber \\
\mcf_2 &=& \frac{1}{8 \pi \sqrt s} \sqrt{R'_\mp R_-}
  \left( C + S_+ D \right) \; , \nonumber \\
\mcf_3 &=& \frac{\avec k'}{8 \pi \sqrt s} \sqrt{R'_\pm R_-}
  \left( -A_{p'} + S_+ B_{p'} \right) \; , 
\label{pivecfi} \\
\mcf_4 &=& \frac{\avec k'}{8 \pi \sqrt s} \sqrt{R'_\mp R_+}
  \left( A_{p'} + S_- B_{p'} \right) \; , \nonumber \\
\mcf_5 &=& - \frac{1}{\avec k'} \widetilde \mcf_4 -
  \frac{1}{8 \pi m_M \sqrt s} \sqrt{R'_\mp R_-}
  \left( S_+ C + m_M^2 D \right) \; , \nonumber \\
\mcf_6 &=& - \frac{1}{\avec k'} \widetilde \mcf_3 -
  \frac{1}{8 \pi m_M \sqrt s} \sqrt{R'_\pm R_+}
  \left( S_- C - m_M^2 D \right) \; , 
\nonumber
\eea
with
\bea
\widetilde \mcf_i &=& \etpp \mcf_i + 
\etp \mcf_i(A_{p'} \ra A_p, B_{p'} \ra B_p) \; , 
\nonumber \\
 \etp &\equiv& \ve_0^\mu p_\mu =
\frac{{\avec k} \sqrt s}{m_M} \; , 
\nonumber \\
\etpp &\equiv& \ve_0^\mu p'_\mu =
\frac{1}{m_M} \left(E_{B'} {\avec k} + {\avec k'} E_M \cos \vt
\right) \; .
\nonumber
\eea
Using Eq. \refe{relphotonsets}, the $\mcf_1$ to $\mcf_4$ of  
Eq. \refe{pivecfi} reduce to the well known photoproduction case
(cf. Eq. (B9) in Ref. \cite{feusti99}\footnote{Note that there are four
  misprints in Eq. (B9) in Ref. \cite{feusti99}: The $A_4$ term in $\mcf_1$
  and $\mcf_2$ should have a $-$ sign, $A_3$ in $\mcf_3$ and $\mcf_4$
  should be an $A_2$.}).

In the c.m. system the $\mcf_i$ are related to the
helicity dependent amplitudes via\footnote{Note that there is a
  misprint in Eq. (B12) in Ref. \cite{feusti99}: The $H_4$ term should
  start with $-\sqrt 2 \sin \fvh$.}
\bea
\mcv_{+\frac{1}{2} +\frac{3}{2}} &= 
  \pm \mcv_{-\frac{1}{2} -\frac{3}{2}} =& 
  f \frac{4 \pi \sqrt s }{\sqrt{m_B m_{B'}}}
  \frac{1}{\sqrt 2} \sin \vartheta \cos \frac{\vartheta}{2} 
  (-\mcf_3 - \mcf_4) \; , \nonumber \\
\mcv_{+\frac{1}{2} - \frac{3}{2}} &= 
  \mp \mcv_{- \frac{1}{2} +\frac{3}{2}} =& 
  f \frac{4 \pi \sqrt s }{\sqrt{m_B m_{B'}}}
  \frac{1}{\sqrt 2} \sin \vartheta \sin \frac{\vartheta}{2} 
  (-\mcf_3 + \mcf_4) \; , \nonumber \\
\mcv_{+\frac{1}{2} +\frac{1}{2}} &= 
  \mp \mcv_{-\frac{1}{2} -\frac{1}{2}} =& 
  f \frac{4 \pi \sqrt s }{\sqrt{m_B m_{B'}}}
  \sqrt 2 \cos \frac{\vartheta}{2} \left[ -\mcf_1 + \mcf_2 
    + \sin^2 \frac{\vartheta}{2} (\mcf_3 - \mcf_4) \right] \; , 
  \nonumber \\
\mcv_{+\frac{1}{2} - \frac{1}{2}} &= 
  \pm \mcv_{- \frac{1}{2} +\frac{1}{2}} =& 
  f \frac{4 \pi \sqrt s }{\sqrt{m_B m_{B'}}}
  \sqrt 2 \sin \frac{\vartheta}{2} \left[ \mcf_1 + \mcf_2 
    + \cos^2 \frac{\vartheta}{2} (\mcf_3 + \mcf_4) \right] \; , 
  \nonumber \\
\mcv_{+\frac{1}{2} +0} &= \mp \mcv_{-\frac{1}{2} -0} =& 
  f \frac{4 \pi \sqrt s }{\sqrt{m_B m_{B'}}}
  \varepsilon^0 \cos \frac{\vartheta}{2} (-\mcf_5 - \mcf_6) \; , 
  \nonumber \\
\mcv_{+\frac{1}{2} -0} &= \mp \mcv_{-\frac{1}{2} +0} =& 
  f \frac{4 \pi \sqrt s }{\sqrt{m_B m_{B'}}}
  \varepsilon^0 \cos \frac{\vartheta}{2} (-\mcf_5 + \mcf_6) \; ,
  \label{fandmpivec}
\eea
where the upper (lower) sign and $f = \mi$ ($f = 1$) hold for
pseudoscalar (scalar) meson production. Here we have used the helicity 
notation introduced in Appendix \ref{applagr}, and in addition (for
electroproduction) $0$: $\lambda_\gamma - \lambda_N = 0 + \foh =
\foh$.

\subsection{Compon scattering and photoproduction of vector mesons}

Replacing the Dirac operator $\Gamma (s, u)$ by $\Gamma_{\mu \nu} (s,
u) \varepsilon^\mu_{\lambda_V} \varepsilon_{\lambda_{V'}}^{\nu^\dagger}$,
$\Gamma_{\mu \nu}$ can be rewritten by 
\bea
\Gamma_{\mu \nu}(s, u) &=& 
      A_{\mu \nu} + B_{\mu \nu} \slash k + 
      C_\nu \gamma_\mu + D_\nu \slash k \gamma_\mu + 
      E_\mu \gamma_\nu + F_\mu \slash k \gamma_\nu + 
      G \gamma_\mu \gamma_\nu + H \slash k \gamma_\mu \gamma_\nu
\hspace*{5mm} \label{gammunuvnvn}
\eea
with
\bea
A_{\mu \nu} &=& A_{pp} p_\mu p_\nu + A_{pp'} p_\mu p'_\nu +
  A_{p'p} p'_\mu p_\nu + 
A_{p'p'} p'_\mu p'_\nu + A_g g_{\mu \nu}, 
\; \mbox{ similarly for } B_{\mu \nu} \; , 
  \nonumber \\
C_\nu &=& C_p p_\nu + C_{p'} p'_\nu, \; \mbox{ similarly for } D_\nu \; , 
  \nonumber \\
E_\mu &=& E_p p_\mu + E_{p'} p'_\mu, \; \mbox{ similarly for } F_\mu 
  \; , \label{minvnvn}
\eea
which underlies six gauge constraints in photoproduction of vector
mesons ($\Gamma_{\mu \nu} k^\mu =0$) and another six in Compton
scattering ($\Gamma_{\mu \nu} {k'}^\nu =0$), reducing the number of
independent operators correctly (see also Ref. \cite{gregiphd}). The
formulas for the calculation of the spin dependent amplitudes
$\mcv_{fi}$ are identical to the calculation for $VN \ra VN$ (see
PMI \cite{pm1}).

\section{Partial Waves and Electromagnetic Multipoles}
\label{apppartials}

The partial-wave decomposition of the photoproduction reactions works
completely analogously to the hadronic reactions, which is discussed
in detail in PMI \cite{pm1}. In this appendix the relations
between our helicity partial waves and the standard photoproduction
multipoles are given. 

Our helicity states are given by
\bea
|J,\lambda;\pm \rangle &\equiv&
\frac{1}{\sqt} 
\left( |J,+\lambda\rangle \pm \eta |J,-\lambda \rangle \right) \; , 
\label{paritystates}
\eea
with parity $P=(-1)^{J\pm \foh}$, $\eta \equiv \eta_k \eta_p
(-1)^{s_k+s_p+\foh}$, and the intrinsic parities ($\eta_k$, $\eta_p$)
and spins ($s_k$, $s_p$) of the two particles. After some
Clebsch-Gordan manipulation, one can extract the relations between the 
helicity states of Eq. \refe{paritystates} and the usual
magnetic ($M$), electric ($E$), and scalar ($S$) photon nucleon
multipole states \cite{gehlen,gregiphd}:
\bea
|J = j_\gamma + \sfoh, M/E \rangle &=& 
\mp \frac{1}{\sqrt{2 (j_\gamma + 1)}} 
\left(
\sqrt{j_\gamma} |J, \sfoh ; \pm \rangle +
\sqrt{j_\gamma + 2} |J, \sfth ; \pm \rangle
\right) \; , 
\nonumber \\
|J = j_\gamma - \sfoh, M/E \rangle &=& 
\mp \frac{1}{\sqrt{2 j_\gamma}} 
\left(
\sqrt{j_\gamma + 1} |J, \sfoh ; \mp \rangle -
\sqrt{j_\gamma - 1} |J, \sfth ; \mp \rangle
\right) \; , 
\nonumber \\
|J = j_\gamma \pm \sfoh, S \rangle &=& 
\pm |J, 0 ; \mp \rangle \; .
\label{photomultiheli}
\eea
Here the photon angular momentum $\ell_\gamma$ and the total photon
spin $j_\gamma$ are given by $j_\gamma = \ell_\gamma$ for the magnetic
and $j_\gamma = 1 \oplus \ell_\gamma$ for the electric and scalar
states. Since the two-particle helicity states $|J,\lambda;\pm
\rangle$ are of parity $P = (-1)^{J \pm \foh}$, this also holds true
for the corresponding nucleon photon multipole states. With this
relation, establishing the connection between the photoproduction
multipoles of any final state and the two-particle helicity amplitudes
is straightforward, and can also be easily achieved for more
complicated reactions such as $\gamma N \ra \pi \Delta$.

\subsubsection{Photoproduction of pions}

Sandwiching the interaction matrix $T$ between the multipole states
[Eq. \refe{photomultiheli}] and the $\pi N$ parity helicity states of 
Eq. \refe{paritystates}: $\langle J,\lambda;\pm |_{\pi N} = 
(\langle J,+\lambda | \pm \langle J,-\lambda |) / \sqt$, one finds the 
electromagnetic partial-wave multipoles:
\bea
\mct^{M/E}_{j_\gamma+} &=&
_{\pi N} \langle J,\lambda;\pm | \; T \; |j_\gamma +, M/E \rangle
=
\mp \frac{1}{\sqrt{2(j_\gamma+1)}}
\left( 
\sqrt{j_\gamma} \; \mct_{\foh \foh}^{J\pm} +
\sqrt{j_\gamma +2} \; \mct_{\foh \fth}^{J\pm}
\right) \; , 
\nonumber \\
\mct^{M/E}_{j_\gamma-} &=& 
_{\pi N} \langle J,\lambda;\mp | \; T \; |j_\gamma -, M/E \rangle
=
\mp \frac{1}{\sqrt{2 j_\gamma}}
\left( 
\sqrt{j_\gamma +1} \; \mct_{\foh \foh}^{J\mp} - 
\sqrt{j_\gamma -1} \; \mct_{\foh \fth}^{J\mp}
\right) \; , 
\nonumber \\
\mct^S_{j_\gamma \pm} &=& 
_{\pi N} \langle J,\lambda;\mp | \; T \; |j_\gamma \pm, S \rangle
=
\pm \mct_{\foh 0}^{J\mp} \; , 
\nonumber 
\label{elecmulti}
\eea
with the notation $j_\gamma \pm$: $J = j_\gamma \pm \foh$. Using the
relations between the above partial-wave multipoles and the standard
CGLN multipoles \cite{cgln,ben95} (cf. Table
\ref{tabangmomgampi})\footnote{Note that $M_{j_\gamma +}$ and
  $E_{j_\gamma+}$ have the wrong sign in Table IX in Ref. \cite{feusti99}.} 
\begin{table}
  \begin{center}
    \begin{tabular}
      {c|c|r|r|l}
      \hhline{=====}
      PW & CGLN mult. & $J\;\;\;$ & $P\;\;\;\;\;$ & $\;\;\;\ell_\pi$ \\ 
      \hline 
      $\mct^M_{j_\gamma+}$ & $+\alpha M_{j_\gamma+}$ & 
      $j_\gamma+\foh$ & $-(-1)^{j_\gamma}$ & $j_\gamma$ \\
      $\mct^M_{j_\gamma-}$ & $-\alpha M_{j_\gamma-}$ & 
      $j_\gamma-\foh$ & $-(-1)^{j_\gamma}$ & $j_\gamma$ \\
      $\mct^E_{j_\gamma+}$ & $-\alpha E_{j_\gamma+}$ & 
      $j_\gamma+\foh$ & $ (-1)^{j_\gamma}$ & $j_\gamma+1$ \\
      $\mct^E_{j_\gamma-}$ & $-\alpha E_{j_\gamma-}$ & 
      $j_\gamma-\foh$ & $ (-1)^{j_\gamma}$ & $j_\gamma-1$ \\
      $\mct^S_{j_\gamma+}$ & $-\beta E_{j_\gamma+}$ & 
      $j_\gamma+\foh$ & $ (-1)^{j_\gamma}$ & $j_\gamma+1$ \\
      $\mct^S_{j_\gamma-}$ & $+\beta' E_{j_\gamma+}$ & 
      $j_\gamma-\foh$ & $ (-1)^{j_\gamma}$ & $j_\gamma-1$ \\
      \hhline{=====}
    \end{tabular}
  \end{center}
  \caption{Relation between the partial-wave multipoles
    $\mct_{j_\gamma+}$ and the CGLN multipoles in pion 
    electroproduction. $\ell_\pi$ denotes the pion angular momentum,
    $j_\gamma$ the total photon spin, $J$ and $P$ the total spin and
    parity of the amplitudes; $\alpha = \sqrt{\avec k {\avec k}'
      j_\gamma (j_\gamma +1)}$, $\beta = \sqrt{\avec k {\avec k}'} 
      (j_\gamma +1)$ and $\beta' = \sqrt{\avec k {\avec k}'} j_\gamma$.
      \label{tabangmomgampi}} 
\end{table}
one finds, setting $J = \ell_\pi + \foh$:
\bea
M_{\ell_\pi+} &=& 
-\frac{1}{\sqrt{2 \avec k {\avec k}'} (\ell_\pi+1)}
\left( 
\mct_{\foh \foh}^{J+} +
\sqrt \frac{\ell_\pi +2}{\ell_\pi} \; \mct_{\foh \fth}^{J+}
\right) \; , 
\nonumber \\
M_{(\ell_\pi+1)-} &=& 
+\frac{1}{\sqrt{2 \avec k {\avec k}'} (\ell_\pi+1)}
\left( 
\mct_{\foh \foh}^{J-} -
\sqrt \frac{\ell_\pi}{\ell_\pi +2} \; \mct_{\foh \fth}^{J-}
\right) \; , 
\nonumber \\
E_{(\ell_\pi+1)-} &=& 
-\frac{1}{\sqrt{2 \avec k {\avec k}'} (\ell_\pi+1)}
\left( 
\mct_{\foh \foh}^{J-} +
\sqrt \frac{\ell_\pi +2}{\ell_\pi} \; \mct_{\foh \fth}^{J-}
\right) \; , 
\nonumber \\
E_{\ell_\pi+} &=& 
-\frac{1}{\sqrt{2 \avec k {\avec k}'} (\ell_\pi+1)}
\left( 
\mct_{\foh \foh}^{J+} -
\sqrt \frac{\ell_\pi}{\ell_\pi +2} \; \mct_{\foh \fth}^{J+}
\right) \; , 
\nonumber \\
S_{(\ell_\pi+1)-} &=& 
-\frac{1}{\sqrt{\avec k {\avec k}'}(\ell_\pi+1)}
\mct_{\foh 0}^{J-} \; , 
\nonumber \\
S_{\ell_\pi+} &=& 
-\frac{1}{\sqrt{\avec k {\avec k}'}(\ell_\pi+1)}
\mct_{\foh 0}^{J+} \; , 
\; .
\label{multipoles}
\eea

\subsubsection{Compton scattering}

Proceeding in the same way as in pion photoproduction, the interaction
matrix $T$ is sandwiched between incoming and outgoing multipole
states of Eq. \refe{photomultiheli} to project out the desired
multipole amplitudes:
\bea
\mct^{\stackrel{MM}{EE}}_{j_\gamma+} &=& 
+\frac{1}{2 (j_\gamma+1)}
\left[ 
  j_\gamma \; \mct_{\foh \foh}^{J\pm} +
  \sqrt{j_\gamma (j_\gamma +2)} 
  \left(
    \mct_{\fth \foh}^{J\pm} + \mct_{\foh \fth}^{J\pm}
  \right) +
  (j_\gamma + 2) \; \mct_{\fth \fth}^{J\pm}
\right] \; , 
\nonumber \\
\mct^{\stackrel{MM}{EE}}_{(j_\gamma+1)-} &=& 
+\frac{1}{2 (j_\gamma +1)}
\left[ 
  (j_\gamma +2) \; \mct_{\foh \foh}^{J\mp} -
  \sqrt{j_\gamma (j_\gamma +2)} 
  \left(
    \mct_{\fth \foh}^{J\mp} + \mct_{\foh \fth}^{J\mp}
  \right) +
  j_\gamma \; \mct_{\fth \fth}^{J\mp}
\right] \; , 
\nonumber \\
\mct^{\stackrel{ME}{EM}}_{j_\gamma+} &=& 
-\frac{1}{2 (j_\gamma +1)}
\left[ 
  \sqrt{j_\gamma (j_\gamma +2)} 
  \left(
  \mct_{\foh \foh}^{J\mp} - \mct_{\fth \fth}^{J\mp}
  \right) -
  j_\gamma \; \mct_{\fth \foh}^{J\mp} + 
  (j_\gamma +2) \; \mct_{\foh \fth}^{J\mp}
\right] \; , 
\nonumber \\
\mct^{\stackrel{ME}{EM}}_{(j_\gamma+1)-} &=& 
-\frac{1}{2 (j_\gamma +1)}
\left[ 
  \sqrt{j_\gamma (j_\gamma +2)} 
  \left(
  \mct_{\foh \foh}^{J\pm} - \mct_{\fth \fth}^{J\pm}
  \right) +
  (j_\gamma +2) \; \mct_{\fth \foh}^{J\pm} -
  j_\gamma \; \mct_{\foh \fth}^{J\pm}
\right] \; ,
\label{comptontmulti}
\eea
where the lower index of $\mct$ characterizes the incoming photon
state and the total spin thus is always chosen to be $J = j_\gamma +
\foh$. With the multipole normalization factor $[\avec k
  \sqrt{j_\gamma (j_\gamma +1)}]^{-1}$ and time reversal
symmetry ($\mct_{\lambda' \lambda}^J = \mct_{\lambda \lambda'}^J$) the 
Compton multipole amplitudes given in Eq. (A.6) of Pfeil {\it et
al.} \cite{pfeil} and Eq. (B16) of Ref. \cite{feusti99} are recovered.

\section{Isospin Decomposition}
\label{appiso}

The isospin decomposition of photon-induced reactions can be realized
by splitting up the photon into an isoscalar
$|I,I_z\rangle=|0,0\rangle$ and the third component of an isovector
$|I,I_z\rangle=|1,0\rangle$ particle. Taking into account this isospin 
ambivalence of the photon, all photon couplings can also be split up
and the isospin operators in the Lagrangians of Appendix \ref{appiso}
are identical to the hadronic reactions given in PMI \cite{pm1}; 
see also Ref. \cite{gregiphd}.

\subsection{Photoproduction of $(I = 1 \oplus \foh)$ final states}

The isospin ambivalence of the photon is introduced into the
isospin decomposition of the amplitude for photoproduction of $I= 1
\oplus \foh$ hadronic final states ($\pi N$, $\zeta N$, $K \Sigma$) by 
combining the equations for the isospin decomposition of $\pi N \ra
\pi N$ and $\pi N \ra \eta N$ (see PMI \cite{pm1}),
\bea
\langle \vp_k ; \; I=\sfoh | \; T_{f\gamma} \; 
| \gamma ; \; I=\sfoh \rangle = 
\langle I=\sfoh | \; \sfot \tau_k \tau_3 T^{\foh}_{f\gamma} + 
(\delta_{k3} - \sfot \tau_k \tau_3 ) T^{\fth}_{f\gamma} -
\smallfrac{1}{\sqth} \tau_k T^0_{f\gamma} \; 
| I=\sfoh \rangle \; , \hspace*{-5mm}
\nonumber \\
\eea
where $\langle \vp_k |$ refers to the outgoing asymptotic isospin-$1$
particle. The meaning of the upper indices is similar to the helicity
notation:
\begin{itemize}
\item{$0$: isoscalar photon coupling with the nucleon (total isospin
    $I=\foh$).}
\item{$\foh$: isovector photon coupling with the nucleon to total 
$I=\foh$.}
\item{$\fth$: isovector photon coupling with the nucleon to total 
$I=\fth$.}
\end{itemize}
This leads explicitly to the following amplitudes:
\bea
\langle 1, \; \; \, 0; \; \sfoh, +\sfoh | \; T_{f\gamma} \; 
| \gamma; \; \sfoh, +\sfoh \rangle &=& 
\fot ( 2 T^\fth_{f\gamma} + T^\foh_{f\gamma} ) - \frac{1}{\sqrt 3}
T^0_{f\gamma} \; ,
\nonumber \\ 
\langle 1, \; \; \, 0; \; \sfoh, -\sfoh | \; T_{f\gamma} \; 
| \gamma; \; \sfoh, -\sfoh \rangle &=& 
\fot ( 2 T^\fth_{f\gamma} + T^\foh_{f\gamma} ) + \frac{1}{\sqrt 3}
T^0_{f\gamma} \; ,
\nonumber \\ 
\langle 1, +1; \; \sfoh, -\sfoh | \; T_{f\gamma} \; 
| \gamma; \; \sfoh, +\sfoh \rangle &=& 
\frac{\sqt}{3} ( T^\fth_{f\gamma} - T^\foh_{f\gamma} ) +
\frac{\sqt}{\sqth} T^0_{f\gamma} \; ,
\nonumber \\ 
\langle 1, -1; \; \sfoh, +\sfoh | \; T_{f\gamma} \; 
| \gamma; \; \sfoh, -\sfoh \rangle &=& 
\frac{\sqt}{3} ( T^\fth_{f\gamma} - T^\foh_{f\gamma} ) -
\frac{\sqt}{\sqth} T^0_{f\gamma} 
\; . \label{ampliisophot32}
\eea
The so-called proton ($T^p_{\pi \gamma}$) and neutron ($T^n_{\pi
  \gamma}$) isospin amplitudes introduced in Ref. \cite{moorhouse} are 
commonly used amplitude combinations with total isospin $I = \foh$ and 
related to the above ones in the following ways:
\bea
T^p_{\pi \gamma} &\equiv& \sfot 
(-\sqt \langle \pi^+ n | T | \gamma p \rangle +
\langle \pi^0 p | T | \gamma p \rangle ) \, = 
+ \sfot T^\foh_{\pi \gamma} - \smallfrac{1}{\sqth} T^0_{\pi \gamma} \; ,
\nonumber \\
T^n_{\pi \gamma} &\equiv& \sfot 
(+\sqt \langle \pi^- p | T | \gamma n \rangle -
\langle \pi^0 n | T | \gamma n \rangle ) =
- \sfot T^\foh_{\pi \gamma} - \smallfrac{1}{\sqth} T^0_{\pi \gamma} 
\; . \nonumber
\eea

\subsection{Photoproduction of $(I = 0 \oplus \foh = \foh)$ final
  states}
\label{appisogampi}

For photoproduction of $I = 0 \oplus \foh = \foh$ hadronic final
states ($\eta N$, $K \Lambda$, $\omega N$) only a total isospin of $I
= \foh$ is allowed, and the result is a slight extension of the isospin
decomposition of $\pi N \ra \eta N$ (cf. PMI \cite{pm1}):
\bea
\langle I = 0 ; \; I=\sfoh | \; T_{f\gamma} \; | \gamma ; \; I=\sfoh \rangle = 
\langle I=\sfoh | \; T^0_{f\gamma}  - \smallfrac{1}{\sqth} \tau_3
T^{\foh}_{f\gamma} \; | 
I=\sfoh \rangle \; .
\nonumber
\eea
The resulting proton ($T^p_{f \gamma}$) and neutron ($T^n_{f \gamma}$)
isospin amplitudes are
\bea
T^p_{f\gamma} &\equiv& 
\langle 0, 0; \; \sfoh, +\sfoh | \; T_{f\gamma} \; | \gamma ; \; \sfoh,
+\sfoh \rangle = 
- \smallfrac{1}{\sqth} T^\foh_{f \gamma} + T^0_{f \gamma} \; ,
\nonumber \\
T^n_{f\gamma} &\equiv& 
\langle 0, 0; \; \sfoh, -\sfoh | \; T_{f\gamma} \; | \gamma ; \; \sfoh,
-\sfoh \rangle = 
+ \smallfrac{1}{\sqth} T^\foh_{f\gamma}+ T^0_{f\gamma} \; .
\label{isogampi}
\eea

\subsection{Compton scattering}

For Compton scattering, the incoming and outgoing photons are
decomposed into their isoscalar and isovector contributions. Thus
the isospin decomposition now reads
\bea
\langle \gamma ; I=\sfoh | \; T_{\gamma \gamma} \; | \gamma ; 
I=\sfoh \rangle =  
\langle I=\sfoh | \; 
T^{00}_{\gamma \gamma} - 
\smallfrac{1}{\sqth} \tau_3 (T^{01}_{\gamma \gamma} + T^{10}_{\gamma
  \gamma}) +
\sfot T^{11,\foh}_{\gamma \gamma} + \sftt T^{11,\fth}_{\gamma \gamma}
| I=\sfoh \rangle \hspace*{5mm}
\label{isocompton}
\eea
because of $\tau_3^2 = 1 \: \! \! \! 1_2$. The upper indices denote
the isospin of the outgoing and incoming photons. For the case when
both photons are isovectors ($^{11}$), the total isospin of the
$\gamma N$ system is also given.

Experimentally, only two amplitudes ($\gamma p \ra \gamma p$ 
and $\gamma n \ra \gamma n$) are accessible. For these cases Eq.
\refe{isocompton} results in
\bea
\langle \gamma ; p | \; T_{\gamma \gamma} \; | \gamma ; p \rangle &=&
T^{00}_{\gamma \gamma} - 
\smallfrac{1}{\sqth} (T^{01}_{\gamma \gamma} + T^{10}_{\gamma \gamma})
+ \sfot T^{11,\foh}_{\gamma \gamma} 
+ \sftt T^{11,\fth}_{\gamma \gamma} \; ,
\nonumber \\
\langle \gamma ; n | \; T_{\gamma \gamma} \; | \gamma ; n \rangle &=&
T^{00}_{\gamma \gamma} +
\smallfrac{1}{\sqth} (T^{01}_{\gamma \gamma} + T^{10}_{\gamma \gamma})
+ \sfot T^{11,\foh}_{\gamma \gamma} 
+ \sftt T^{11,\fth}_{\gamma \gamma} \; .
\label{isocomptonpn}
\eea

\section{Observables}
\label{appobs}

\subsection{Cross sections}

The differential cross section
\bea
\frac{d \sigma}{d \Omega} &=& 
\frac{(4\pi)^2}{\avec k^2} \frac{1}{s_i}
\sum \limits_{\lambda , \lambda'} 
\left| \mct_{\lambda' \lambda} (\vt) \right|^2
\label{difcross}
\eea
with (e.g., for $\lambda,\lambda' > 0$)
\bea
\mct_{\lambda' \lambda} (\vt) 
&=& \frac{1}{4 \pi}
\sum \limits_J (J + \sfoh) d^J_{\lambda \lambda'} (\vt) 
\left( 
\mct_{\lambda' \lambda}^{J+} + \mct_{\lambda' \lambda}^{J-}
\right)
\label{amplitdif}
\eea
and total cross section formulae
\bea
\sigma = \frac{4 \pi}{\avec k^2} 
\frac{1}{s_i} \sum_{J,P} \sum_{\lambda ,\lambda'} (J + \sfoh)
\left| \mct_{\lambda' \lambda}^{JP} \right|^2
\label{totalcross}
\eea
are completely identical to the hadronic reactions given in PMI
\cite{pm1}. $s_i$ in Eq. \refe{totalcross} is the usual spin
averaging factor for the initial state. Note that while in
Eq. \refe{difcross} the sum runs over all values for $\lambda$ and
$\lambda'$, in Eq. \refe{totalcross} the second sum extends only over
positive $\lambda$ and $\lambda'$.

The reduced cross section in $\eta$ photoproduction is:
\bea
\sigma_{red} = \sqrt{\frac{\sigma}{4 \pi }\frac{\avec k}{\avec k'}} = 
\sqrt {
\frac{1}{\avec k \avec k'} \frac{1}{s_i} 
\sum_{J,P} \sum_{\lambda ,\lambda'} (J + \sfoh)
\left| \mct_{\lambda' \lambda}^{JP} \right|^2 } \; .
\nonumber
\eea

\subsection{Polarization observables}

With the cross section intensity
\bea
\mci (\vt) \equiv 
\foh \sum \limits_{\lambda , \lambda'} 
\left| \mct_{\lambda' \lambda} (\vt) \right|^2 \; ,
\label{crossint}
\eea
where the sum extends over all possible values for $\lambda$ and 
$\lambda'$, the polarization observables are given in the following
way: 

\subsubsection{Photoproduction of (pseudo-) scalar mesons}

The single polarization observables are given by
\bea
\mci (\vt) \Sigma &=&
2 \mbox{Re} \left( 
\mct_{\foh \fth} \mct_{\foh -\foh}^* +
\mct_{\foh \foh} \mct_{\foh -\fth}^*
\right)
\hspace{3mm} \mbox{photon asym.} \; ,
\nonumber \\
\mci (\vt) \mcp &=& 
2 \mbox{Im} \left( 
\mct_{\foh \fth} \mct_{\foh -\fth}^* -
\mct_{\foh \foh} \mct_{\foh -\foh}^*
\right)
\hspace{3mm} \mbox{recoil asym.} \; ,
\\
\mci (\vt) \mct &=&
2 \mbox{Im} \left( 
\mct_{\foh \fth} \mct_{\foh \foh}^* -
\mct_{\foh -\fth} \mct_{\foh -\foh}^*
\right)
\hspace{3mm} \mbox{target asym.} \; .
\nonumber
\eea

\subsubsection{Compton scattering}

The single polarization observables are given by
\bea
\mci (\vt) \Sigma &=&
2 \mbox{Re} \left( 
\left( \mct_{\fth \fth} + \mct_{\foh \foh} \right)^* \mct_{\foh -\fth}
+
\left( \mct_{\fth -\fth} - \mct_{\foh -\foh} \right)^* \mct_{\fth \foh}
\right) \; ,
\nonumber \\
\mci (\vt) \mct &=& 
2 \mbox{Im} \left( 
\left( \mct_{\fth \fth} + \mct_{\foh \foh} \right)^* \mct_{\fth \foh}
-
\left( \mct_{\fth -\fth} - \mct_{\foh -\foh} \right)^* \mct_{\foh -\fth}
\right)
= \mci (\vt) \mcp
\eea
for the photon and target/recoil asymmetry, respectively.

\subsubsection{Photoproduction of vector mesons}

The single polarization observables are given by
\bea
\mci (\vt) \Sigma &=&
2 \mbox{Re} \left( +
\mct_{\fth \fth}^* \mct_{\fth -\foh} +
\mct_{\foh \foh}^* \mct_{\foh -\fth} + 
\mct_{\fth -\fth}^* \mct_{\fth \foh} +
\mct_{\foh -\foh}^* \mct_{\foh \fth} +
\mct_{0 \fth}^* \mct_{0 -\foh} +
\mct_{0 -\fth}^* \mct_{0 \foh}
\right) \; ,
\nonumber \\
\mci (\vt) \mct &=& 
2 \mbox{Im} \left( +
\mct_{\fth \fth}^* \mct_{\fth \foh} -
\mct_{\foh \foh}^* \mct_{\foh \fth} -
\mct_{\fth -\fth}^* \mct_{\fth -\foh} +
\mct_{\foh -\foh}^* \mct_{\foh -\fth} +
\mct_{0 \fth}^* \mct_{0 \foh} -
\mct_{0 -\fth}^* \mct_{0 -\foh}
\right) \; ,
\\
\mci (\vt) \mcp &=& 
\nonumber
2 \mbox{Im} \left( -
\mct_{\fth \fth}^* \mct_{\foh \fth} +
\mct_{\foh \foh}^* \mct_{\fth \foh} -
\mct_{\fth -\fth}^* \mct_{\foh -\fth} +
\mct_{\foh -\foh}^* \mct_{\fth -\foh} -
\mct_{0 \fth}^* \mct_{0 -\fth} +
\mct_{0 \foh}^* \mct_{0 -\foh}
\right)
\eea
for the photon and target/recoil asymmetry, respectively. The vector
meson and some double polarization observables can be found in
Appendix B of Ref. \cite{titov98}.

\end{appendix}

\end{document}